\documentclass[12pt]{article}
\usepackage{amsmath}
\usepackage{graphicx}
\usepackage{enumerate}
\usepackage{natbib}
\usepackage{url} % not crucial - just used below for the URL 
\usepackage{hyperref}
\usepackage{caption}
\usepackage{subcaption}
\usepackage{footnote}
\usepackage{float}
\usepackage{bm}
\usepackage{authblk}
\usepackage{enumitem}
\usepackage{adjustbox}
\usepackage{placeins}
\usepackage{setspace}
\usepackage{algorithm}
\usepackage{algpseudocode}
\usepackage{booktabs} 
\usepackage{xurl}
\usepackage{xr}
\makeatletter

\newcommand*{\addFileDependency}[1]{% argument=file name and extension
\typeout{(#1)}% latexmk will find this if $recorder=0
\@addtofilelist{#1}
\IfFileExists{#1}{}{\typeout{No file #1.}}
}\makeatother

\newcommand*{\myexternaldocument}[1]{%
\externaldocument{#1}%
\addFileDependency{#1.tex}%
\addFileDependency{#1.aux}%
}
\myexternaldocument{JASA_supplement}

\usepackage{array}
\newcommand{\blind}{1}

\RequirePackage{amsthm,amsmath,amsfonts,amssymb,mathtools}
\newtheorem{Theorem}{Theorem}%[section]

\DeclareMathOperator*{\argmax}{arg\,max}
\DeclarePairedDelimiter\ceil{\lceil}{\rceil}
\DeclarePairedDelimiter\floor{\lfloor}{\rfloor}

{
    \theoremstyle{remark} 
    
}

\def\ind[#1]{\mathbb{I}\left\lbrace#1\right\rbrace}
\def\Fnull[#1#2]{{F}_{Y_{#1}}(#2)}
\def\prehatF[#1#2]{\hat{F}^{(u)}_{Y_{#1}}(#2)}
\def\posthatF[#1#2]{\hat{F}^{(1-u)}_{Y_{#1}}(#2)}
\def\denom[#1]{\frac{1}{#1}}
\def\add[#1#2]{\sum_{#1=1}^#2}
\def\addneq[#1#2]{\sum_{#1\neq #2}}
\def\Fone[#1]{{F}^{(1)}_{{#1}}}
\def\Ftwo[#1]{{F}^{(2)}_{{#1}}}
\newcommand{\M}{\mathcal{M}}
\newcommand{\idx}{\mathcal{I}}
\newcommand{\R}{\mathbb{R}}
\newcommand{\intgrt}{\int_0^{\mathcal{M}}}
\newcommand{\rem}{O\left(\frac{1}{n}\right)}
\def\prob{\mathbb{P}}
\def\E{\mathbb{E}}

\def\diam{\mathrm{diam}}

\newcommand{\Break}{\State \textbf{break} }

\begin{document}

\def\spacingset#1{\renewcommand{\baselinestretch}%
{#1}\small\normalsize} \spacingset{1}

%%%%%%%%%%%%%%%%%%%%%%%%%%%%%%%%%%%%%%%%%%%%%%%%%%%%%%%%%%%%%%%%%%%%%%%%%%%%%%

\if1\blind
{
  \title{\bf Change Point Inference for Non-Euclidean Data Sequences using Distance Profiles}

  \author[1]{Paromita Dubey\thanks{Paromita Dubey gratefully acknowledges support from NSF grant DMS-2311034.}}
  % \author[2]{Minxing Zheng}
  \author[2]{Minxing Zheng\thanks{This work was primarily conducted while Minxing Zheng was at the University of Southern California.}}
  \affil[1]{Department of Data Sciences and Operations, University of Southern California}
  \affil[2]{Heinz College of Information Systems and Public Policy, Carnegie Mellon University}

  \date{}
  \maketitle
} \fi

\if0\blind
{
  \bigskip
  \bigskip
  \bigskip
  \begin{center}
    {\LARGE\bf Change Point Detection for Random Objects\\
    using Distance Profiles}
  \end{center}
  \medskip
} \fi
\vspace{-2em}
\begin{abstract}
We introduce a powerful scan statistic and the corresponding test for detecting the presence and pinpointing the location of a change point within the distribution of a data sequence with the data elements residing in a separable metric space $(\Omega, d)$. These change points mark abrupt shifts in the distribution of the data sequence as characterized using distance profiles, where the distance profile of an element $\omega \in \Omega$ is the distribution of distances from $\omega$ as dictated by the data. This approach is tuning parameter free, fully non-parametric and universally applicable to diverse data types, including distributional and network data, as long as distances between the data objects are available. We obtain an explicit characterization of the asymptotic distribution of the test statistic under the null hypothesis of no change points, rigorous guarantees on the consistency of the test in the presence of change points under fixed and local alternatives and  near-optimal convergence of the estimated change point location, all under practicable settings. To compare with state-of-the-art methods we conduct simulations covering multivariate data, bivariate distributional data and sequences of graph Laplacians, and illustrate our method on real data sequences of the U.S. electricity generation compositions and Bluetooth proximity networks.
\end{abstract}

\noindent%

\spacingset{1.75} % DON'T change the spacing!

\vspace{-1em}

\section{Introduction}

\vspace{-1em}

With origins dating back to the 1950s \citep{page:1954,page:1955}, change point analysis has remained a thriving research area in statistics fueled by its burgeoning relevance in diverse domains such as biology and medicine \citep{erdm:2008, Mugg:2010, chun:06}, economics and finance \citep{lavi:07,thie:18}, neurociences \citep{stoe:21,crib:17}, social sciences \citep{koss:06,shar:16}, climate and environmental studies \citep{lund:23}, and more recently in the context of COVID-19 \citep{dehn:20,jian:23}, to name but a few; see \citep{chen:12, amin:17,truo:20} for recent reviews.  The primary objective of change point detection is to identify and locate any abrupt alteration in the data generating mechanism within an observed data sequence indexed by time or another meaningful order, to be represented as ${Y_1,\dots,Y_n}$. We consider the offline scenario, where the data sequence is of fixed length. A change point, denoted as $n_\tau$, is characterized by the transition from the distribution $P_1$ governing ${Y_1,\dots,Y_{n_\tau}}$ to another distribution $P_2$ governing ${Y_{n_\tau+1},\dots,Y_n}$ where $P_1 \neq P_2$.

In cases where the observations $Y_i$ reside in the Euclidean space $\mathbb{R}^p$, the problem's intricacies are influenced by the choice of the dimensionality $p$.
While the univariate scenario has been thoroughly studied \citep{carl:94,niu:16,wang:20}, additional challenges encountered in the multivariate case have triggered developments in both parametric \citep{jame:87, James:1992, zhan:10} as well as non-parametric frameworks \citep{matt:2014, lung:2015, jira:15, Padilla:2022}.
In the high-dimensional setting, the problem becomes significantly more intricate and elusive due to the curse of dimensionality. 
Specialized investigations targeting high-dimensional scenarios have emerged, including works by \citet{wang:18,enik:19,liu:21,wang:22}.

In modern data science, it is becoming increasingly common to encounter data that do not lie in an Euclidean space. Such data elements, often referred to as ``random objects", extend the traditional concept of random vectors into the realm of general metric spaces. Common examples include brain networks \citep{spor:22}, gene regulation networks \citep{nie:2017}, linguistic object data \citep{tava:19}, distributional data \citep{mata:21}, compositional data \citep{gloo:16}, phylogenetic trees datasets \citep{holm:03,kim:20} and many more. The bottleneck in working with such data lies in the absence of standard vector space operations, leaving us primarily with pairwise distances between these objects as our basis for analysis. 

Methods to tackle change point analysis in these settings have evolved simultaneously. In addition to approaches designed for special cases like network data \citep{wang:21}, distributional sequences \citep{horv:21}, compositional data \citep{kj:21} etc, which are not applicable more generally, there has been a surge in fully non-parametric approaches that can be placed in one of many broad categories such as distance-based \citep{matt:2014}, kernel-based \citep{harc:08, li:2015, Arlo:2019} and graph-based \citep{shi:17,chen:23}. Nevertheless, each category of methods has its own set of limitations. Theoretical guarantees in \cite{matt:2014} are derived for multivariate data sequences.  In the case of kernel-based methods, critical decisions such as choosing the appropriate kernel and setting parameters like bandwidth and penalty constants can significantly impact their result but are often challenging to determine in practice.  On the other hand, the performance of graph-based methods is highly contingent on choosing from various graph construction methods, a choice that is often difficult to make. Recently, \citet{Dube:2020} proposed a tuning-free approach based on Fr\'echet means and variances in general metric spaces, however this test is not powerful against changes beyond Fr\'echet means and variances of the data.

In this paper, we propose an off-the-shelf, tuning parameter-free (except for a cut-off interval at the end-points where change points are assumed not to occur) non-parametric change point detection method which comes with rigorous type I error control guarantees, even when using permutation cutoffs, guaranteed consistency under fixed and local alternatives and an optimal rate of convergence (up to $\log$ terms) for the estimated change point. Our approach hinges on distance profiles \citep{dube:24}, where the distance profile of a data element is the distribution of distances from that element as governed by the data.  We demonstrate the broad applicability and exceptional finite-sample performance of our method through extensive simulations covering various types of multivariate data, bivariate distributional data, and network data across a variety of scenarios. We illustrate our method on two real-world datasets: Bluetooth proximity networks in the MIT reality mining study and U.S. electricity generation compositional data. 

The organization of the paper is as follows. In Section \ref{sec:pre}, we first describe the problem's setup, introducing the distance profiles and presenting our scan statistic. The rest of Section \ref{sec:pre} is dedicated to laying out the methodology and the theoretical foundations of our proposed test, including a precise characterization of the asymptotic distribution of the scan statistic under the null hypothesis of no change points, analysis of the power of the test under local alternatives and establishing rates of convergence for the estimated change point. Moving on to Section \ref{sec: simulation}, we introduce the various simulation settings including exploration of different types of random objects and change points scenarios. The performance of our test is illustrated in real-world applications, namely, the MIT Reality Mining networks and the U.S. electricity generation compositional data, in Section \ref{sec: applicaiton}. In Section \ref{sec:multCP} we explore the adaptation of our method utilizing seeded binary segmentation \citep{kova:2023} to scenarios involving multiple change points, which we illustrate using the set up of stochastic block models with multiple change points. We conclude with Section \ref{sec:discussion} where we discuss the capacities of this new method, limitations and avenues for future extensions. 

%This is an example of a new parapgraph with a numbered footnote\footnote{\url{https://data.gov.uk/}} and a second footnote marker.\footnote{Example of footnote text.}
\vspace{-1em}

\section{Methodology} \label{sec:pre}

\vspace{-1em}

\subsection{Distance profiles of random objects}

\vspace{-0.5em}

Distance profile, introduced in \citep{dube:24}, is a simple yet powerful device for analyzing random objects in metric spaces. Let $(\Omega,d)$ be a separable metric space. Consider a probability space $(S,\mathcal{S},\mathbb{P})$, where $\mathcal{S}$ is the Borel sigma algebra on a domain $S$ and $\mathbb{P}$ is a probability measure. A random object $X$ is a measurable function, $X: S \to \Omega$ and $P$ is a Borel probability measure on $\Omega$ that is induced by $X$, i.e. $P(A) = \mathbb{P}(\{s \in S: X(s)\in A\})=:\mathbb{P}(X \in A)=\mathbb{P}(X^{-1}(A))=:\mathbb{P} \circ X^{-1}(A)$, for any Borel measurable set $A \subseteq \Omega$. For any point $\omega \in \Omega$ its distance profile is the cumulative distribution function (cdf) of the distance between $\omega$ and the random object $X$ that is distributed according to $P$. Formally we define the distance profile at $\omega$ as
\begin{equation*}
    F_\omega(t) = \mathbb{P}(d(\omega,X) \leq t), \quad t \in \mathbb{R}.
\end{equation*}
We suppress the dependence of $F_\omega$ on $P$ to keep the notation simple. Intuitively, if an element $\omega$ is more centrally located, i.e. closer to most other elements, it will have a distance profile with more mass near 0 unlike points which are distantly located from the data. With a sequence of independent observations $X_1,\dots,X_n$ from $P$, we estimate the distance profile at $\omega$ as 
\begin{equation*}
    \hat{F}_{\omega}(t)=\frac{1}{n} \sum_{j=1}^n \ind[d(\omega,X_j) \leq t], \quad t \in \mathbb{R}.
\end{equation*} 
The collection of the distance profiles $\{F_\omega:\omega \in \Omega\}$ comprises the one-dimensional marginals of the stochastic process $\{d(\omega,X)\}_{\omega \in \Omega}$ and serve as distinctive descriptors of the underlying Borel probability measure $P$ whenever $P$ can be characterized uniquely by open balls. Under special conditions on $(\Omega,d)$, for example if $d^r$ is of strong negative type for some $r>0$ \citep{lyon:13}, this unique characterization holds for all Borel probability measures; see Proposition 1 in \cite{dube:24}. Motivated by the framework introduced in \cite{dube:24} our objective is to leverage these elementary distance profiles for detecting change points in the intricate distribution of a random object sequence.

\vspace{-1em}

\subsection{Change point detection problem}

\vspace{-0.5em}

\label{sec:problem_setup}
Let $Y_1,Y_2,\dots,Y_n$ be a sequence of random objects taking values in a separable metric space $(\Omega,d)$ with a finite covering number. Given two different Borel probability measures $P_1$ and $P_2$ on $\Omega$, we will test the null hypothesis,
\begin{equation}
\label{eq: null}
    H_0: Y_1,Y_2,\dots,Y_n \sim P_1
\end{equation}
against the single change point alternative
\iffalse
\begin{equation}
    \begin{aligned}
\label{eq: alt}
    H_1: & \exists \ \tau \  \in (0,1) \textup{ such that} \\ & \Bigl\{ 
                 \begin{array}{ll}
                  Y_1,Y_2,\dots,Y_{[n\tau]} \sim P_1 \\
                  Y_{[n\tau]+1},Y_{[n\tau]+2},\dots,Y_{[n]} \sim P_2\\
                \end{array}
              \Bigr.
    \end{aligned}
\end{equation}
\fi
\begin{equation}
    \begin{aligned}
\label{eq: alt}
    H_1: & \exists \ \tau \  \in (0,1) \textup{ such that}  
    & \begin{cases}
        Y_1,Y_2,\dots,Y_{[n\tau]} \sim P_1 \\
                  Y_{[n\tau]+1},Y_{[n\tau]+2},\dots,Y_{n} \sim P_2\\
    \end{cases}
    \end{aligned}
\end{equation}
where $\tau$ denotes the change point. Our aim is to test the above hypothesis and accurately identify $\tau$ when it exists. In keeping with traditional change point methods, we will employ a scan statistic that involves dividing the data sequence into two segments, one before and one after the potential change points. In this process, the test statistic seeks to maximize the dissimilarities between these segments enabling subsequent inference and estimation of the change points if any. %We quantify the dissimilarity between the data segments in a tuning free way and target a divergence between the underlying population distributions in large samples (refer to the quantity $\Delta$ as defined in equation \eqref{eq: divergence} in \citep{dube:24}) whenever the distributions are such that they can be uniquely identified using the distance profiles corresponding to all the elements in $\Omega$, thereby capturing a broad range of alternatives. 
To ensure the validity of large-sample analysis, it is important that both segments contain a minimum number of observations so that we can accurately capture the dissimilarity between them.  Consequently, we make the assumption that the change point $\tau$ lies in a compact interval $\mathcal{I}_c=[c,1-c] \subset [0,1]$, for some $c>0$.

\vspace{-1em}

\subsection{Scan statistic and type I error control} 
\label{sec: test_construction}

\vspace{-0.5em}

\begin{figure*}[!h]%
    \centering
    \includegraphics[width = \textwidth]{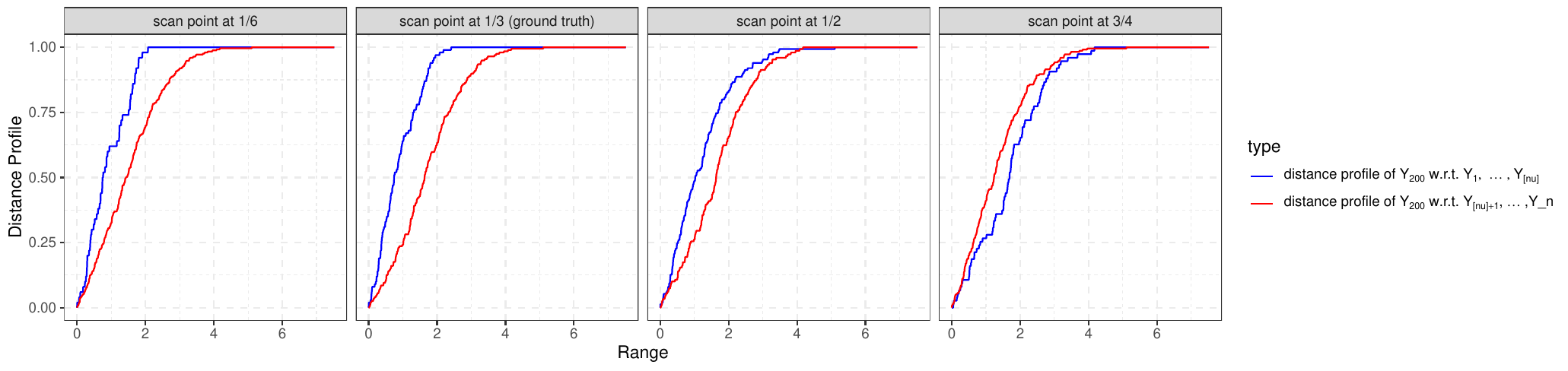}
    \caption{{The distance profiles of $Y_{200}$ in the sequence of observations $Y_i, \ i=1,\dots,300$, where $Y_i \sim N(0,1),\ i=1,\dots,100$ and $Y_i \sim N(2,1),\ i=101,\dots,300$ with respect to $Y_1,\dots,Y_{[nu]}$ and $Y_{[nu]+1},\dots,Y_{n}$ at different scan points $u=\frac{1}{6}, \frac{1}{3} \text{(the change point)}, \frac{1}{2}, \frac{3}{4}$.}}
    \label{fig:toy_example}
\end{figure*}

While scanning the data sequence segmented at $u \in \idx_c$, let $\prehatF[it]$ be the estimated distance profile of the observation $Y_i$ with respect to the data segment $Y_1,\dots,Y_{[nu]}$ given by
\begin{equation*}
    \prehatF[it] = \frac{1}{[nu]} \sum_{j=1}^{[nu]} \ind[d(Y_i,Y_j) \leq t], \quad t \in \R.
\end{equation*}
Similarly, $\posthatF[it]$, defined with respect to the data segment $Y_{[nu]+1},\dots,Y_{n}$, is given by
\begin{equation*}
\begin{split}
    \posthatF[it] = \frac{1}{(n-[nu])} \sum_{j=[nu]+1}^{n} \ind[d(Y_i,Y_j) \leq t],  \quad t \in \R.
\end{split}
\end{equation*} 
To capture the discrepancy between the data segments $Y_1,\dots,Y_{[nu]}$ and $Y_{[nu]+1},\dots,Y_n$ we propose the statistic given by
\begin{equation}
    \label{eq: scan_statistic}
    \begin{split}
        \hat{T}_n(u) =  \frac{[nu](n-[nu])}{n^2}  \left\lbrace \frac{1}{n} \add[in] \int_0^\M (\prehatF[it]  - \posthatF[it])^2 dt \vphantom{\int_0^\M} \right\rbrace
    \end{split}
\end{equation}
where $\M=\diam(\Omega)$ is the diameter of $\Omega$. The motivation to investigate this scan statistic is that if the two segments $Y_1,\dots,Y_{[nu]}$ and $Y_{[nu]+1},\dots,Y_n$ have different distributions, then the centrality of an observation $Y_i$, as encoded in the distance profiles, will be different across the two segments, and as a result $\hat{T}_n(u)$ tends to be large when $u$ is close to $\tau$ when the change point $\tau$ exists. {We demonstrate this using a toy example in Figure \ref{fig:toy_example}.}
Hence to test the null hypothesis $H_0$ \eqref{eq: null}, we will use the test statistic
\begin{equation} 
\label{eq: test_stat}
    \hat{T}_n = \sup_{u \in \mathcal{I}_c} n\hat{T}_n(u) = \max_{[nc]\leq k \leq n-[nc]} n\hat{T}_n\left(\frac{k}{n}\right).
\end{equation}

In order to construct an asymptotic level $\alpha$ test we derive the distribution of the test statistic \eqref{eq: test_stat} under $H_0$ \eqref{eq: null} in Theorem~\ref{thm: asymptotic_distribution} under the following assumptions. 

\begin{enumerate}[label = (A\arabic*)]
    \item  \label{ass:dpfctn} Let $F_\omega^{(1)}(t)=\prob(d(\omega,Y)\leq t)$, where $Y \sim P_1$, and $F_\omega^{(2)}(t)=\prob(d(x,Y')\leq t)$, where $Y' \sim P_2$. Assume that $F_\omega^{(1)}(t)$ and $F_\omega^{(2)}(t)$ are absolutely continuous for each $\omega \in \Omega$, with densities given by $f_\omega^{(1)}(t)$ and $f_\omega^{(2)}(t)$ respectively. Assume that there exists $L_1, L_2 > 0$ such that
$\sup_{\omega \in \Omega} \sup_{t \in \mathbb{R}} |f_\omega^{(1)}(t)| \leq L_1$ and $\sup_{\omega \in \Omega} \sup_{t \in \mathbb{R}} |f_\omega^{(2)}(t)| \leq L_2$. Moreover assume that $\inf_{t \in \mathrm{supp}(f^{(1)}_\omega) }f^{(1)}_\omega(t) > 0$ and $\inf_{t \in \mathrm{supp}(f^{(2)}_\omega) }f^{(2)}_\omega(t) > 0$ for each $\omega \in \Omega$.

\item \label{ass:entropy} Let $N(\epsilon, \Omega, d)$ be the covering number of the space $\Omega$ with balls of radius $\epsilon$ and $\log N(\epsilon, \Omega, d)$ the corresponding metric entropy, which satisfies 
    \begin{equation} \label{entropy} 
        \epsilon \log N(\epsilon, \Omega, d) \rightarrow 0\quad  \text{as} \quad  \epsilon \rightarrow 0.
    \end{equation}
\end{enumerate}

Assumption \ref{ass:dpfctn} imposes regularity conditions on the distance profiles under the distributions $P_1$ and $P_2$ and covers a wide array of data generating processes in general metric spaces. While assumption \ref{ass:dpfctn} does not accommodate discrete distributions on metric spaces, addressing the particular challenges associated with discrete settings is beyond the scope of this paper and presents avenues for future research. Assumption \ref{ass:entropy} constrains the complexity of the metric space $(\Omega,d)$ and is applicable to a broad range of spaces including any space $\Omega$ which can be represented as a subset of elements in a finite dimensional Euclidean space, for example networks \citep{kola:20,gine:17}, simplex valued objects in a fixed dimension \citep{jeon:20,chen:12:2} and the space of phylogenetic trees with the same number of tips \citep{kim:20,bill:01}. Assumption \ref{ass:entropy} holds for any $\Omega$ which is a VC-class of sets or a VC-class of functions \citep[Theorems~2.6.4 and 2.6.7,][]{well:96}, for  $p$-dimensional smooth function classes $C_1^\alpha(\mathcal{X})$ \citep[page 155,][]{well:96} on bounded convex sets $\mathcal{X}$ in $\mathbb{R}^p$ equipped with the $\|\cdot\|_\infty$-norm \citep[Theorem~2.7.1,][]{well:96} and $\|\cdot\|_{r,Q}$-norm %which is the $L_r(Q)$ 
for any probability measure $Q$ on $\mathbb{R}^p$ \citep[Corollary~2.7.2,][]{well:96} if $\alpha \geq p+1$ and for the case when $\Omega$ is the space of one-dimensional distributions on some compact interval $I \subset \mathbb{R}$ that are absolutely continuous with respect to the Lebesgue measure on $I$ with smooth uniformly bounded densities and $d=d_W$ with $d_W$ being the 2-Wasserstein metric \citep{dube:24}. In fact Assumption \ref{ass:entropy} is satisfied when $\Omega$ is the space of $p$-dimensional distributions on a compact convex set $I \subset \mathbb{R}^p$, represented using their distribution functions endowed with the {$L_{r}$} metric with respect to the Lebesgue measure on $I$ if $\Omega \subset C_1^\alpha(I)$ for $\alpha \geq p+1$. Next, we present Theorem \eqref{thm: asymptotic_distribution} which establishes the null distribution of the test statistic \eqref{eq: test_stat} as $n \rightarrow \infty$ as the law of the random variable $\mathcal{T}$ as introduced in the same theorem. 
 
\begin{Theorem}
\label{thm: asymptotic_distribution}
Under $H_0$ and assumptions \ref{ass:dpfctn} and \ref{ass:entropy}, as $n \rightarrow \infty$, $\hat{T}_n$ converges in distribution to the law of a random variable $ \mathcal{T}=\sup_{u \in \idx_c} \sum_{j=1}^\infty \mathbb{E}_Y\{\lambda_j^Y\}\mathcal{G}_j^2(u)$, where $Y \sim P_1$, $\lambda^x_1 \geq \lambda^x_2 \geq \dots$ correspond to the eigenvalues of the covariance function given by $C_x(t_1,t_2) = \mathrm{Cov} \left( \ind[d(x,Y)\leq t_1],\ind[d(x,Y)\leq t_2]\right)$ and $\mathcal{G}_1, \mathcal{G}_2, \dots$ are independent zero mean Gaussian processes with covariance given by  $c(u_1,u_2)=\sqrt{\frac{(1-u_1)(1-u_2)}{u_1u_2}} \min(u_1,u_2)$+ $\sqrt{\frac{u_1u_2}{(1-u_1)(1-u_2)}} \min(1-u_1,1-u_2)$.
\end{Theorem}

 Theorem \ref{thm: asymptotic_distribution} presents significant technical challenges beyond the results established in \citep{dube:24} for two sample testing. Here we discuss the major bottleneck that we needed to overcome in proving Theorem \ref{thm: asymptotic_distribution} and relegate the technical details to the Supplement. Since the goal is to have a sample-splitting free scan statistic,  we use the same observations $Y_1, \dots, Y_n$ to estimate the distance profiles and thereafter to estimate the scan statistic that makes each summand in \eqref{eq: scan_statistic} significantly dependent on each other. To obtain the null distribution in Theorem \ref{thm: asymptotic_distribution} we decompose the test statistic \eqref{eq: test_stat} into several parts, some of which contribute to the asymptotic null distribution while we show that the others are asymptotically negligible using L\'evy-type maximal inequalities \citep{eich:01} and decoupling results for $U$-processes \citep{gine:97}.

Under $H_0$, the test statistic \eqref{eq: test_stat} converges to the supremum of a weighted infinite sum of independent squared Gaussian processes. While the supremum of sums of independent squared Brownian Bridges has appeared in the null asymptotics for detecting change points in functional data sequences \citep{asto:12}, our scenario involves a weighted sum,  where the weights are contingent on the data distribution. In fact, the null distribution is a large sample approximation of finite maximums of weighted infinite sums of chi-squares given by the law of the random variable $\max_{[nc] \leq k \leq (n-[nc])}\sum_{j=1}^\infty \E_{Y}\{\lambda_j^Y\} \xi^2_{jk}$ as $n \rightarrow \infty$, where $\mathbf{\xi_j}=(\xi_{j[nc]}, \dots, \xi_{j(n-[nc])})$, $j=1,2,\dots$ are independent zero mean Gaussian random vectors with $\mathrm{Cov}(\xi_{jr},\xi_{js})=c\left(\frac{r}{n},\frac{s}{n}\right)$ (see page 18 of the Supplement for further details) and $c(\cdot,\cdot)$ as defined in Theorem \ref{thm: asymptotic_distribution}. 

Ideally for a level $\alpha$ test, one should propose to reject $H_0$ \eqref{eq: null} if $\hat{T}_n > q_{\alpha}$ where $q_\alpha$ is the $(1-\alpha)$-quantile of $\mathcal{T}$. Equivalently one must reject $H_0$ when ${p} \leq \alpha$ where ${p}=\prob_{H_0}(\mathcal{T} \geq \hat{T}_n)$ is the asymptotic p value of the test. Since the law of $\mathcal{T}$ depends on the underlying data distribution and the rejection region, either using the critical value $q_\alpha$ or the p value ${p}$ must be approximated in practice. Staying in line with adopted conventions in the literature, we adopt a random permutation scheme to obtain this approximation that is described next and later design a framework in Section \ref{sec: power} to study the power of the test using the practicable permutation approximation scheme in large samples.

Let $\Pi$ denote the collection of $n!$ permutations of $\{1,2, \dots, n\}$. Let $\pi_0=(\pi_0(1), \dots, \pi_0(n))$ be the identity permutation such that $\pi_0(j)=j$ for $j=1,\dots,n$ and $\pi_1, \pi_2, \dots, \pi_K$ be $K$ i.i.d samples from the uniform distribution over $\Pi$ where each $\pi_k=(\pi_k(1), \dots, \pi_k(n))$ is a permutation of $\{1,\dots,n\}$. For each $k=0,1,2,\dots$, let $\hat{T}^{\pi_k}_n$ be the test statistic evaluated on a reordering of the data given by $Y_{\pi_k(1)}, \dots, Y_{\pi_k(n)}$. Then the permutation p value is given by $\hat{p}_K = \frac{1}{K+1} \sum_{k=0}^K \ind[\hat{T}^{\pi_k}_n \geq \hat{T}_n]$ and the test is rejected at level $\alpha$ when $\hat{p}_K \leq \alpha$. It is easy to see that under $H_0$ \eqref{eq: null}, $\hat{p}_K$ as an approximation of $p$ controls the type I error of the test at level $\alpha$ with high probability for sufficiently large $K$ (see \cite{chun:16}). We will show in Theorem \ref{thm: consistency_of_test} that as $n \rightarrow \infty$, the rejection region determined by  $\hat{p}_K$, the permutation p value, leads to a consistent test even under local alternatives close to $H_0$.

\vspace{-1em}

\subsection{Power analysis under local alternatives}
\label{sec: power}

\vspace{-0.5em}

We will study the large sample power of the test, first assuming that the oracle asymptotic critical value $q_\alpha$ is available, and thereafter using the practicable permutation scheme for the test. A few definitions are in order before we can state our results. For $t \in \mathbb{R}$ and a random object $Y$, where it is possible that either $Y \sim P_1$ or $Y \sim P_2$, let $\Fone[Y](t)= \prob_{Y'} \left( d(Y, Y') \leq t\right)$ with $Y' \sim P_1$ and $\Ftwo[Y](t)= \prob_{Y''} \left( d(Y, Y'') \leq t\right)$ with $Y'' \sim P_2$ and both $Y'$ and $Y''$ are independent of $Y$. Define the quantity $\Delta=\Delta(P_1, P_2)$ as
\begin{equation}
\begin{split}
    \label{eq: divergence}
    \Delta = \E_{Y \sim P_1} \left( \int_0^\mathcal{M} \left\lbrace \Fone[Y](t) - \Ftwo[Y](t)\right\rbrace^2 dt \right)+  \E_{Y \sim P_2} \left( \int_0^\mathcal{M} \left\lbrace \Fone[Y](t) - \Ftwo[Y](t)\right\rbrace^2 dt \right)
\end{split}
\end{equation}
%\end{align*}
where $\M=\diam(\Omega)$ is the diameter of $\Omega$. Immediately one sees that under $H_0$ \eqref{eq: null}, $\Delta = 0$. In fact %$\Delta$ corresponds to the quantity $D^w_{XY}$ introduced in \cite{dube:24} operated with uniform weights and 
under  mild conditions $(\Omega,d)$, $P_1$ and $P_2$, $\Delta=0$ if and only if $P_1 = P_2$.

Let $\mathcal{P}_{(\Omega,d)}$ denote the class of all Borel probability measures on $(\Omega,d)$ which are uniquely determined by open balls, that is, for $Q_1, Q_2 \in \mathcal{P}_{(\Omega,d)}$, $Q_1 = Q_2$ if and only if $F^{Q_1}_\omega(t) = F^{Q_2}_\omega(t)$ for all $\omega \in \Omega$ and $t \in \mathbb{R}$. In fact $\mathcal{P}_{(\Omega,d)}$ contains all Borel probability measures on $(\Omega,d)$ under special conditions on $(\Omega,d)$ such as, for example, when $(\Omega,d^k)$ is of strong negative type \citep{lyon:13} for some $k > 0$ %(see Proposition 1 in \cite{dube:24}). 
Then $\Delta=0$ implies that $F_\omega^{P_1}(u)=F_\omega^{P_2}(u)$ for almost any $t \in \mathbb{R}$ and for any $\omega$ in the union of the supports of $P_1$ and $P_2$. Hence if $\Omega$ is contained in the union of the supports of $P_1$ and $P_2$, then $\Delta=0$ implies that $P_1=P_2$ whenever $P_1,P_2 \in \mathcal{P}_{(\Omega,d)}$ in which case $\Delta$ serves as a divergence measure between $P_1$ and $P_2$ and can be used to measure the discrepancy between $H_1$ \eqref{eq: alt} and $H_0$ \eqref{eq: null}.

To investigate the power of the test we consider the challenging case, a sequence of local alternatives $H_{1,n}$ that shrinks to $H_0$ where 
\begin{equation}
\label{eq: cont_alt}
H_{1,n}=\left\lbrace (P_1, P_2): \Delta = a_n \right\rbrace
\end{equation}
with $a_n \rightarrow 0$ as $n \rightarrow \infty$. In this framework, the oracle asymptotic power of a level $\alpha$ test is quantified using 
\begin{equation*}
\beta^\alpha_n=\prob_{H_{1,n}} \left( {p} \leq \alpha \right)
\end{equation*}
where $p$ is the oracle asymptotic p-value and depends on the null distribution of the test described by the law of $\mathcal{T}$ in Theorem \ref{thm: asymptotic_distribution}. Since the law of $\mathcal{T}$ is unknown, we obtain a tractable estimator of $p$ given by $\hat{p}_K$ using the permutation scheme, and the practicable power is quantified as 
\begin{equation}
\label{eq: power_perm}
\tilde{\beta}^\alpha_n=\prob_{H_{1,n}} \left( \hat{p}_K \leq \alpha \right).
\end{equation}
Theorem \ref{thm: consistency_of_test} gives the asymptotic consistency for any level $\alpha$ test, by showing that the power of the test in both the oracle and the practicable regimes converge to one as $n \rightarrow \infty$ in the hard to detect scenario of local alternatives provided that $n a_n \rightarrow \infty$, i.e., $a_n$ does not decay too fast as $n \rightarrow \infty$.

\begin{Theorem}
\label{thm: consistency_of_test}
Under assumptions \ref{ass:dpfctn} and \ref{ass:entropy}, for any $\alpha \in (0,1)$ and as $n \rightarrow \infty$,  $\beta^\alpha_n \rightarrow 1$ as $n \rightarrow \infty$.  Moreover if $n a_n \rightarrow \infty$ and $K \geq \frac{1}{\alpha}$, where $K$ is the total number of permutations used to estimate $\hat{p}_K$ in \eqref{eq: power_perm} then $\tilde{\beta}^\alpha_n \rightarrow 1$ as $n \rightarrow \infty$. 
\end{Theorem}

To understand the intuition behind the power results in Theorem \ref{thm: consistency_of_test}, begin by defining the function $b(u,\tau)$ as follows: $$b(u,\tau)=\frac{(1-\tau)^2}{(1-u)^2} \ind[u \leq \tau] + \frac{\tau^2}{u^2} \ind[u > \tau]. $$ Next, introduce an oracle process $T_n(u)$ defined by:
\begin{equation}
\begin{split}
    \label{eq: oracle}
    T_n(u) = a(u) b(u,\tau) \left( \denom[n] \add[in] \intgrt  \left\lbrace \Fone[Y_i](t) -\Ftwo[Y_i](t)\right\rbrace^2 dt \vphantom{\int_0^\M} \right).
\end{split}
\end{equation}
Lemma 6in section S2 of the online supplement demonstrates that $\sup_{u \in \idx_c} \lvert \hat{T}_n(u) -T_n(u) \rvert =o_P(1)$ as $n \rightarrow \infty$. Specifically under $H_{1,n}$ and assumptions \ref{ass:dpfctn} and \ref{ass:entropy} , it is established that
\begin{equation}
\label{eq: bound_oh_That}
      \sup_{u \in \idx_c} \lvert \hat{T}_n(u) -T_n(u) \rvert = O_P\left(\frac{1}{n}+\sqrt{\frac{a_n}{n}}\right).
\end{equation}
Consequently, the oracle process $T_n(u)$ characterizes the uniform large sample limit of the scan function $\hat{T}_n(u)$. Under the alternative, the scan function is upper and lower bounded by factors of $\Delta$ as defined in equation \eqref{eq: divergence}. Utilizing equation \eqref{eq: bound_oh_That}, it is demonstrated that as $n \rightarrow \infty$, the probability under $H_{1,n}$ of $\sup_{u \in \idx_c} \hat{T}_n(u) > T_n(\tau)/2$ tends to 1, whereas the probability of $\sup_{u \in \idx_c} \hat{T}^{\pi_k}_n(u) \leq T_n(\tau)/2$ for all $k=1, \dots, K$ tends to 1. This is achieved through the exchangeability of random permutations conditional on the data and the coupling construction of an i.i.d data sequence from the mixture distribution $\bar{P}=\tau P_1 +(1-\tau) P_2$ using the observations $Y_1, \dots, Y_n$  as described in section 5.3 of \cite{chun:13}.  Therefore the permutation p value $\hat{p}_K$ leads to a consistent test under $H_{1,n}$ as $n \rightarrow \infty$. 

\vspace{-1em}

\subsection{Rates of convergence of the estimated change point}

\vspace{-0.5em}

In this section we describe the estimation of $\tau$ under the alternative. As a straightforward consequence of our test statistic \eqref{eq: test_stat} we estimate $\tau$ as
\begin{equation}
\label{eq: chpt}
    \hat{\tau} = \argmax_{u \in \mathcal{I}_c} \hat{T}_n(u).
\end{equation}
Next, we explore the asymptotic behavior of $\hat{\tau}$ around $\tau$. We examine two scenarios: first, when the alternative is fixed, a common focus in nonparametric change point analysis, and second, when the alternative is such that $(P_1, P_2) \in H_{1,n}$ with $H_{1,n}$ defined in \eqref{eq: cont_alt}. In this case, $\Delta(P_1,P_2)=a_n$ with $a_n \rightarrow 0$ as $n \rightarrow \infty$, presenting a challenging setup for accurately detecting $\tau$. Theorem \ref{thm: consistency_of_chpt} establishes the asymptotic near-optimal rate of convergence for the change point in the first scenario with fixed alternatives when $P_1 \neq P_2$ \citep{madr:21}. In the second scenario, the change point estimator $\hat{\tau}$ remains a consistent estimator of $\tau$ only if $na_n \rightarrow \infty$ as $n \rightarrow \infty$, a condition also crucial for the test's consistency as outlined in Theorem \ref{thm: consistency_of_test}.

\begin{Theorem}
\label{thm: consistency_of_chpt}
For any fixed alternative when $P_1 \neq P_2$, under assumptions \ref{ass:dpfctn} and \ref{ass:entropy}, as $n \rightarrow \infty$,  
\begin{equation*}
    \lvert \hat{\tau} - \tau \rvert = O_P\left( \frac{\log(n)}{n} \right).
\end{equation*}
For the local alternatives $H_{1,n}$, there exists $L > 0$ such that $\prob_{H_{1,n}} \left( {na_n} \lvert \hat{\tau} - \tau \rvert \geq L \right) \rightarrow 0$  as $n \rightarrow \infty$. 
\end{Theorem}

To grasp the intuition behind Theorem \ref{thm: consistency_of_chpt}, observe that $\hat{\tau}$ is a maximizer of $\hat{T}_n(\cdot)$ while $\tau$ is the unique maximizer of $T_n(\cdot)$ \eqref{eq: oracle} almost surely. Moreover by the definition in \eqref{eq: oracle} and the bounds in equation 25 in Lemma 6 in the online supplement, one has $T_n(\tau) -T_n(\hat{\tau}) \geq C \lvert \hat{\tau}-\tau \rvert \Delta$ for some constant $C > 0$. Hence one has almost surely that 
\begin{equation*}
    \{\hat{T}_n(\hat{\tau}) -  \hat{T}_n(\tau)\} - \{T_n(\hat{\tau})-T_n(\tau) \} \geq  C \lvert \hat{\tau}-\tau \rvert \Delta. 
\end{equation*}
By leveraging convergence rates for $$\E \left( \sup_{|u-\tau| \leq \delta_n} \lvert \hat{T}_n(u) - \hat{T}_n(\tau)-T_n(u)+ \\ T_n(\tau) \rvert \right)$$ for sequences $\delta_n \rightarrow 0$ with $n\delta_n \rightarrow \infty$ as $n \rightarrow \infty$, one can deduce the convergence rates of $\lvert \hat{\tau} - \tau \rvert$ as outlined in Theorem \ref{thm: consistency_of_chpt}.

\vspace{-1em}

\section{Simulations} \label{sec: simulation}

\vspace{-0.5em}

We assess the finite sample performance of our approach\footnotemark across various frameworks involving three different random object spaces. First, we carry out simulations capturing diverse change point scenarios in sequences of Euclidean random vectors, then we study change points in non-Euclidean settings involving bivariate distributional data sequences where the bivariate distributions are equipped with the $L^2$ metric between corresponding the cdfs, and finally move on to change points in random network sequences where the networks are represented as graph Laplacians with the Frobenius metric between them and are generated from a preferential attachment model \citep{bara:1999}. For each random object we explore a range of alternatives beyond location shifts and scale shifts, such as, sudden changes in the tails of the population distribution, the population distribution abruptly switching to a mixture distribution with two components and abrupt changes in network node attachment mechanisms. 
\footnotetext{Code for implementing our method is available at \url{https://github.com/mzheng-usc/cpd_dist_profile/tree/main}}
We generate random object sequences of length $n =300$ and set the level $\alpha$ of the test to be $\alpha=0.05$. In each scenario, we first calibrate the type I error of the new test under $H_0$ and then evaluate the empirical power of the test for a succession of alternatives that capture increasing discrepancies between the distributions of the two data segments before and after the change point with $\tau=1/3$. We employ our scan statistic on the interval $\mathcal{I}_c=[0.1,0.9]$. We conduct 500 Monte Carlo runs and set the empirical power to be the proportion of the rejections at $\alpha=0.05$ out of the $500$ Monte Carlo runs. The critical value of is approximated using the permutation scheme described in Section \ref{sec: test_construction} with $1000$ permutations within each Monte Carlo run. To demonstrate our findings we present empirical power plots where we expect that the empirical power will be maintained at the level $\alpha=0.05$ under $H_0$ and will increase with increasing disparities between the the two data segments. Along with the power trends we also investigate the accuracy of the estimated change point locations using their mean absolute deviation (MAE), calculated as 
\[
\textup{MAE}=\frac{1}{500}\sum_{i=1}^{500}|\hat{\tau}_i-\tau|,
\]
where $\hat{\tau}_i$ is the estimated change point location in the $i$-th Monte Carlo run. We expect that a successful method will have better accuracy in detecting change points, and therefore, have lower MAE with higher distributional differences between the data segments.

We compare the power and the MAE of our test statistic, which we will refer to as ``dist-CP" here on, with the energy-based change point detection test (``energy-CP'')\citep{matt:2014}, graph-based change point detection test (``graph-CP'')\citep{chu:2019}, and kernel-based change point detection test (``kernel-CP'') \citep{harc:08, Jone:2020}. 
%And add kernel-based change point detection \citep{Arlo:2019} for the estimated location accuracy comparison. 
We adopt the same configuration whenever needed for each of these tests, for example, we set the number of permutations to be 1000 to obtain the p-value approximations. For the graph-based test, we apply the generalized edge-count scan statistic \citep{chu:2019} and as recommened, choose 5-MST (minimum spanning tree) to construct the initial similarity graph. {For the kernel-CP, we use the Gaussian kernel and select the bandwidth to be the median of the input pairwise distances. In terms of the implementation, we use the R packages ``gSeg''\citep{chen:2020} and  ``ecp''\citep{Jame:2015}  for graph-CP and energy-CP separately, and Python package ``Chapydette''\citep{Jone:2020} for kernel-CP.}
%Minimum spanning tree is a tree graph that connects all the observations minimizing the total distance across edges. $k$th MST is the spanning tree that is orthogonal to all $i$th MST, for $i<k$, while minimizing the total distance across edges. And $k$-MST is the union of all $i$th MST, for $i=1,\dots,k$. For the kernel-based test, we conducted 1000 permutations to calculate the p values and took the $\argmax$ of the object value to estimate the change point location.
%we choose the penalty constant using ``slope heuristics'' method and set the maximum number of segments $D_{max}=n/\sqrt{\log{n}}$ as both suggested in \citet{Arlo:2019}. 
%In the figures below, ``dist-CP'' stands for our proposed test result, ``energy-CP'' stands for the result of the energy-based method, ``graph-CP'' stands for the result of the graph-based method and ``kernel-CP'' stands for the kernel-based method. We will also refer to them as the corresponding method in the following contexts. 

%\textcolor{red}{1. Describe the packages (R or python) that were used for the other methods. 2. Describe the kernel that was defined based on the distances for kernel-CP.}

\vspace{-1em}

\subsection{Multivariate data}

\vspace{-0.5em}

\begin{figure*}[htbp]
     \centering
     \begin{subfigure}[b]{\textwidth}
         \centering
         \includegraphics[width=1\textwidth]{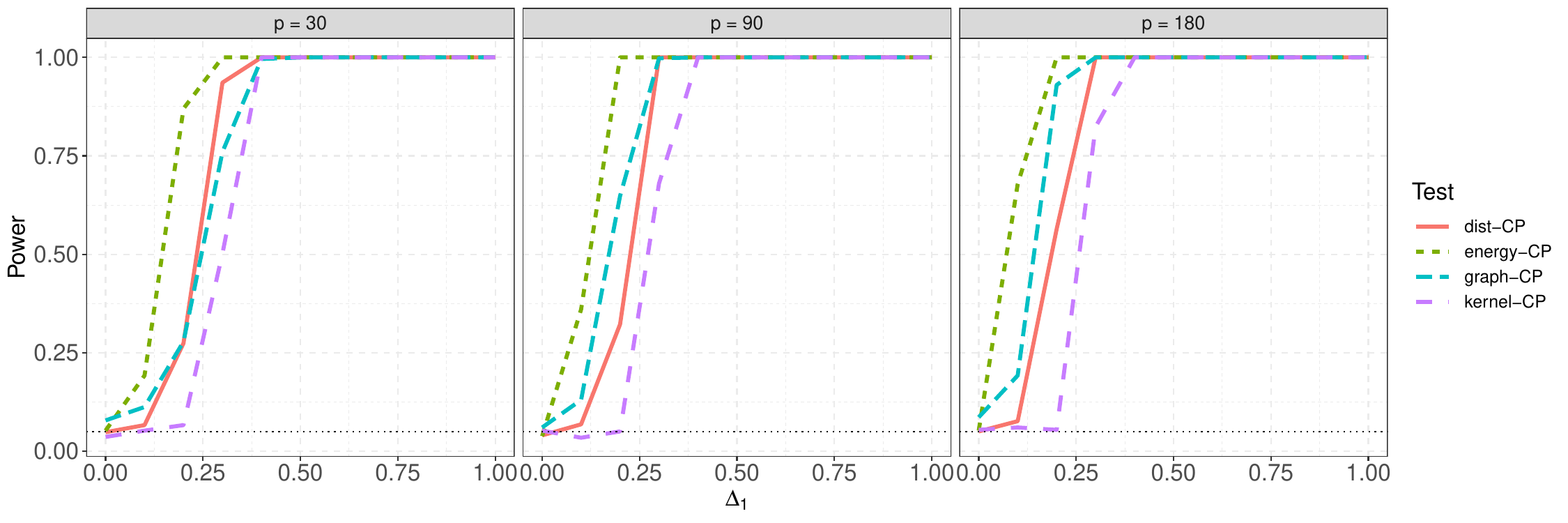}
         \caption{}
         \label{fig:normal_mean_power}
     \end{subfigure}
     \hfill
     \begin{subfigure}[b]{\textwidth}
         \centering
         \includegraphics[width=1\textwidth]{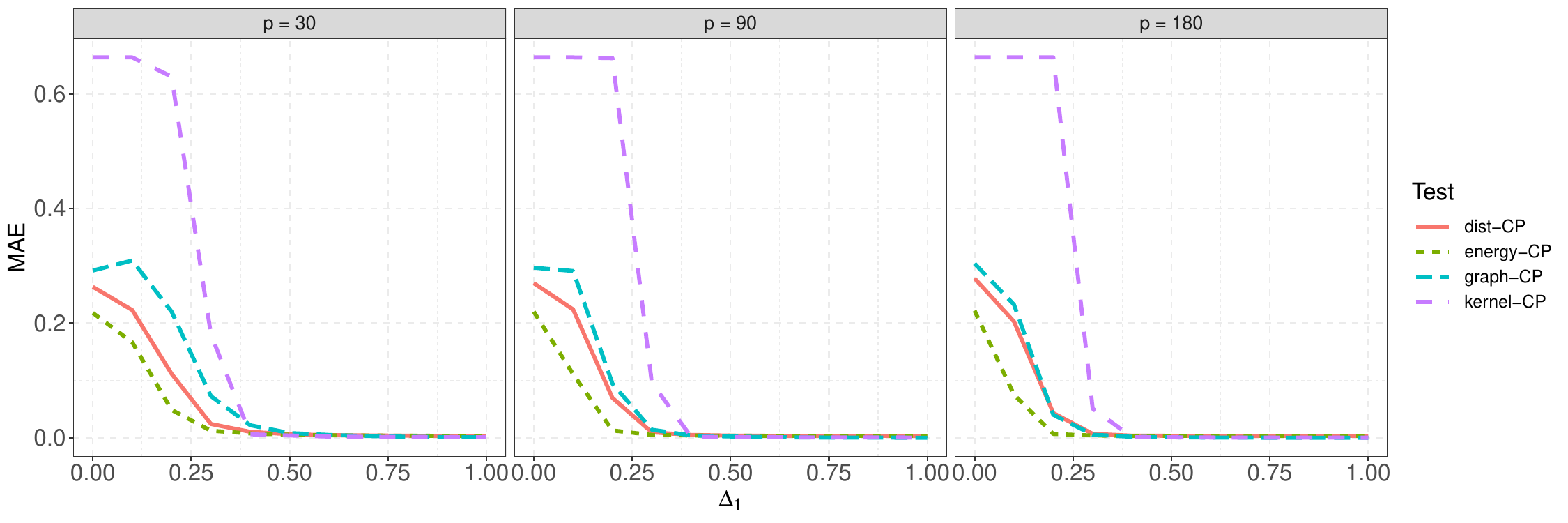}
         \caption{}
         \label{fig:normal_mean_loc}
     \end{subfigure}
     \caption{In Figure \ref{fig:normal_mean_power}, we present the power comparisons with respect to $\Delta_1$ for a sequence of $p$-dimensional random vectors sampled from $N(\mu,\Sigma)$, where $\mu=\mathbf{0}_p=(0,0,\dots,0)^T$ for $Y_i, \ i=1,\dots,100$ and $\mu=\Delta_1 \mathbf{1}_p=\Delta_1(1,1,\dots,1)^T$ for $Y_i, \ i=101,\dots,300$. $\Sigma$ is held fixed for the whole sequence as $\Sigma=U\Lambda U^T$, where $\Lambda$ is a diagonal matrix with $k$th diagonal entry being $\textup{cos}(k\pi/p)+1.5$ for $k=1,\dots,p$, and $U$ is an orthogonal matrix with the first columns being $p^{-1/2}(1,1,\dots,1)^T$. The dotted black line indicates the significance level of 0.05. Figure \ref{fig:normal_mean_loc} presents the MAE of the estimated change points with respect to $\Delta_1$.}
     \label{fig:normal_mean}
\end{figure*}

\begin{figure*}[htbp]
     \centering
     \begin{subfigure}[b]{1\textwidth}
         \centering
         \includegraphics[width=1\textwidth]{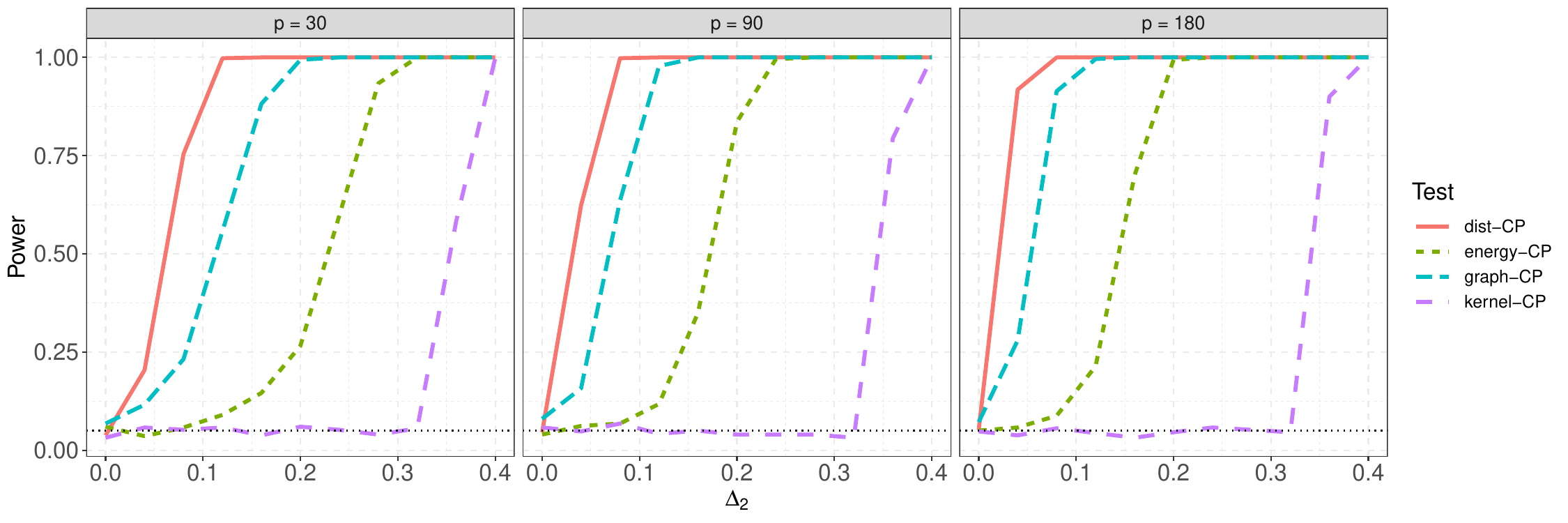}
         \caption{}
         \label{fig:normal_scale_power}
     \end{subfigure}
     \hfill
     \begin{subfigure}[b]{1\textwidth}
         \centering
         \includegraphics[width=1\textwidth]{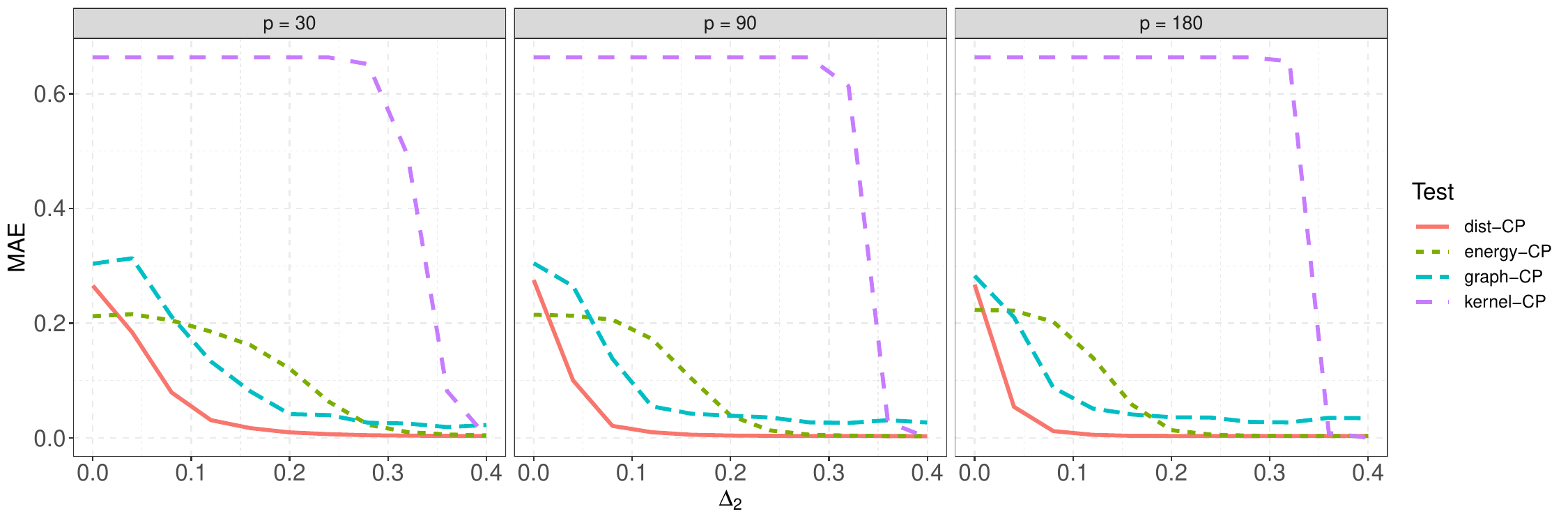}
         \caption{}
         \label{fig:normal_scale_loc}
     \end{subfigure}
     \caption{In Figure \ref{fig:normal_scale_power}, we present the power comparisons with respect to $\Delta_2$ for a sequence of $p$-dimensional random vectors sampled from $N(\mu,\Sigma)$, where $\mu=\mathbf{0}_p$ for the whole sequence. $\Sigma=0.8 \mathbf{I}_p$ for $Y_i, \ i=1,\dots,100$ and $\Sigma=(0.8-\Delta_2)\mathbf{I}_p$ for $Y_i, \ i=101,\dots,300$. The dotted black line indicates the significance level of 0.05. Figure \ref{fig:normal_scale_loc} presents the MAE of the estimated change points with respect to $\Delta_2$.}
     \label{fig:normal_scale}
\end{figure*}

In this setting the data elements are random vectors endowed with $l^2$ metric and are generated from the  $p$-dimensional Gaussian distribution $N(\mu,\Sigma)$ with mean $\mu$ and covariance matrix given by $\Sigma$. We explore four different types of changes to the population distributions. First we consider location and scale changes, in dimensions $p = 30, 90$ and $180$ where to study location change,  we set $\mu=\mathbf{0}_p=(0,0,\dots,0)^T$ for $Y_i, \ i=1,\dots,100$ and $\mu=\Delta_1 \mathbf{1}_p=\Delta_1(1,1,\dots,1)^T$ for $Y_i, \ i=101,\dots,300$, where we let $\Delta_1$ range from 0 to 1. The covariance matrix is fixed and constructed as $\Sigma=U\Lambda U^T$, where $\Lambda$ is a diagonal matrix with $k$-th diagonal entry being $\textup{cos}(k\pi/p)+1.5$ for $k=1,\dots,p$, and $U$ is an orthogonal matrix with the first columns being $p^{-1/2}(1,1,\dots,1)^T$, such that the mean change loads along the first eigenvector of the data. To investigate scale change, we fix the $\mu=\mathbf{0}_p$ and set $\Sigma=0.8 \mathbf{I}_p$ for $Y_i, \ i=1,\dots,100$ and $\Sigma=(0.8-\Delta_2)\mathbf{I}_p$ for $Y_i, \ i=101,\dots,300$, where we let $\Delta_2$ range from 0 to 0.4. We present the results of location and scale shifts in Figure \ref{fig:normal_mean}, where Figure \ref{fig:normal_mean_power} illustrates the empirical power performances of the tests and Figure \ref{fig:normal_mean_loc} shows the MAE of estimated change points. In Figure \ref{fig:normal_scale}, we present the corresponding results for scale change. In Figure \ref{fig:normal_mean_power}, we see that all the methods maintain type I error control. %Dist-CP performs similarly with graph-CP under low dimension $p=30$ and the power of graph-CP is boosted when the dimension increases. 
Energy-CP outperforms the other methods in terms of power and MAE but dist-CP has competitive power performance across all different settings in the mean change scenario. For scale changes, dist-CP has the best performance in terms of both empirical power and MAE across all settings as illustrated in Figure \ref{fig:normal_scale}. %It is also noticed that graph-CP has a biased estimated location shown in \ref{fig:normal_scale_loc}. 

\begin{figure*}[hbt!]
     \centering
     \begin{subfigure}[b]{\textwidth}
         \centering
         \includegraphics[width=1\textwidth]{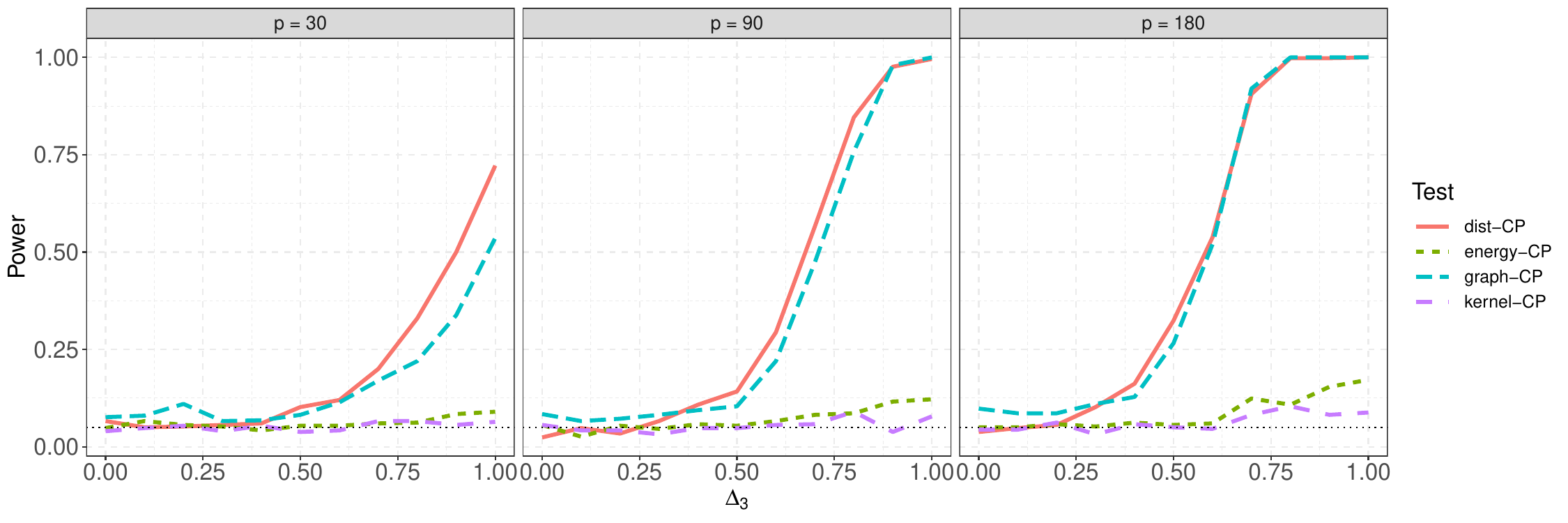}
         \caption{}
         \label{fig:normal_mix_power}
     \end{subfigure}
     \hfill
     \begin{subfigure}[b]{\textwidth}
         \centering
         \includegraphics[width=1\textwidth]{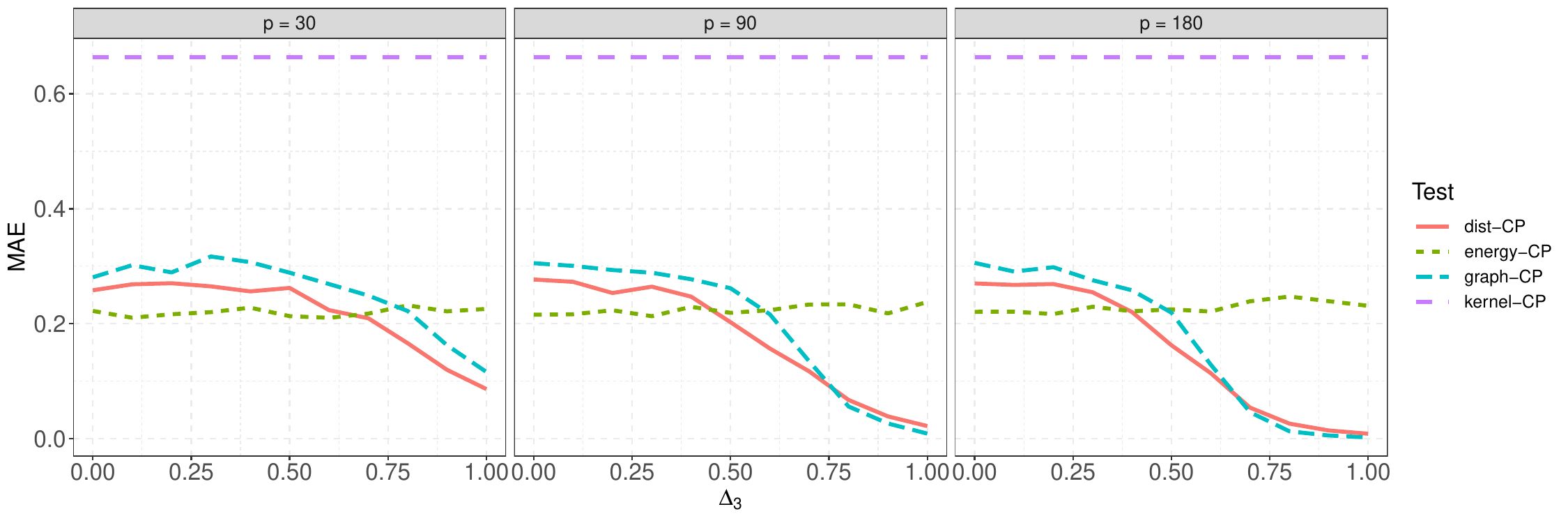}
         \caption{}
         \label{fig:normal_mix_loc}
     \end{subfigure}
     \caption{In Figure \ref{fig:normal_mix_power}, we present the power comparisons with respect to $\Delta_3$ for a sequence of $p$ dimensional random vectors. Here, $Y_1,\dots,Y_{100}$ are generated from the standard $p$ dimensional Gaussian distribution $N(\mathbf{0}_p,\mathbf{I}_p)$. $Y_{101},\dots,Y_{300}$ are constructed with independent samples of $AZ_1+(1-A)Z_2$, where $A \sim \textup{Bernoulli(0.5)}$, $Z_1 \sim N(-\mu,\mathbf{I}_p)$, $Z_2 \sim N(\mu,\mathbf{I}_p)$,  where $\mu = (\Delta_3 \mathbf{1}_{0.1p},\mathbf{0}_{0.9p})^T$, and $A$, $Z_1$, and $Z_2$ are independent. The dotted black line indicates the significance level of 0.05. Figure \ref{fig:normal_mix_loc} presents the MAE of the estimated change points with respect to $\Delta_3$}
     \label{fig:normal_mix}
\end{figure*}

Next we study sudden splitting of a homogeneous population into a mixture of two components with different means. To study this we take $Y_1,\dots,Y_{100}$ to be generated from the standard $p$-dimensional Gaussian distribution $N(\mathbf{0}_p,\mathbf{I}_p)$ and let $Y_{101},\dots,Y_{300}$ be generated from a mixture of two Gaussian distributions with the overall population mean same as $Y_1,\dots,Y_{100}$. To be more specific, $Y_{101},\dots,Y_{300}$ are constructed with independent samples of $AZ_1+(1-A)Z_2$, where $A \sim \textup{Bernoulli(0.5)}$, $Z_1 \sim N(-\mu,\mathbf{I}_p)$, $Z_2 \sim N(\mu,\mathbf{I}_p)$,  where $\mu = (\Delta_3 \mathbf{1}_{0.1p},\mathbf{0}_{0.9p})^T$, and $A$, $Z_1$, and $Z_2$ are independent. Here, $\Delta_3 \in [0,1]$. In Figure \ref{fig:normal_mix} we illustrate that dist-CP outperforms all other approaches both in terms of empirical power and MAE in this complex change point scenario.

\begin{figure*}[htbp]
     \centering
     \begin{subfigure}[b]{1\textwidth}
         \centering
         \includegraphics[width=1\textwidth]{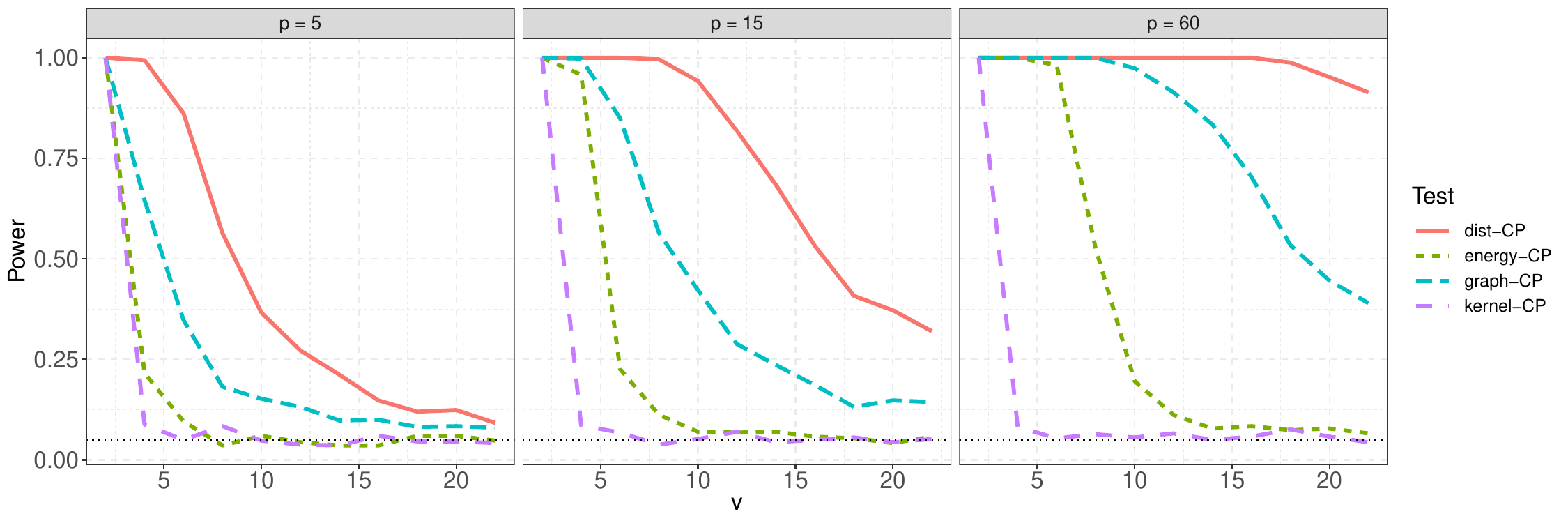}
         \caption{}
         \label{fig:normal_tail_power}
     \end{subfigure}
     \hfill
     \begin{subfigure}[b]{1\textwidth}
         \centering
         \includegraphics[width=1\textwidth]{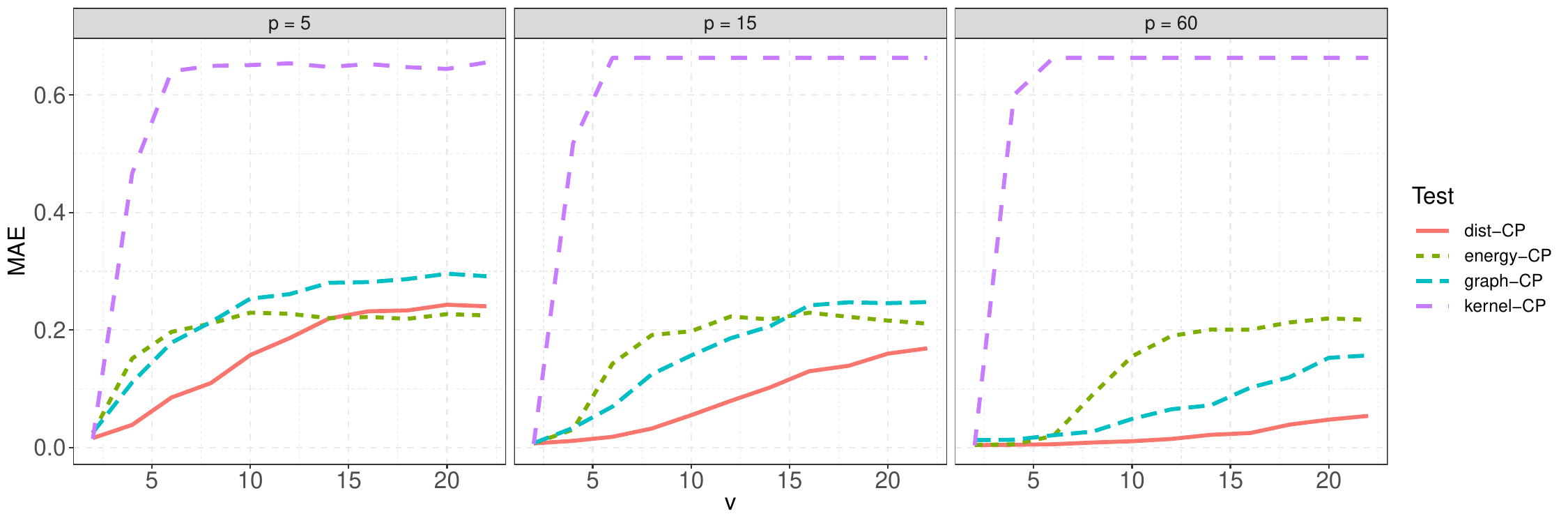}
         \caption{}
         \label{fig:normal_tail_loc}
     \end{subfigure}
     \caption{In Figure \ref{fig:normal_tail_power}, we present the power comparison for increasing values of $v$ for a sequence of $p$ dimensional random vectors. Here, $Y_i \sim N(\mathbf{0}_p,\mathbf{I}_p)$ with $p \in \{5,15,60\}$ for $i=1,\dots,100$ and $Y_i \sim t_v$ for $i=101,\dots,300$, where $t_v$ stands for $t$ distribution with $v$ degrees of freedom. The dotted black line indicates the significance level of 0.05. Figure \ref{fig:normal_tail_loc} presents the MAE of the estimated change points with respect to $v$.}
     \label{fig:normal_tail}
\end{figure*}

Finally, we experiment with changes to the tail of the population distribution. We generate $Y_i \sim N(\mathbf{0}_p,\mathbf{I}_p)$ with $p \in \{5,15,60\}$ for $i=1,\dots,100$ and {$Y_i$ as a $p$-dimensional random vector whose components are independent and identically distributed as the t-distribution with $v$ degrees of freedom for $i=101,\dots,300$}
%, where $t_v$ stands for $t$ distribution with $v$ degrees of freedom 
We let $v$ range in $\{2,\dots,22\}$ to reflect the change from Gaussian tails to successively heavier tails. The results as shown in Figure \ref{fig:normal_tail} demonstrate that dist-CP has the best power and MAE performance across all settings.

\vspace{-1em}

\subsection{Bivariate distributional data}

\begin{figure*}[htb!]
\centering
\begin{adjustbox}{minipage=0.8\linewidth}
     \begin{subfigure}[b]{0.49\textwidth}
         \centering
         \includegraphics[width=\textwidth]{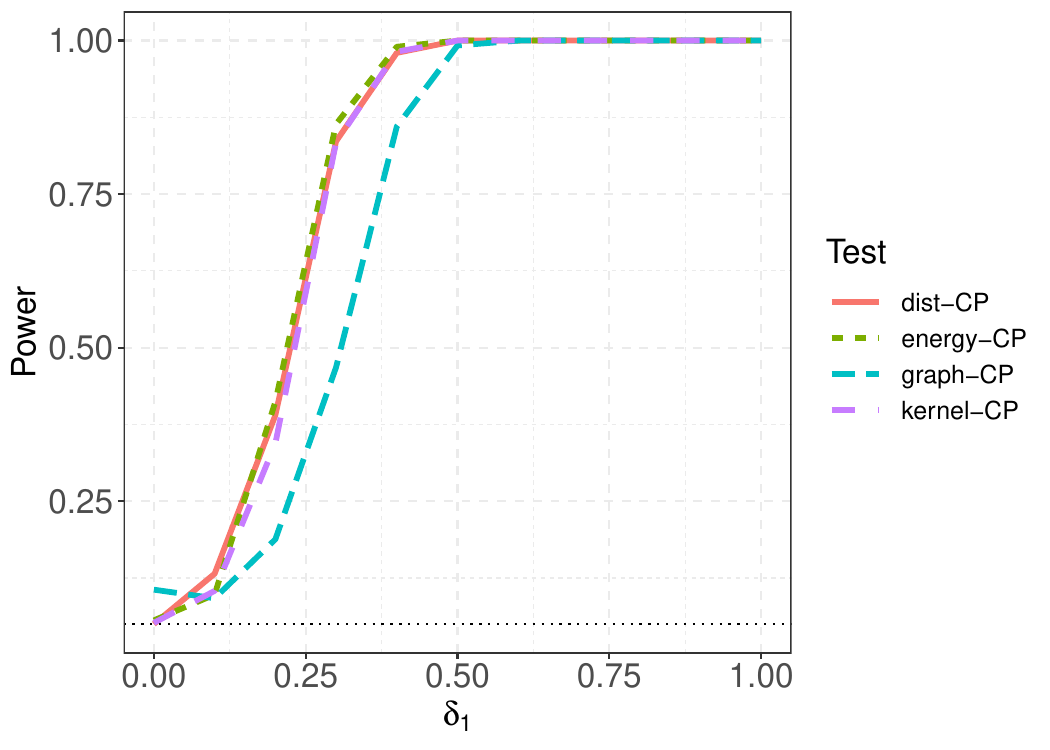}
         \caption{}
         \label{fig:dist_mean_power}
     \end{subfigure}
     \hfill
     \begin{subfigure}[b]{0.49\textwidth}
         \centering
         \includegraphics[width=\textwidth]{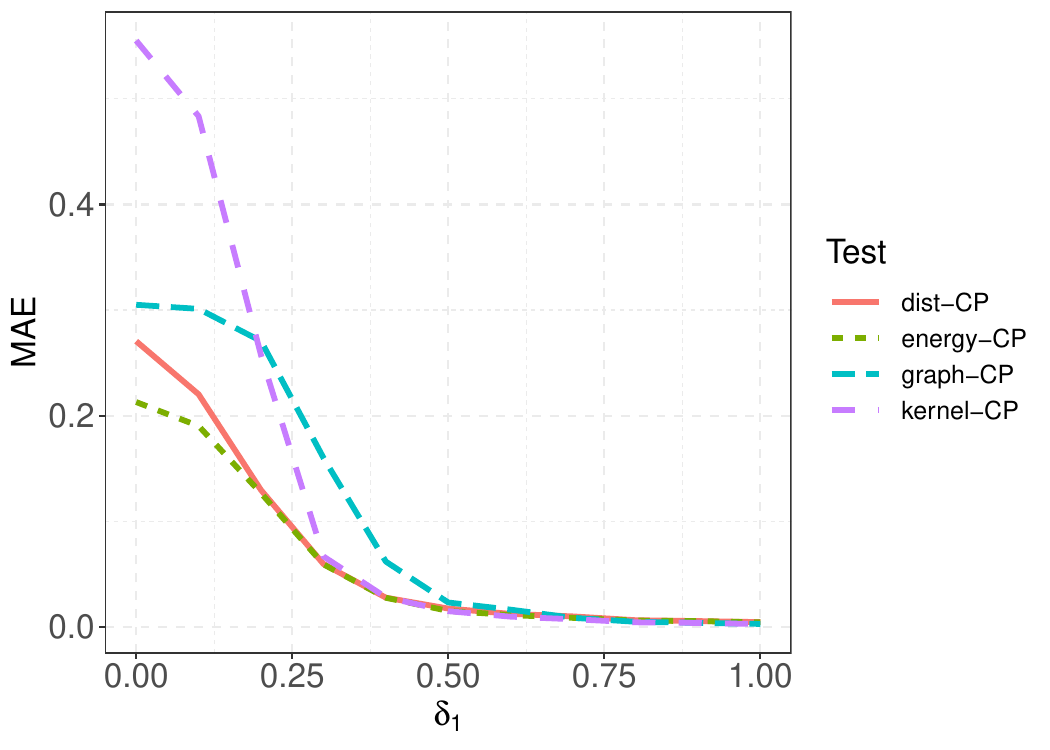}
         \caption{}
         \label{fig:dist_mean_loc}
     \end{subfigure}
\end{adjustbox}
     \caption{In Figure \ref{fig:dist_mean_power}, we present the power comparisons with respect to $\delta_1$ for a sequence of bivariate Gaussian distributional data. Here, $Y_i = N(Z_i,0.25\mathbf{I}_2)$, where $Z_i \sim N(\mathbf{0}_2,0.25\mathbf{I}_2)$ for $i=1,\dots,100$, and $Z_i \sim N((\delta_1,0)^T,0.25\mathbf{I}_2)$ for $i=101,\dots,300$. The dotted black line indicates the significance level of 0.05. Figure \ref{fig:dist_mean_loc} presents the MAE of the estimated change points with respect to  $\delta_1$.}
     \label{fig:dist_mean}
\end{figure*}

\begin{figure*}[htb!]
     \centering
     \begin{adjustbox}{minipage=0.8\linewidth}
     \begin{subfigure}[b]{0.49\textwidth}
         \centering
         \includegraphics[width=\textwidth]{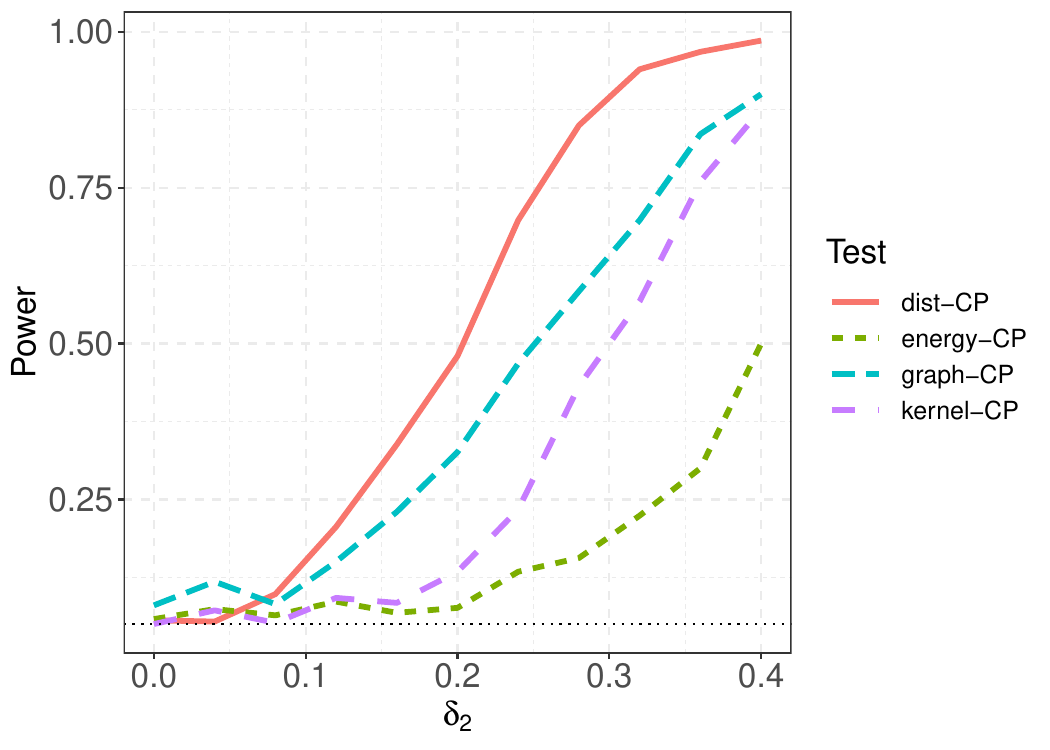}
         \caption{}
         \label{fig:dist_scale_power}
     \end{subfigure}
     \hfill
     \begin{subfigure}[b]{0.49\textwidth}
         \centering
         \includegraphics[width=\textwidth]{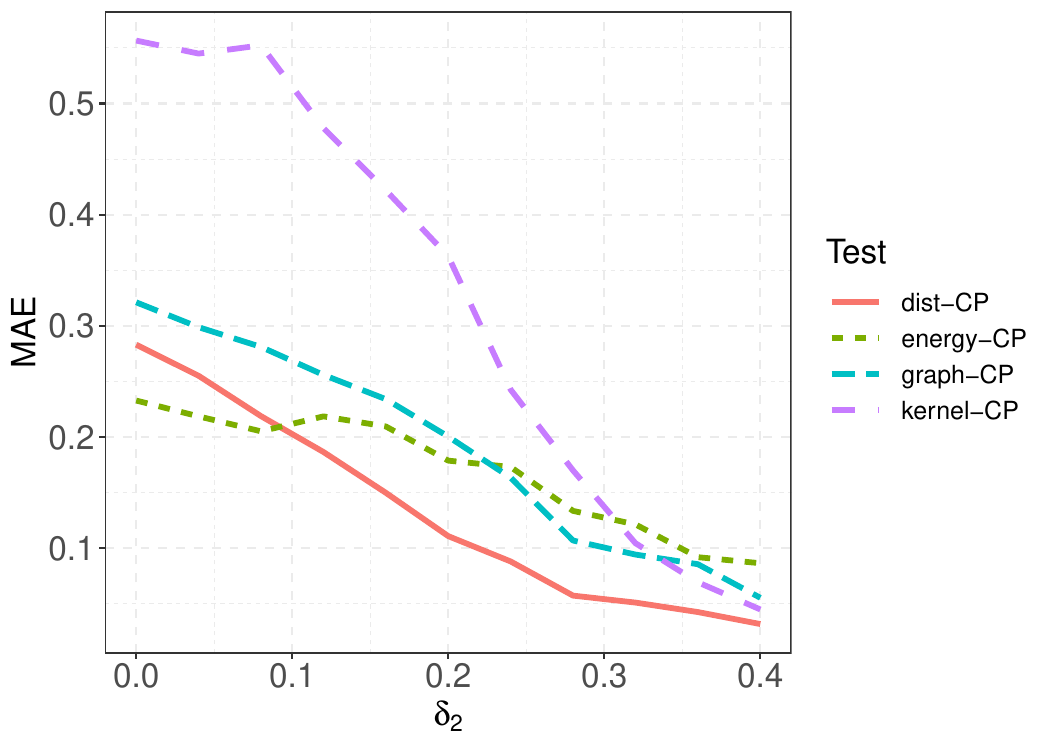}
         \caption{}
         \label{fig:dist_scale_loc}
     \end{subfigure}
\end{adjustbox}
     \caption{In Figure \ref{fig:dist_scale_power}, we present the power comparison for increasing values of $\delta_2$ for a sequence of bivariate Gaussian distributional data. Here, $Y_i = N(Z_i,0.25\mathbf{I}_2)$, where $Z_i \sim N(\mathbf{0}_2,0.4^2\mathbf{I}_2)$ for $i=1,\dots,100$, and $Z_i \sim N(\mathbf{0}_2,\textup{diag}((0.4+\delta_2)^2,0.4^2))$ for $i=101,\dots,300$. The dotted black line indicates the significance level of 0.05. Figure \ref{fig:dist_scale_loc} presents the MAE of the estimated change points with respect to  $\delta_2$.}
     \label{fig:dist_scale}
     
\end{figure*}

Here the random objects are random bivariate probability distributions equipped with the metric between corresponding cumulative distribution function representations defined as $d(x,y)=\int_\mathbb{R}\int_\mathbb{R}  |F_x(u,v)-F_y(u,v)|\ \,du\,dv$, where $F_x(u,v)$ is the bivariate cdf of $x$. Each observation $Y_i$ for $i=1,\dots,300$ is itself a bivariate distribution with a cdf representation where we explore two types of changes: changes in the process generating the means of $Y_i$ and changes in the process generating the variances in the covariance structure of $Y_i$. Specifically, for the first scenario, $Y_i = N(Z_i,0.25\mathbf{I}_2)$, where $Z_i \sim N(\mathbf{0}_2,0.25\mathbf{I}_2)$ for $i=1,\dots,100$, and $Z_i \sim N((\delta_1,0)^T,0.25\mathbf{I}_2)$ for $i=101,\dots,300$ where $\delta_1 \in [0,1]$. For changes in the scales of the random distributions we generate $Y_i = N(Z_i,0.25\mathbf{I}_2)$, where $Z_i \sim N(\mathbf{0}_2,0.4^2\mathbf{I}_2)$ for $i=1,\dots,100$, and $Z_i \sim N(\mathbf{0}_2,\textup{diag}((0.4+\delta_2)^2,0.4^2))$ for $i=101,\dots,300$ and let $\delta_2 \in [0,4]$. We illustrate our findings in \ref{fig:dist_mean} and \ref{fig:dist_scale}. In Figure \ref{fig:dist_mean}, dist-CP, energy-CP, and kernel-CP have similar performance and are better than graph-CP in terms of both power and MAE. In the second case, dist-CP dominates the performance across all settings.

\vspace{-1em}

\subsection{Network data}
\begin{figure*}[htb]
     \centering
     \begin{adjustbox}{minipage=0.8\linewidth}
     \begin{subfigure}[b]{0.49\textwidth}
         \centering
         \includegraphics[width=\textwidth]{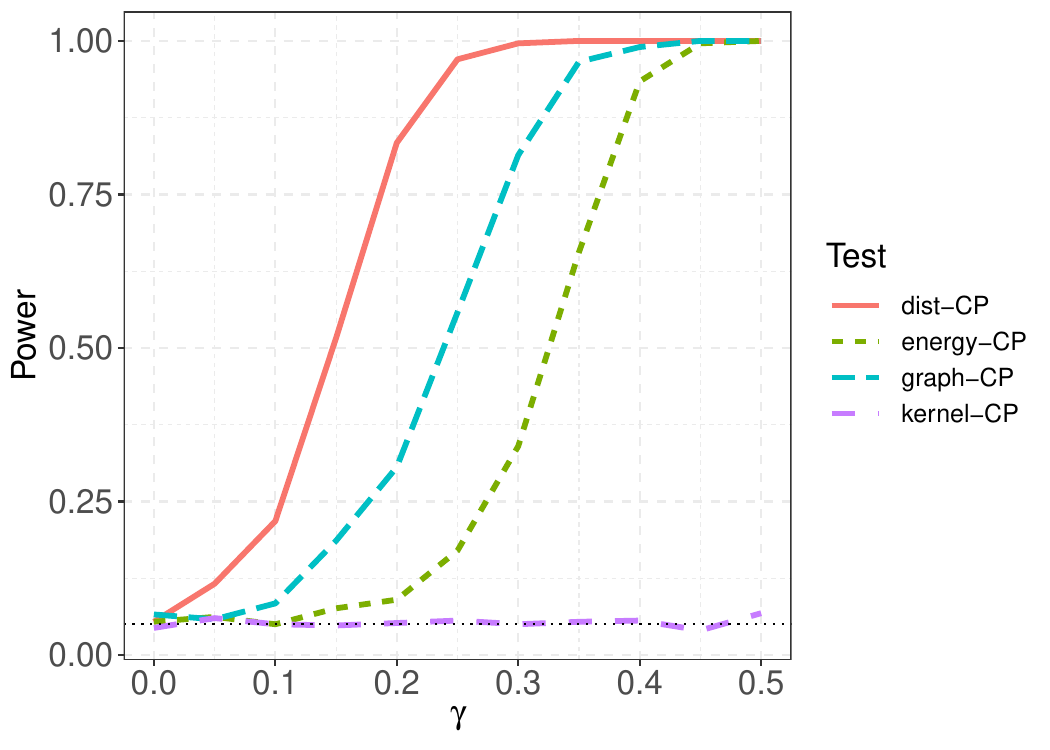}
         \caption{}
         \label{fig:net_power}
     \end{subfigure}
     \hfill
     \begin{subfigure}[b]{0.49\textwidth}
         \centering
         \includegraphics[width=\textwidth]{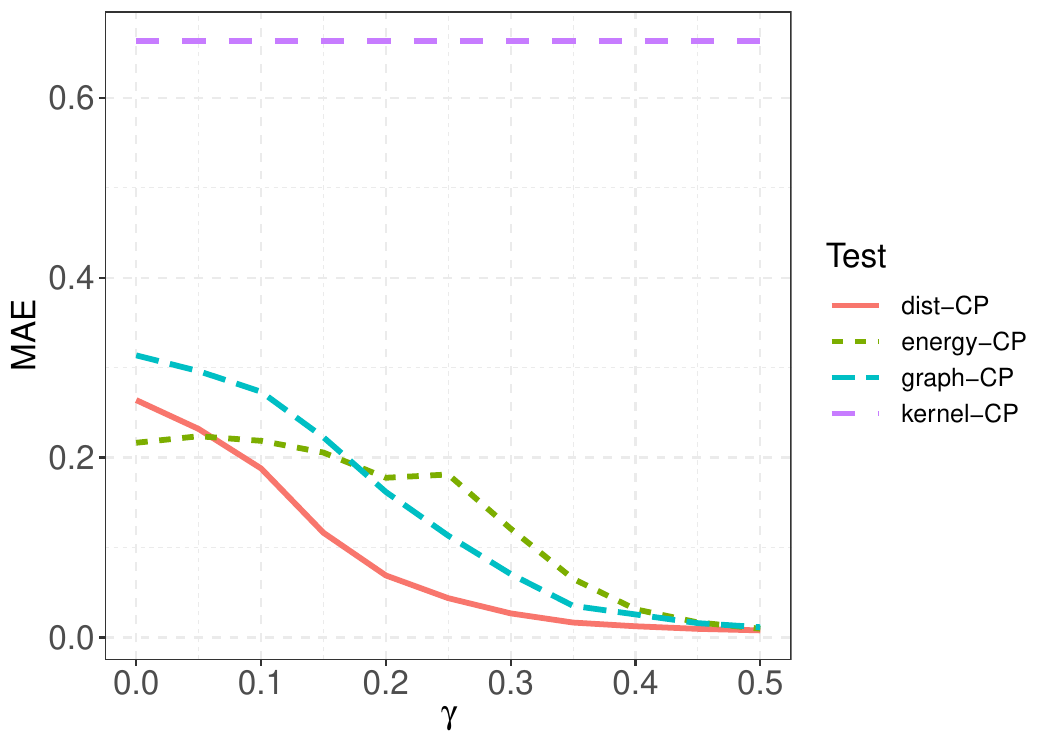}
         \caption{}
         \label{fig:net_loc}
     \end{subfigure}
     \end{adjustbox}
     \caption{In Figure \ref{fig:net_power}, we present the power comparisons with respect to increasing values $\gamma$ for a sequence of random networks with 200 nodes generated by the preferential attachment model. For a node with degree $k$, its attachment function is proportional to $k^\gamma$. $Y_i$ is generated with $\gamma=0$ for $i=1,\dots,100$, and $Y_i$ is generated with $\gamma \in [0,0.5]$ for $i=101,\dots,300$. The dotted black line indicates the significance level of 0.05. Figure \ref{fig:net_loc} presents the MAE of the estimated change points with respect to $\gamma$.}
     \label{fig:net}
\end{figure*}

Lastly, we consider random object sequences where the data elements are random networks endowed with the Frobenius metric between the corresponding Laplacian matrix \footnote{The graph Laplacian for a network is defined $L=D-A$ where $D$ is the degree matrix (diagonal matrix spanned by node degrees) and $A$ is the adjacency matrix.}. Each $Y_i$ is a network with $200$ nodes generated from the preferential attachment model \citep{bara:1999} where for a node with degree $k$, its attachment function is proportional to $k^\gamma$. $Y_i$ is generated with $\gamma=0$ for $i=1,\dots,100$, and $Y_i$ is generated with $\gamma$ from 0 to 0.5 for $i=101,\dots,300$. %We vary $\gamma$ to change the density of the network and distinguish two segments. 
We present the simulation results in Figure \ref{fig:net}. It shows that dist-CP outperforms the other methods in both power and change point location estimation MAE.

\vspace{-2em}

\section{Data applications} \label{sec: applicaiton}

\vspace{-1em}

 \subsection{U.S. electricity generation dataset} 

 \vspace{-0.5em}

We analyze the monthly U.S. electricity generation compositions obtained from \url{https://www.eia.gov/electricity/data/browser/}. We preprocess the data elements into a compositional form so that each entry of the compositional vector represents the percentage of net generation contribution from a specific source. During the preprocessing, we merge some similar categories of resources together and end up with 7 categories: Coal; Petroleum (petroleum liquids and petroleum coke); Gas (natural and other gases); Nuclear; Conventional hydroelectric; Renewables (wind, geothermal, biomass (total) and other); Solar (small-scale solar photovoltaic and all utility-scale solar). We obtain a sample of $n=264$ observations starting from Jan 2001 to Dec 2022. Each $Y_i$ takes values in a 6-simplex $\Delta^6=\{\bm{x}\in \mathbb{R}^7:\bm{x}^T\mathbf{1}_7=1\}$, where $\mathbf{1}_7=(1,1,\dots,1)^T$. The metric we apply between each object is 
\[d(\bm{x},\bm{y})=\textup{arccos}(\sqrt{\bm{x}}^T\sqrt{\bm{y}}),\quad \bm{x},\bm{y} \in \mathbb{R}^7 ,\] where $\sqrt{\bm{x}}$ is the component-wise square root, i.e., $\sqrt{\bm{x}}=(\sqrt{x_1},\sqrt{x_2},\dots,\sqrt{x_7})$. 
% \begin{figure}[H]%
%     \centering
%     \includegraphics[width = 0.9\linewidth]{images/real_data/elec_preview.pdf}
%     \caption{Preprocessed U.S. Electricity Generation Dataset in compositional vector form}
%     \label{fig:energy_preview}
% \end{figure}  

%\blu{show the result of reality mining and compare it with ecp? iilustration}

\begin{figure}[!htb]%
    \centering
    \includegraphics[width = 0.6\textwidth]{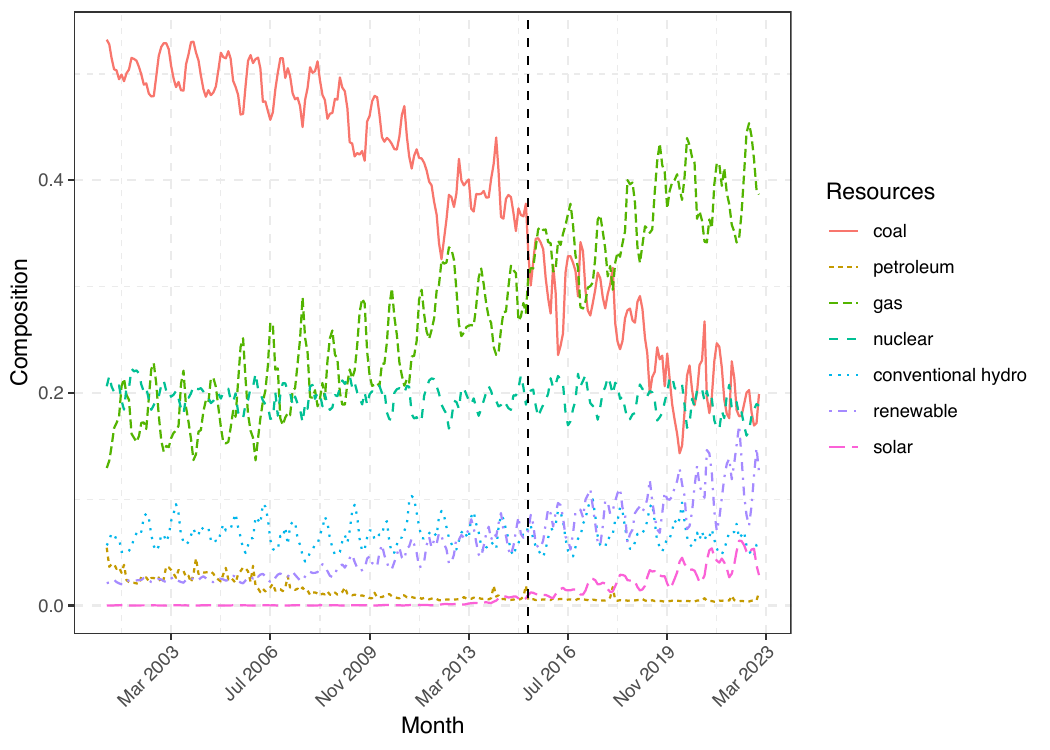}
    \caption{The timeline of U.S. electricity compositions evolution and the change point February 2015 indicated by the vertical dashed black line.}
    \label{fig:energy_cpd}
\end{figure}

We apply dist-CP to the preprocessed data %and present the plot of the scan statistic in Figure \ref{fig:energy_scan}
and note that the scan statistic peaks in month of February 2015. In Figure \ref{fig:energy_cpd}, we present the statistically significant estimated change point location with a vertical dash line. At the change point, we can observe that the percentage of contributions from solar and renewable sources is increasing more rapidly while petroleum contributions drop rapidly.  U.S. reached new milestones in renewables electricity generation in the year 2015 (see \url{https://obamawhitehouse.archives.gov/blog/2016/01/13/renewable-electricity-progress-accelerated-2015})%\citep{hold:2016} 
 \ which explains the detected change point. 
 \iffalse
\begin{figure}[!htb]%
    \centering
    \includegraphics[width = 0.4\textwidth]{images/real_data/elec_scan_stat.pdf}
    \caption{The scan statistic for the monthly U.S. electricity compositions data. The vertical red line indicates the location of the estimated change point in the month of February 2015.}
    \label{fig:energy_scan} 
\end{figure} 
\fi

\vspace{-1em}

\subsection{MIT reality mining dataset}

\vspace{-0.5 em}

The MIT Media Laboratory conducted the reality mining experiment from 2004 to 2005 on students and faculty at MIT \citep{Eagl:2006} in order to explore human interactions based on Bluetooth and other phone applications' activities. We study the participants' Bluetooth proximity networks, where {nodes represent participants} and the edges represent whether the relevant participants had at least one physical interaction within the time interval. 

%\subsubsection{External reference preprocessed data}
We use the reality mining 1392 built-in dataset in the R package ``GreedySBTM'' \citep{Rast:2019} (see \url{https://github.com/cran/GreedySBTM/blob/master/data/reality_mining_1392.RData}). The time frames corresponding to intervals of 4 hours, starting from September 14 2004 to May 5 2005. At each time point the network has $=96$ nodes. We further compressed the data by merging the time windows with a day to obtain 232 daily networks. We work with the pairwise Frobenius metric between the graph Laplacian representation of the networks. Using dist-CP the estimated change point is 2004-12-15, which is during the finals week and close to the start of the winter break. %We present the scan statistics for the whole sequence in Figure \ref{fig:MIT_scan}. %The energy-CP and kernel-CP are consistent with our results. And this result is also consistent with previous work \citep{chen:2015,nie:2021,peel:2015}, where they also detect a similar change point location for the MIT reality mining dataset.

\iffalse
\begin{figure}[htb!]%
    \centering
    \includegraphics[width = 0.4\linewidth]{images/real_data/mit_scan_stat.pdf}
    \caption{The scan statistic for MIT reality mining dataset. The vertical red line indicates the estimated change point 2004-12-15 which is during the finals week and around the beginning of the winter break.}
    \label{fig:MIT_scan}
\end{figure} 

\fi

\vspace{-1em}

\section{Multiple change points} 

\label{sec:multCP}

\vspace{-0.5 em}

In this section, we investigate the extension of our scan statistic to the task of detecting multiple change points by combining it with the recently proposed seeded binary segmentation algorithm \citep{kova:2023}, a deterministic variant of the wild binary segmentation \citep{Fryz:2014}. Seeded binary segmentation creates a collection of seeded intervals $\mathcal{I}_\gamma$ which removes unnecessarily long intervals in wild binary segmentation that might contain multiple change points. As introduced in \citet{kova:2023}, for a sequence of length $n$, the collection of intervals $\mathcal{I}_\gamma$ is given by  
\begin{equation}
\label{eq_seed_interval}
    \mathcal{I}_\gamma=\bigcup_{k=1}^{\ceil{\log_{1/\gamma}(n)}}\mathcal{I}_k
\end{equation}
where $\gamma \in [1/2,1)$ is a decay parameter, $\mathcal{I}_k=\bigcup_{i=1}^{n_k}\{(\floor{(i-1)s_k}),\floor{(i-1)s_k+l_k}\}$ and $n_k=2\ceil{(1/\gamma)^{k-1}}-1$. Each $\mathcal{I}_k$ is also a collection of intervals of length $l_k$ that are evenly shifted by $s_k$, where $l_k=n\gamma^{k-1}$, $s_k=(T-l_k)/(n_k-1)$. 
In algorithm \ref{alg:seed_multiple_cpd}, we describe the detailed implementation of the Mcpd\_DP, the multiple change point detection algorithm based on the combination of distance profiles and seeded binary segmentation. {Once we obtain $\mathcal{I}_\gamma$, we conduct the proposed single change point detection on each of the inner intervals thus deriving a set of potential change point locations. We compare the test statistic in each inner interval of $\mathcal{I}_\gamma$ with a threshold which is set at 90\%-quantile of the permutation null distribution of the test statistic derived on the entire sequence in the current implementation. The selection rule is to keep intervals in which the test statistic is greater than the above threshold and remove all the other intervals sequentially until the maximum test statistic in the entire collection of remaining intervals does not exceed the threshold.} In the experiments, we set $\gamma={1/2}^{1/2}$ as suggested by \citet{kova:2023} and set minLen in algorithm \ref{alg:seed_multiple_cpd} to be 10. %The threshold $\Delta$ is calculated based on the permutation test. 

% switch off footnote line locally
\renewcommand*\footnoterule{}
% collect footnotes within algorithm environment
\begin{savenotes}
\begin{algorithm}[!htbp]
    %\algsetup{linenosize=\tiny}
    \normalsize
    \caption{Multiple change point detection based on distance profile: \textsc{Mcpd\_DP}}
    \label{alg:seed_multiple_cpd}
    \textbf{Input:} $l$ and $u$: lower and upper boundaries of the original sequence, \textup{minLen}: minimum length of each interval, $\gamma$: decay parameter, initial change point set $\hat{\tau}=\emptyset$  \\
    \textbf{Output:} detected change point set $\hat{\tau}$
    \begin{algorithmic}[1]
        \Procedure{Mcpd\_DP}{$l$,$u$,$\gamma$,$\hat{\tau}$,minLen}
        \State \textbf{if} $u-l<$\textup{minLen} \textbf{then} STOP.
        \State calculate the seeded interval collection $\mathcal{I}_\gamma$ as described in equation \ref{eq_seed_interval}.
        \For {i in 1 to $|\mathcal{I}_\gamma|$}
            \State denote $u_i$ and $l_i$ the upper and lower boundaries of the $i$th interval in $\mathcal{I}_\gamma$.
            \State denote the proposed test statistics on interval $[l_i,u_i]$ as $\hat{T}_i$ and record its corresponding estimated change point location as $\hat{t}_i$
        \EndFor
         %\State $i' \leftarrow \textup{argmin}_{i \in \{1,\dots,|\mathcal{I}_\gamma|\}} \ p_i$ \footnote{if there are multiple $i'$ candidates, we choose the one with the narrowest corresponding interval.}
        \State Let $\Delta$ be the 90\% quantile of the approximated permutation distribution based on the entire sample $Y_1,\dots,Y_n$. 
        \If{$\max_i \hat{T}_i>=\Delta$}
            \State $i' \leftarrow \textup{argmax}_{i \in \{1,\dots,|\mathcal{I}_\gamma|\}} \hat{T}_i $
            \State $\hat{\tau} \leftarrow \hat{\tau} \cup \hat{t}_{i'}$ 
            \State conduct \textsc{Mcpd\_DP}($l$,$\hat{t}_{i'}$,$\gamma$,$\hat{\tau}$,minLen) and \textsc{Mcpd\_DP}($\hat{t}_{i'}$,$u$,$\gamma$,$\hat{\tau}$,minLen)
        \Else 
            \Break
        \EndIf
        %\State $i' \leftarrow \textup{argmax}_{i \in \{1,\dots,|\mathcal{I}_\gamma|\}} $
        %\State $\hat{t}  \leftarrow \textup{argmax}_{t \in \{l_{i'}+[n_{i'}c],\dots,u_{i'}-[n_{i'}c]}\} \ \hat{T}_{n_{i'}}(\frac{t}{n_{i'}})$ \footnote{$n_i=u_i-l_i$ denotes the length of the $i$th interval and $c$ is the parameter described in section \ref{sec:problem_setup}}
        %\State \textbf{if} $u-l<$\textup{minLen} \textbf{then} 
        %\If{$p_{i'}<\alpha$}
                %\State $\hat{\tau} \leftarrow \hat{\tau} \cup \hat{t}$
                
        %\EndIf
        \EndProcedure
    \end{algorithmic}
\end{algorithm}
\end{savenotes}
% footnotes are displayed at the end of the savenotes environment

\vspace{-1em}

\subsection{Simulations on network sequences generated using the Stochastic Block Model}

To demonstrate the practical efficacy of Mcpd\_DP in Algorithm \ref{alg:seed_multiple_cpd}, we conduct an experiment on sequences of networks generated according to the stochastic block model (SBM). SBM generates networks with adjacency matrices $A=[A_{ij}]\in \mathbb{R}^{n \times n}$ with $K$ communities, where  $A_{ij} \sim \textup{Bernoulli}(P_{ij})$ with {$P=[P_{ij}]=\Pi B \Pi^T-\text{diag}(\Pi B \Pi^T)$}. Here $\Pi \in \mathbb{R}^{n \times K}$ encodes the community membership of each node, that is, $\Pi_{ik}=1$ if node $i$ belongs to community $k$ and $\Pi_{ik}=0$ otherwise. Here $B \in \mathbb{R}^{K \times K}$ is the connection probability matrix, where $B_{ij}$ indicates the connection probability between $i$-th and $j$-th community. We let $\pi \in \mathbb{R}^K$ to be such that $\pi_i$ indicates the number of nodes in $i$-th community. 

To incorporate different kinds of changes, we focus on the diverse characteristics of the SBM such as changes in the number of nodes in each community and changes in the number of communities. We generate a sequence of networks of length $n=400$ in the SBM framework with change points evenly spaced at $\tau=\{\frac{1}{4}, \frac{2}{4}, \frac{3}{4}\}$. Specifically $Y_i$ are generated with $B=$
 $\begin{pmatrix}
  0.2 & 0.001 & 0.001\\ 
  0.001 & 0.2 & 0.001\\
  0.001 & 0.001 & 0.2\\
\end{pmatrix}$,
and $\pi=(100, 100 ,100)$ for $1 \leq i \leq 100$. For $101 \leq i \leq 200$, we change the connection probability matrix to $B=$
  $\begin{pmatrix}
  0.8 & 0.001 & 0.001\\ 
  0.001 & 0.2 & 0.001\\
  0.001 & 0.001 & 0.8\\
\end{pmatrix}$, and keep $\pi=(100, 100 ,100)$ the same as that for  $1 \leq i \leq 100$. For $201 \leq i \leq 300$, we change $\pi=(200, 50 ,50)$ and retain the same $B$ as $101 \leq i \leq 200$. Finally, for $301 \leq i \leq 400$, we change the number of communities, we set $B=$
 $\begin{pmatrix}
  0.5 & 0.01\\ 
  0.01 & 0.5\\
\end{pmatrix}$ and $\pi=(200,100)$.

 We apply Mcpd\_DP \ref{alg:seed_multiple_cpd} on the network sequences generated according to the SBM framework described above. % and compare the performance with the multiple change points version of energy-CP \citep{matt:2014,Jame:2015} and kernel-CP \citep{Arlo:2019}. 
 We conduct 500 Monte Carlo runs and evaluate Algorithm \ref{alg:seed_multiple_cpd} on two aspects; first, whether the correct number of change points can be detected and second, conditioning on the correctly estimating the number of change points, what is the MAE of the estimated change points. Out of 500 Monte Carlo runs, Mcpd\_DP has a 100\% success rate in correctly estimating the 3 change points %are 100\%, 95.8\% and 100\% for the proposed algorithm Mcpd\_DP, energy-CP and kernel-CP and all of the three methods 
 and achieves zero MAE conditioning on correctly estimating the number of 3 points.

\vspace{-1em}

\subsection{Multiple change points in real data}
\begin{figure}[!h]%
    \centering
    \includegraphics[width = 0.6\linewidth]{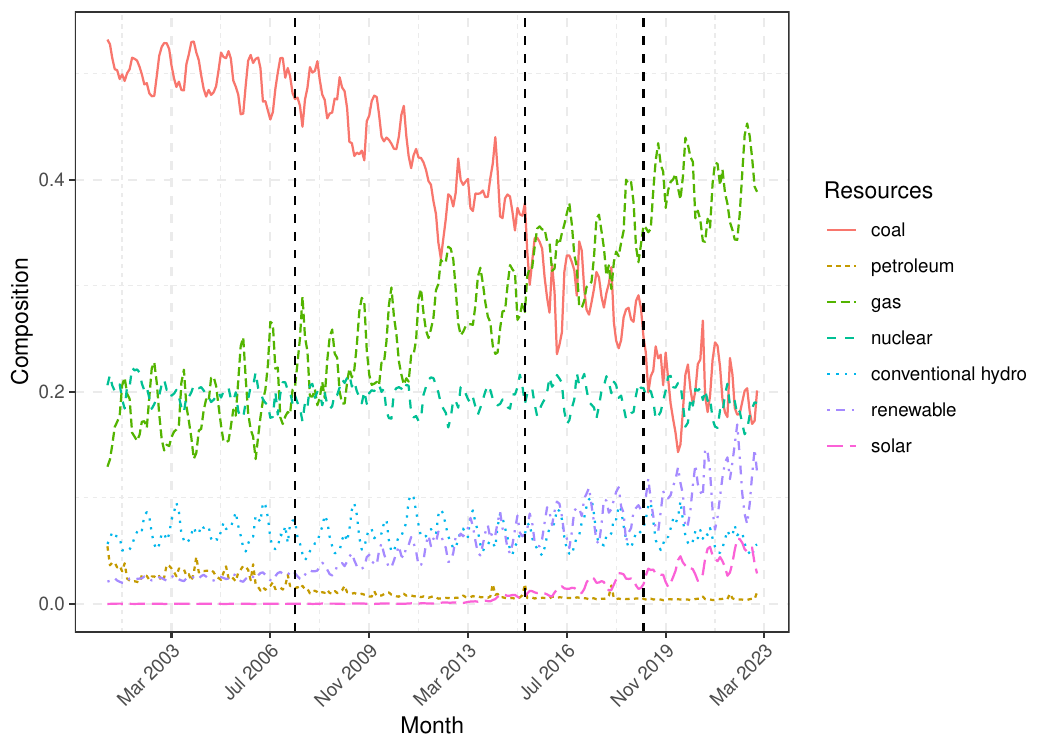}
    \caption{{Multiple change points detected for the U.S. electricity generation data. The change points, indicated by the vertical black dashed lines, are at months: May 2007, February 2015 and February 2019.}}
    \label{fig:elec_multiple}
\end{figure}

Next we illustrate the implementation of Mcpd\_DP \ref{alg:seed_multiple_cpd} on the real data examples investigated in Section \ref{sec: applicaiton}. For the U.S. electricity generation data, Mcpd\_DP detects May 2007, February 2015 and February 2019 as the change points as illustrated in Figure \ref{fig:elec_multiple}. In May 2007, while renewable sources pick up suddenly, the petroleum component starts a rapid downward trend alongside abrupt changes in the trends of the coal component. In February 2015 the solar component starts to accelerate along with continuing changes in the renewable component with similar change patterns again in February 2019. %In comparison, the kernel-CP and the energy-CP detects changepoints at  month 63, month 99, month 130, month 154, month 171, month 194, month 219 and month 243. \textcolor{red}{write the months instead of month numbers.}
{For MIT reality mining data, Mcpd\_DP \ref{alg:seed_multiple_cpd} detects change points at 2004-10-16 (beginning of sponsor week), 2004-12-16 (around the end of the finals week and beginning of the winter break), 2005-01-01 (starting of the independent activities period) and  2005-03-10 (right after the exam week) according to events labeled in \citet{peel:2015}.}

\vspace{-1em}

\section{Discussion}
\label{sec:discussion}

\vspace{-0.5em}

We introduce a nonparametric change point detection method %, denoted as dist-CP, 
designed for random objects taking values in general metric spaces. The proposed scan statistic is based solely on pairwise distances and is tuning parameter-free, making it an appealing choice for practitioners working with complex data who seek a straightforward application without the burden of specifying additional parameters except for interval cut-offs where change points are presumed absent. This stands in contrast to other existing approaches for change point detection in metric space data, which typically necessitate parameter choices such as similarity graph selection in graph-based methods and kernel and bandwidth selection in kernel-based methods. We derive the asymptotic distribution of the proposed  test statistic under $H_0$ to ensure type I error control and establish the large sample consistency of the test and the estimated change points under local alternatives. We extend all rigorous theoretical guarantees to the practicable permutation approximations of the null distribution. Comprehensive simulations across various scenarios, spanning random vectors, distributional data, and networks, showcase the efficacy of the new test in many challenging settings, including scale and tail probability changes in Gaussian random vectors and preferential attachment changes in random networks. We study the extensions to multiple change point scenarios by combining the single change point detection algorithm with seeded binary segmentation. 
The data applications lead to insightful findings in the U.S. electricity generation compositions timeline and in the bluetooth proximity networks of the MIT reality mining experiment.  With its versatility and minimal parameter requirements the proposed tool has the potential for widespread application across various domains as long as distances can be defined between the data elements. Future research could explore several interesting directions, such as understanding the choice of metric in the implementation of dist-CP, adapting dist-CP for discrete distributions on metric spaces, analyzing dist-CP on metric space-valued time series with serial dependence, and conducting a rigorous theoretical analysis of dist-CP when combined with the seeded binary segmentation algorithm in the presence of multiple change points.

\iffalse

\section{Disclosure Statement}
No potential conflict of interest was reported by the authors.

\section{Funding}
Paromita Dubey was supported in part by NSF grant DMS-2311034.

\section{Supplementary Materials}
Section \ref{sec:supp_proof} provides the proofs for Theorem \ref{thm: asymptotic_distribution}, \ref{thm: consistency_of_test} and \ref{thm: consistency_of_chpt}. Section \ref{sec: aux} provides some auxiliary results along with their proofs.
\fi

\singlespacing

\bibliographystyle{apalike}

\bibliography{JASA_template/reference}
\end{document}

\def\spacingset#1{\renewcommand{\baselinestretch}%
{#1}\small\normalsize} \spacingset{1}

%%%%%%%%%%%%%%%%%%%%%%%%%%%%%%%%%%%%%%%%%%%%%%%%%%%%%%%%%%%%%%%%%%%%%%%%%%%%%%

% \if1\blind
% {
%   \title{\bf Supplement for ``Change Point Inference for Non-Euclidean
% Data Sequences using Distance Profiles''}
%   \author{Paromita Dubey\thanks{
%     Paromita Dubey gratefully acknowledges support from NSF grant DMS-2311034. }\hspace{.2cm}
%     and
%     Minxing Zheng \\ Department of Data Sciences and Operations, University of Southern California}
%   \maketitle
% } \fi

% \if0\blind
% {
%   \bigskip
%   \bigskip
%   \bigskip
%   \begin{center}
%     {\LARGE\bf Supplement for ``Change Point Detection for Random Objects using Distance Profiles''}
% \end{center}
%   \medskip
% } \fi

\if1\blind
{
  \title{\bf Supplement for ``Change Point Inference for Non-Euclidean Data Sequences using Distance Profiles''}

  \author[1]{Paromita Dubey\thanks{Paromita Dubey gratefully acknowledges support from NSF grant DMS-2311034. This work was primarily conducted at the University of Southern California.}}
  \author[2]{Minxing Zheng\thanks{This work was primarily conducted while Minxing Zheng was at the University of Southern California.}}

  \affil[1]{Department of Data Sciences and Operations, University of Southern California}
  \affil[2]{Heinz College of Information Systems and Public Policy, Carnegie Mellon University}

  \date{}
  \maketitle
} \fi

\if0\blind
{
  \bigskip
  \bigskip
  \bigskip
  \begin{center}
    {\LARGE\bf Supplement for ``Change Point Detection for Random Objects using Distance Profiles''}
  \end{center}
  \medskip
} \fi
\bigskip
%\newpage
\spacingset{1.9} % DON'T change the spacing!

In the following, some notations have been repeated for constants appearing in multiple proof sub-parts but they are made clear from the context. Moreover, all supremums over uncountable spaces can be easily argued to be equal to supremums over underlying countable dense sets thus avoiding measurability issues wherever necessary.

\section{Proofs of main results}
\label{sec:supp_proof}

\subsection*{Proof of Theorem 1} Since we are operating under $H_0$, we have $P_1=P_2$. Let $Y'$ be an $\Omega$-valued random object generated according to $P_1$ and which is independent of the data. Then for each observation $Y_i$, we define the random distribution function $\Fnull[it]=\prob_{Y'} \left( d(Y_i, Y') \leq t \mid Y_i\right)$ for $t \in \R$. Next for each $u \in \idx_c$ and $t \in [0,\M]$, we define 
\begin{equation*}
    \hat{T}_n(t,u) = \frac{[nu](n-[nu])}{n} \left\lbrace \frac{1}{n} \add[in] (\prehatF[it]-\posthatF[it])^2 \right\rbrace.
\end{equation*}
Since the test statistic $\hat{T}_n = \sup_{u \in \idx_c} \int_0^{\M} \hat{T}_n(t,u) \ dt=\max_{[nc] \leq k \leq n-[nc]} \int_0^{\M} \hat{T}_n\left(t,\frac{k}{n}\right) \ dt$ is a continuous functional of the process $(t,u) \mapsto \hat{T}_n(t,u)$ in the space $\ell^\infty([0,\mathcal{M}] \times \idx_c)$, we will first derive the asymptotic weak convergence of the process $(t,u) \mapsto \hat{T}_n(t,u)$ in the space $\ell^\infty([0,\mathcal{M}] \times \idx_c)$ under $H_0$ and then use the continuous mapping theorem to get the limiting distribution of $\hat{T}_n$ under $H_0$. Let $b_n(u) = \frac{[nu](n-[nu])}{n^2}$. Observe that
\begin{align*}
     & \hat{T}_n(t,u) \\ = & b_n(u) \left\lbrace \add[in] (\prehatF[it]-\posthatF[it])^2 \right \rbrace \\ =  & b_n(u) \left\lbrace  \add[in] \left(  \frac{1}{[nu]} \sum_{j=1}^{[nu]}  \lbrace \ind[d(Y_i,Y_j) \leq t]-\Fnull[it] \rbrace \right. \right. \\  & \hspace{3em} -\left. \left. \frac{1}{(n-[nu])} \sum_{j=[nu]+1}^{n}  \lbrace \ind[d(Y_i,Y_j) \leq t]-\Fnull[it] \rbrace \right)^2 \right\rbrace \\ = & \hat{T}^1_n(t,u) + \hat{T}^2_n(t,u)
\end{align*}
where 
\begin{align*}
    & \hat{T}^1_n(t,u) \\= & b_n(u) \left\lbrace \add[in] \left( \frac{1}{[nu]} \sum_{j=1}^{[nu]}  \lbrace \ind[d(Y_i,Y_j) \leq t]-\Fnull[it] \rbrace -\frac{1}{(n-[nu])} \sum_{j=[nu]+1}^{n}  \lbrace \ind[d(Y_i,Y_j) \leq t]-\Fnull[it] \rbrace \right)^2 \right. \\  - & \left. n b_n(u) \E_{Y} \left( \frac{1}{[nu]} \sum_{j=1}^{[nu]}  \lbrace \ind[d(Y,Y_j) \leq t]-F_{Y}(t) \rbrace -\frac{1}{(n-[nu])} \sum_{j=[nu]+1}^{n}  \lbrace \ind[d(Y,Y_j) \leq t]-F_{Y}(t) \rbrace \right)^2 \right\rbrace
\end{align*}
and 
\begin{align*}
    & \hat{T}^2_n(t,u) \\ = & n b_n(u)  \E_{Y} \left( \frac{1}{[nu]} \sum_{j=1}^{[nu]}  \lbrace \ind[d(Y,Y_j) \leq t]-F_{Y}(t) \rbrace -\frac{1}{(n-[nu])} \sum_{j=[nu]+1}^{n}  \lbrace \ind[d(Y,Y_j) \leq t]-F_{Y}(t) \rbrace \right)^2
\end{align*}
where $Y$ is a random object following $P_1$ that is independent of the data and $Y'$. Next we will show that as $n \rightarrow \infty$, $\sup_{u \in \idx_c, t \in [0,\M]} \hat{T}^1_n(t,u)=o_P(1)$ which by using the Slutsky's theorem implies that the weak limit of the process $(t,u) \mapsto \hat{T}^2_n(t,u)$ determines the weak limit of $(t,u) \mapsto \hat{T}_n(t,u)$ as $n \rightarrow \infty$. 

\noindent \textbf{\underline{Part I:}} $\sup_{u \in \idx_c, t \in [0,\M]} \hat{T}^1_n(t,u)=o_P(1)$ as $n \rightarrow \infty$.
\\ Observe that 
\begin{align*}
    & \left( \frac{1}{[nu]} \sum_{j=1}^{[nu]}  \lbrace \ind[d(Y_i,Y_j) \leq t]-\Fnull[it] \rbrace -\frac{1}{(n-[nu])} \sum_{j=[nu]+1}^{n}  \lbrace \ind[d(Y_i,Y_j) \leq t]-\Fnull[it] \rbrace \right)^2 \\ & =\frac{1}{([nu])^2} \sum_{j=1}^{[nu]} \sum_{k=1}^{[nu]} \lbrace \ind[d(Y_i,Y_j) \leq t]-\Fnull[it] \rbrace\lbrace \ind[d(Y_i,Y_k) \leq t]-\Fnull[it] \rbrace \\ & + \frac{1}{(n-[nu])^2}  \sum_{j=[nu]+1}^{n} \sum_{k=[nu]+1}^{n} \lbrace \ind[d(Y_i,Y_j) \leq t]-\Fnull[it] \rbrace\lbrace \ind[d(Y_i,Y_k) \leq t]-\Fnull[it] \rbrace \\ & - 2 \frac{1}{[nu]} \frac{1}{(n-[nu])} \sum_{j=1}^{[nu]} \sum_{k=[nu]+1}^{n} \lbrace \ind[d(Y_i,Y_j) \leq t]-\Fnull[it] \rbrace\lbrace \ind[d(Y_i,Y_k) \leq t]-\Fnull[it] \rbrace. 
\end{align*}
Define $h_{t}(Y_i,Y_j,Y_k)= \lbrace \ind[d(Y_i,Y_j) \leq t]-\Fnull[it] \rbrace\lbrace \ind[d(Y_i,Y_k) \leq t]-\Fnull[it] \rbrace - \E_{Y_i} ( \lbrace \ind[d(Y_i,Y_j) \leq t]-\Fnull[it] \rbrace\lbrace \ind[d(Y_i,Y_k) \leq t]-\Fnull[it] \rbrace )$ for $t \in [0,\M]$. Then one has
\begin{align*}
    &  \hat{T}^1_n(t,u) =  b_n(u) \{U^1_n(t,u) + U^2_n (t,u)+ U^3_n(t,u)\}
\end{align*}
where 
\begin{equation*}
    U^1_n(t,u)= \frac{1}{([nu])^2}  \add[in] \sum_{j=1}^{[nu]} \sum_{k=1}^{[nu]} h_{t}(Y_i,Y_j,Y_k),
\end{equation*}
\begin{equation*}
    U^2_n(t,u)=  \frac{1}{(n-[nu])^2}  \add[in]  \sum_{j=[nu]+1}^{n} \sum_{k=[nu]+1}^{n}  h_{t}(Y_i,Y_j,Y_k),
\end{equation*}
and
\begin{equation*}
    U^3_n(t,u)= - 2 \frac{1}{[nu]} \frac{1}{(n-[nu])} \add[in]  \sum_{j=1}^{[nu]} \sum_{k=[nu]+1}^{n} h_{t}(Y_i,Y_j,Y_k).
\end{equation*}
Note that $\sup_{u \in \idx_c} \lvert b_n(u)-u(1-u)\rvert| = o_P(1)$ as $n \rightarrow \infty$. Hence to complete part I, we will show that the individual terms $U^1_n(t,u)$, $U^2_n(t,u)$ and $U^3_n(t,u)$ are each asymptotically negligible as $n \rightarrow \infty$. 

\noindent For handling $U^1_n(t,u)$ we divide it as
\begin{align*}
      U^1_n(t,u) = & \frac{1}{([nu])^2} \sum_{i=1}^{[nu]}  \sum_{j=1}^{[nu]} \sum_{k=1}^{[nu]} h_{t}(Y_i,Y_j,Y_k) + \frac{1}{([nu])^2} \sum_{i=[nu]+1}^{n}  \sum_{j=1}^{[nu]} \sum_{k=1}^{[nu]} h_{t}(Y_i,Y_j,Y_k) \\ =  & I+ II + III+IV+V+VI+VII
\end{align*}
where 
\begin{equation*}
    I = \frac{1}{([nu])^2} \sum_{i,j=1, j \neq i}^{[nu]}   h_{t}(Y_i,Y_j,Y_j), \quad II = \frac{1}{([nu])^2} \sum_{i,j=1, j \neq i}^{[nu]}   h_{t}(Y_i,Y_i,Y_j),
\end{equation*}
\begin{equation*}
    III = \frac{1}{([nu])^2} \sum_{i,j=1, j \neq i}^{[nu]}   h_{t}(Y_i,Y_j,Y_i), \quad  IV = \frac{1}{([nu])^2} \sum_{i=1}^{[nu]}   h_{t}(Y_i,Y_i,Y_i),
\end{equation*}
\begin{equation*}
    V = \frac{1}{([nu])^2} \sum_{i,j,k=1, i\neq j \neq k}^{[nu]}   h_{t}(Y_i,Y_j,Y_k), \quad   VI = \frac{1}{([nu])^2} \sum_{i=[nu]+1}^{n}  \sum_{j=1}^{[nu]}  h_{t}(Y_i,Y_j,Y_j),
\end{equation*}
and 
\begin{equation*}
    VII = \frac{1}{([nu])^2} \sum_{i=[nu]+1}^{n}  \sum_{j,k=1,j \neq k}^{[nu]}  h_{t}(Y_i,Y_j,Y_k).
\end{equation*}
Without loss of generality $h_t(x,y,z)$ can be symmetrized in $(x,y,z)$ effortlessly, which allows us to use tools from U-process theory for establishing asymptotic convergence of terms $I$ to $VII$.
\\ \underline{Term $I$:} Observe that term I is asymptotically equivalent to a partial sum U-process of order 2 indexed by $t \in [0,\M]$. Let $\{Y^k_i\}_{i \in \mathbb{N}}$, $k=1,2,3$ be $3$ independent copies of the data $\{Y_i, i \in \mathbb{N}\}$. Let $b_n=\frac{1}{n(n-1)}$. By using the L\'evy-type maximal inequality in Lemma 2.8 together with Corollary 2.15 in \citep{eich:01}, we have for any $\varepsilon > 0$ and for universal constants $C_0, C_1, K_0, K_1 > 0$ that
\begin{align}
\label{eq: levy_I}
    & \prob \left( \sup_{u \in \idx_c} \sup_{t \in [0,\M]} \left \lvert b_n \sum_{i,j=1, j \neq i}^{[nu]}   h_{t}(Y_i,Y_j,Y_j)  \right \rvert  > \varepsilon \right) \nonumber \\ \leq & C_0 \prob \left( \sup_{t \in [0,\M]} \left \lvert b_n \sum_{i,j=1, j \neq i}^{n}   h_{t}(Y_i,Y_j,Y_j) \right \rvert  > C_1 \varepsilon \right) + K_0 \prob \left( \sup_{t \in [0,\M]} \left \lvert b_n \sum_{i=1}^{n}   h_{t}(Y^1_i,Y^2_i,Y^2_i) \right \rvert  > K_1 \varepsilon \right). 
\end{align}
By Lemma \ref{lma: 2sample_1} in  Section \ref{sec: aux} we conclude that under $H_0$ and assumption A1,
\begin{equation}
\label{eq: part1_I}
    \sup_{t \in [0,\M]} \left \lvert b_n \sum_{i,j=1, j \neq i}^{n}   h_{t}(Y_i,Y_j,Y_j) \right \rvert = o_P(1)
\end{equation}
as $n \rightarrow \infty$. Next see that $\sup_{t \in [0,\M]} \sup_{x,y,z \in \Omega} \lvert h_{t}(x,y,z) \rvert \leq 1$. Hence as $n \rightarrow \infty$
\begin{equation}
\label{eq: part2_I}
    \sup_{t \in [0,\M]} \left \lvert b_n \sum_{i=1}^{n}   h_{t}(Y^1_i,Y^2_i,Y^2_i) \right \rvert = o_P(1).
\end{equation}
Therefore combining \eqref{eq: part1_I} and \eqref{eq: part2_I} in conjunction with \eqref{eq: levy_I}, it turns out that as $n \rightarrow \infty$ one has $$\sup_{u \in \idx_c} \sup_{t \in [0,\M]} \left \lvert b_n \sum_{i,j=1, j \neq i}^{[nu]}   h_{t}(Y_i,Y_j,Y_j)  \right \rvert = o_P(1).$$ Since $\sup_{u \in \idx_c} \frac{1}{b_n ([nu])^2}=O(1)$ as $n \rightarrow \infty$, one has that term $\sup_{u \in \idx_c} \sup_{t \in [0,\M]} I = o_P(1)$. 
\\ \underline{Terms $II$ and $III$:} Observe that by symmetry Term $II$ is equal to Term $III$. Hence we only deal with Term $II$, which similar to term I is asymptotically equivalent to a partial sum U-process of order 2 indexed by $t \in [0,\M]$. We borrow notations from the arguments used for Term $I$, and use Lemma 2.8 together with Corollary 2.15 in \citep{eich:01} to have for any $\varepsilon > 0$ that
\begin{align}
\label{eq: levy_II}
    & \prob \left( \sup_{u \in \idx_c} \sup_{t \in [0,\M]} \left \lvert b_n \sum_{i,j=1, j \neq i}^{[nu]}   h_{t}(Y_i,Y_i,Y_j)  \right \rvert  > \varepsilon \right) \nonumber \\ \leq & C_0 \prob \left( \sup_{t \in [0,\M]} \left \lvert b_n \sum_{i,j=1, j \neq i}^{n}   h_{t}(Y_i,Y_i,Y_j) \right \rvert  > C_1 \varepsilon \right) + 
    K_0 \prob \left( \sup_{t \in [0,\M]} \left \lvert b_n \sum_{i=1}^{n}   h_{t}(Y^1_i,Y^2_i,Y^2_i) \right \rvert  > K_1 \varepsilon \right). 
\end{align}
By Lemma \ref{lma: 2sample_1} one has $\sup_{t \in [0,\M]} \left \lvert b_n \sum_{i,j=1, j \neq i}^{n}   h_{t}(Y_i,Y_i,Y_j) \right \rvert = o_P(1)$ as $n \rightarrow \infty$ under $H_0$ and assumption A1. This together with \eqref{eq: part2_I} and \eqref{eq: levy_II} implies that as $n \rightarrow \infty$ one has $\sup_{u \in \idx_c} \sup_{t \in [0,\M]} \left \lvert b_n \sum_{i,j=1, j \neq i}^{[nu]}   h_{t}(Y_i,Y_i,Y_j)  \right \rvert = o_P(1)$. Since $\sup_{u \in \idx_c} \frac{1}{b_n ([nu])^2}=O(1)$ as $n \rightarrow \infty$, one has that term $\sup_{u \in \idx_c} \sup_{t \in [0,\M]} II = o_P(1)$. 
\\ \underline{Term $IV$:} Since $\sup_{t \in [0,\M]} \sup_{x,y,z \in \Omega} \lvert h_{t}(x,y,z) \rvert \leq 1$, one has term $\sup_{u \in \idx_c} \sup_{t \in [0,\M]} IV \leq \frac{1}{[nu]}$, and therefore $\sup_{u \in \idx_c} \sup_{t \in [0,\M]} IV=o_P(1)$ as $n \rightarrow \infty$. 
\\ \underline{Term $V$:} Similar to Term $I$, by using Lemma 2.8 together with Corollary 2.15 in \citep{eich:01}, we have for any $\varepsilon > 0$ and for universal constants $B_0,\dots, B_{15} > 0$ that
\begin{align}
\label{eq: levy_V}
    & \prob \left( \sup_{u \in \idx_c} \sup_{t \in [0,\M]} \left \lvert b_n \sum_{i,j,k=1, i\neq j \neq k}^{[nu]}   h_{t}(Y_i,Y_j,Y_k)  \right \rvert  > \varepsilon \right) \nonumber \\ \leq & B_0 \prob \left( \sup_{t \in [0,\M]} \left \lvert b_n \sum_{i,j,k=1, i\neq j \neq k}^{n}   h_{t}(Y_i,Y_j,Y_k)  \right \rvert  > B_1 \varepsilon \right) + \Delta_1  + \Delta_2
\end{align}
where 
\begin{equation*}
    \Delta_1 =  B_2 \prob \left( \sup_{t \in [0,\M]} \left \lvert b_n \sum_{i=1}^{n}   h_{t}(Y^1_i,Y^2_i,Y^3_i) \right \rvert  > B_3 \varepsilon \right),
\end{equation*}
and 
\begin{align*}
    \Delta_2  =  & B_4 \prob \left( \sup_{t \in [0,\M]} \left \lvert b_n \sum_{i,j=1, j \neq i}^{n}   h_{t}(Y^1_i,Y^2_i,Y^3_j) \right \rvert  > B_5 \varepsilon \right) \\ & + B_6 \prob \left( \sup_{t \in [0,\M]} \left \lvert b_n \sum_{i,j=1, j \neq i}^{n}   h_{t}(Y^1_i,Y^2_j,Y^3_i) \right \rvert  > B_7 \varepsilon \right) \\ & + B_8 \prob \left( \sup_{t \in [0,\M]} \left \lvert b_n \sum_{i,j=1, j \neq i}^{n}   h_{t}(Y^1_j,Y^2_i,Y^3_i) \right \rvert  > B_9 \varepsilon \right)  \\ & + B_{10} \prob \left( \sup_{t \in [0,\M]} \left \lvert b_n \sum_{i,j=1, j \neq i}^{n}   h_{t}(Y^1_i,Y^2_j,Y^3_j) \right \rvert  > B_{11} \varepsilon \right)
     \\ & + B_{12} \prob \left( \sup_{t \in [0,\M]} \left \lvert b_n \sum_{i,j=1, j \neq i}^{n}   h_{t}(Y^1_j,Y^2_i,Y^3_j) \right \rvert  > B_{13} \varepsilon \right) \\ & +  B_{14} \prob \left( \sup_{t \in [0,\M]} \left \lvert b_n \sum_{i,j=1, j \neq i}^{n}   h_{t}(Y^1_j,Y^2_j,Y^3_i) \right \rvert  > B_{15} \varepsilon \right).
\end{align*}
Note that as $n \rightarrow \infty$, $nb_n \rightarrow 0$ and since $\sup_{t \in [0,\M]} \sup_{x,y,z \in \Omega} \lvert h_{t}(x,y,z) \rvert \leq 1$, one has that $\Delta_1 \rightarrow 0$ as $n \rightarrow \infty$ for any $\varepsilon > 0$. Since without loss of generality $(x,y,z) \mapsto h_t(x,y,z)$ can be considered symmetric in $(x,y,z)$, the inequality 2.10 in Theorem 2.2 of \cite{gine:97}, there exists constants $D_0, \dots, D_3$ such that
\begin{align*}
    \Delta_2 \leq & D_0 \left\lbrace  \prob \left( \sup_{t \in [0,\M]} \left \lvert b_n \sum_{i,j=1, j \neq i}^{n}   h_{t}(Y_i,Y_i,Y_j) \right \rvert  > D_1 \varepsilon \right) \right. \\ + &  \left. \prob \left( \sup_{t \in [0,\M]} \left \lvert b_n \sum_{i,j=1, j \neq i}^{n}   h_{t}(Y_i,Y_j,Y_i) \right \rvert  > D_2 \varepsilon \right) + \prob \left( \sup_{t \in [0,\M]} \left \lvert b_n \sum_{i,j=1, j \neq i}^{n}   h_{t}(Y_j,Y_i,Y_i) \right \rvert  > D_3 \varepsilon \right) \right\rbrace. 
\end{align*}
Using the arguments for terms $I$, $II$ and $III$ we conclude that $\Delta_2 \rightarrow 0$ as $n \rightarrow \infty$.

By Lemma \ref{lma: 2sample_2} one has under $H_0$ and assumption A1,
\begin{equation}
\label{eq: part1_V}
    \sup_{t \in [0,\M]} \left \lvert b_n \sum_{i,j,k=1, i\neq j \neq k}^{n}   h_{t}(Y_i,Y_j,Y_k)  \right \rvert = o_P(1)
\end{equation}
as $n \rightarrow \infty$. Then \eqref{eq: part1_V} together with the fact that $\Delta_1 \rightarrow 0$ and $\Delta_2 \rightarrow 0$ as $n \rightarrow \infty$ implies that as $n \rightarrow \infty$ one has $\sup_{u \in \idx_c} \sup_{t \in [0,\M]} \left \lvert b_n \sum_{i,j,k=1, i\neq j \neq k}^{[nu]}   h_{t}(Y_i,Y_j,Y_k)  \right \rvert  = o_P(1)$. Since $\sup_{u \in \idx_c} \frac{1}{b_n ([nu])^2}=O(1)$ as $n \rightarrow \infty$, one has that term $\sup_{u \in \idx_c} \sup_{t \in [0,\M]} V = o_P(1)$. 
\\ \underline{Term $VI$:} Let $Z_1, \dots, Z_n$ be an enumeration of $Y_1, \dots, Y_n$ in the reverse order, that is, $Z_i=Y_{n-i+1}$. With this observe that $Z_1, \dots, Z_{(n-[nu])}$ are independent and also identically distributed as $Y_1, \dots, Y_{[nu]}$ under $H_0$. Since $\sup_{u \in \idx_c} \frac{n^2}{ ([nu])^2}=O(1)$ as $n \rightarrow \infty$, it is enough to show that a rescaled version of $VI$ given by $\widetilde{VI}$ is asymptotically negligible, where
\begin{equation*}
    \widetilde{VI}=\frac{1}{n^2} \sum_{i=1}^{(n-[nu])} \sum_{j=1}^{[nu]} h_t(Z_i,Y_j,Y_j).
\end{equation*}
Observe that 
\begin{equation*}
    \widetilde{VI}=\frac{1}{n} \sum_{j=1}^{[nu]} g^Z_{t,u}(Y_j,Y_j)
\end{equation*}
where $g^Z_{t,u}(x,y)=\frac{1}{n}\sum_{i=1}^{(n-[nu])} h_t(Z_i,x,y)$ for $x,y \in \Omega$. By L\'evy's inequality (refer to page 431 in \cite{well:96}) one has for any $\varepsilon > 0$ that 
\begin{equation}
\label{eq: levy_VI}
    \prob \left( \sup_{u \in \idx_c} \sup_{t \in [0,\M]} \sup_{x,y \in \Omega} \left\lvert \frac{1}{n}\sum_{i=1}^{(n-[nu])} h_t(Z_i,x,y) \right\rvert > \varepsilon \right) \leq 2 \prob \left( \sup_{t \in [0,\M]} \sup_{x,y\in \Omega} \left\lvert \frac{1}{n}\sum_{i=1}^{n} h_t(Z_i,x,y) \right\rvert > \varepsilon \right).
\end{equation}
By Lemma \ref{lma: 2sample_3} in conjunction with \eqref{eq: levy_VI} one has as $n \rightarrow \infty$,
\begin{equation}
\label{eq: inner}
    \sup_{u \in \idx_c} \sup_{t \in [0,\M]} \sup_{x,y \in \Omega} \left\lvert g^Z_{t,u}(x,y) \right\rvert = o_P(1).
\end{equation}
Since $\sup_{u \in \idx_c} \sup_{t \in [0,\M]} \widetilde{VI} \leq  \sup_{u \in \idx_c} \sup_{t \in [0,\M]} \sup_{x,y \in \Omega} \left\lvert g^Z_{t,u}(x,y) \right\rvert $ almost surely, one has that $\sup_{u \in \idx_c} \sup_{t \in [0,\M]} \widetilde{VI} = o_P(1)$ as $n \rightarrow \infty$. 
\\ \underline{Term $VII$:} Similar to term $VI$, it is enough to show that the rescaled term
\begin{equation*}
     \widetilde{VII}=\frac{1}{n} \sum_{j,k=1, j \neq k}^{[nu]} g^Z_{t,u}(Y_j,Y_k)
\end{equation*}
is asymptotically negligible.  By the independence of $\{Z_1, \dots, Z_{n-[nu]}\}$ and $Y_1, \dots, Y_{[nu]}$, it turns out that  for any $\varepsilon > 0$, 
\begin{equation*}
    \prob \left( \sup_{u \in \idx_c} \sup_{t \in [0,\M]} \left\lvert \widetilde{VII} \right\rvert > \varepsilon \right) = \prob \left( \sup_{u \in \idx_c} \sup_{t \in [0,\M]} \left\lvert \widetilde{VII}' \right\rvert > \varepsilon \right) 
\end{equation*}
where 
\begin{equation*}
     \widetilde{VII}' = \frac{1}{n} \sum_{j,k=1, j \neq k}^{[nu]} g^{Z'}_{t,u}(Y_j,Y_k)
\end{equation*}
with $Z'_1, \dots, Z'_n$ being an independent copy of $Z_1, \dots, Z_n$, and $g^{Z'}_{t,u}(x,y)=\frac{1}{n}\sum_{i=1}^{(n-[nu])} h_t(Z'_i,x,y)$ for $x,y \in \Omega$. Moreover since 
\begin{equation*}
    \sup_{u \in \idx_c} \sup_{t \in [0,\M]} \left\lvert \widetilde{VII}' \right\rvert \leq \sup_{u \in \idx_c} \sup_{v \in \idx_c} \sup_{t \in [0,\M]} \left\lvert \widetilde{VII}'' \right\rvert
\end{equation*}
where $\widetilde{VII}''=\frac{1}{n} \sum_{j,k=1, j \neq k}^{[nu]} g^{Z'}_{t,v}(Y_j,Y_k)$, it is enough to show that $\sup_{u \in \idx_c} \sup_{v \in \idx_c} \sup_{t \in [0,\M]} \left\lvert \widetilde{VII}'' \right\rvert = o_P(1)$ as $n \rightarrow \infty$. Let $A_t$ denote the event $A_t=\{\sup_{\omega_1,\omega_2 \in \Omega} \sup_{v \in \idx_c}\sup_{t \in [0,\M]} g^{Z'}_{t,v}(\omega_1,\omega_2) \leq t\}$. Note that for any $\epsilon > 0$
\begin{equation*}
    \prob \left( \sup_{u \in \idx_c} \sup_{v \in \idx_c} \sup_{t \in [0,\M]} \left\lvert \widetilde{VII}'' \right\rvert > \epsilon \right) \leq  \prob \left( \sup_{u \in \idx_c} \sup_{v \in \idx_c} \sup_{t \in [0,\M]} \left\lvert \widetilde{VII}'' \right\rvert > \epsilon, A_t \right) + P(A_t^C). 
\end{equation*}
By \eqref{eq: inner} $P(A_t^C) \rightarrow 0$ as $n \rightarrow \infty$ and therefore it is enough to show that as $n \rightarrow \infty$ for any $\epsilon> 0$ one has $\prob \left( \sup_{u \in \idx_c} \sup_{v \in \idx_c} \sup_{t \in [0,\M]} \left\lvert \widetilde{VII}'' \right\rvert > \epsilon, A_t \right) \rightarrow 0$. Next see that 
\begin{align}
    & \prob \left( \sup_{u \in \idx_c} \sup_{v \in \idx_c} \sup_{t \in [0,\M]} \left\lvert \widetilde{VII}'' \right\rvert > \epsilon, A_t \right) \nonumber \\ = & \E_{Z'_1, \dots, Z'_n} \left( \prob_{|\mathcal{Z}'} \left( \sup_{u \in \idx_c} \sup_{v \in \idx_c} \sup_{t \in [0,\M]} \left\lvert \widetilde{VII}'' \right\rvert > \epsilon , A_t \right)\right) \prob(A_t) \label{eq: exp_bound}
\end{align}
By Lemma 2.8 together with Corollary 2.15 in \citep{eich:01} conditionally on $\{Z_1', \dots, Z'_{n}\}$ and $A_t$ one has for any $\epsilon > 0$ and constants $C_0,C_1,K_0,K_1 > 0$ that
\begin{align*}
    \prob_{|\mathcal{Z}'} \left( \sup_{u \in \idx_c} \sup_{v \in \idx_c} \sup_{t \in [0,\M]} \left\lvert \widetilde{VII}'' \right\rvert > \epsilon ,A_t \right) \leq & C_0 \prob_{|\mathcal{Z}'}\left(  \sup_{t \in [0,\M]} \sup_{v \in \idx_c} \left\lvert \frac{1}{n} \sum_{j,k=1, j \neq k}^{n} g^{Z'}_{t,v}(Y_j,Y_k) \right\rvert > C_1 \epsilon , A_t  \right) \\ + & K_0 \prob_{|\mathcal{Z}'}\left(  \sup_{t \in [0,\M]} \sup_{v \in \idx_c} \left\lvert \frac{1}{n} \sum_{j=1}^{n} g^{Z'}_{t,v}(Y^1_j,Y^2_j) \right\rvert > K_1 \epsilon , A_t \right)
\end{align*}
where $\prob_{|\mathcal{Z}'}(\cdot)=\prob(\cdot |Z'_1, \dots, Z'_n)$. Using the inequality 2.10 in Theorem 2.2 of \cite{gine:97} for some constant $D>0$
\begin{align*}
    & \prob_{|\mathcal{Z}'}\left(  \sup_{t \in [0,\M]} \sup_{v \in \idx_c} \left\lvert \frac{1}{n} \sum_{j=1}^{n} g^{Z'}_{t,v}(Y^1_j,Y^2_j) \right\rvert > K_1 \epsilon \mid A_t \right) \\
    \leq &  D \prob_{|\mathcal{Z}'}\left(  \sup_{t \in [0,\M]} \sup_{v \in \idx_c} \left\lvert \frac{1}{n} \sum_{j=1}^{n} g^{Z'}_{t,v}(Y_j,Y_j) \right\rvert > K_1 \epsilon \mid A_t \right).
\end{align*}
See that $\prob_{|\mathcal{Z}'}\left(  \sup_{t \in [0,\M]} \sup_{v \in \idx_c} \left\lvert \frac{1}{n} \sum_{j=1}^{n} g^{Z'}_{t,v}(Y_j,Y_j) \right\rvert > K_1 \epsilon \mid A_t \right)=0$ almost surely is implied by $A_t$ by taking $t \leq K_1\epsilon$. For any such $t$ using \eqref{eq: exp_bound} one has
\begin{align*}
    & \E_{Z'_1, \dots, Z'_n} \left( \prob_{|\mathcal{Z}'} \left( \sup_{u \in \idx_c} \sup_{v \in \idx_c} \sup_{t \in [0,\M]} \left\lvert \widetilde{VII}'' \right\rvert > \epsilon , A_t \right)\right)\\ \leq & C_0 \prob \left(  \sup_{t \in [0,\M]} \sup_{v \in \idx_c} \left\lvert \frac{1}{n} \sum_{j,k=1, j \neq k}^{n} g^{Z'}_{t,v}(Y_j,Y_k) \right\rvert > C_1 \epsilon , A_t  \right).
\end{align*}
Using L\'evy's inequality from page 431 of \cite{well:96} first conditionally on $Y_1, \dots, Y_n$ and then taking expectation over $Y_1, \dots, Y_n$ one has 
\begin{align*}
    &\prob \left(  \sup_{t \in [0,\M]} \sup_{v \in \idx_c} \left\lvert \frac{1}{n} \sum_{j,k=1, j \neq k}^{n} g^{Z'}_{t,v}(Y_j,Y_k) \right\rvert > C_1 \epsilon , A_t  \right) \\=& \prob \left(  \sup_{t \in [0,\M]} \sup_{v \in \idx_c} \left\lvert \frac{1}{n} \sum_{i=1}^{(n-[nv])} \frac{1}{n} \sum_{j,k=1, j \neq k}^{n} h_t(Z'_i,Y_j,Y_k) \right\rvert > C_1 \epsilon, A_t  \right) \\ \leq & 2 \prob \left(  \sup_{t \in [0,\M]} \left\lvert \frac{1}{n} \sum_{j,k=1, j \neq k}^{n} g^{Z'}_{t,0}(Y_j,Y_k) \right\rvert > C_1 \epsilon, A_t  \right)
\end{align*}
and therefore 
\begin{align*}
      & \E_{Z'_1, \dots, Z'_n} \left( \prob_{|\mathcal{Z}'} \left( \sup_{u \in \idx_c} \sup_{v \in \idx_c} \sup_{t \in [0,\M]} \left\lvert \widetilde{VII}'' \right\rvert > \epsilon , A_t \right)\right) \leq \\ &  \E_{Z'_1, \dots, Z'_n} \left( \prob_{|\mathcal{Z}'} \left(  \sup_{t \in [0,\M]} \left\lvert \frac{1}{n} \sum_{j,k=1, j \neq k}^{n} g^{Z'}_{t,0}(Y_j,Y_k) \right\rvert > C_1 \epsilon, A_t  \right)\right).
\end{align*}
By Lemma \ref{lma: 2sample_4} one has for some $0 < \alpha < 1$ that
\begin{equation*}
    \prob_{|\mathcal{Z}'}\left(  \sup_{t \in [0,\M]}  \left\lvert \frac{1}{n} \sum_{j,k=1, j \neq k}^{n} g^{Z'}_{t,0}(Y_j,Y_k) \right\rvert > C_1 \epsilon, A_t  \right) \leq const. \frac{t^\alpha}{\epsilon}.
\end{equation*}
By letting $t \rightarrow 0$ and $n \rightarrow \infty$ and using \eqref{eq: exp_bound} one has $\prob \left( \sup_{u \in \idx_c} \sup_{v \in \idx_c} \sup_{t \in [0,\M]} \left\lvert \widetilde{VII}'' \right\rvert > \epsilon, A_t \right) \rightarrow 0$ as $n \rightarrow \infty$ thereby completing the proof.

Observe that $U^2_n(t,u)$ is analogous to $U^1_n(t,1-u))$ evaluated on $Z_1, \dots, Z_n$, hence the steps to establish the asymptotic negligibility of $U^1_n(t,u)$ carry over directly to establish  $\sup_{t \in [0,\M]}\sup_{u \in \idx_c} \lvert U^2_n(t,u)\rvert=o_P(1)$ as $n \rightarrow \infty$.

Finally see that
\begin{align*}
    U^3_n(t,u)= VIII + IX
\end{align*}
where 
\begin{equation*}
    VIII =  \frac{-2}{[nu](n-[nu])} \sum_{i=1}^{[nu]}  \sum_{j=1}^{[nu]} \sum_{k=[nu]+1}^{n} h_{t}(Y_i,Y_j,Y_k),
\end{equation*}
and
\begin{equation*}
    IX =  \frac{-2}{[nu](n-[nu])} \sum_{i=[nu]+1}^{n}  \sum_{j=1}^{[nu]} \sum_{k=[nu]+1}^{n} h_{t}(Y_i,Y_j,Y_k).
\end{equation*}
Observe that since $IX$ can be expressed as
\begin{equation*}
    IX =  \frac{-2}{[nu](n-[nu])} \sum_{i=1}^{(n-[nu])}   \sum_{j=1}^{(n-[nu])} \sum_{k=[nu]+1}^{n} h_{t}(Z_i,Z_j,Z_k),
\end{equation*}
it is enough to deal with term $VIII$ as $IX$ is analogous to $VIII$ evaluated at $(1-u)$ with $Z_1, \dots, Z_n$. Hence to complete part I, it remains to show that $\sup_{t \in [0,\M]} \sup_{u \in \idx_c} \lvert VIII \rvert =o_P(1)$ as $n \rightarrow \infty$.  Since $\sup_{u \in \idx_c} \frac{n^2}{ ([nu])(n-[nu])}=O(1)$ as $n \rightarrow \infty$, it is enough to show that a rescaled version of $VIII$ given by $\widetilde{VIII}$ is asymptotically negligible, where
\begin{equation*}
    \widetilde{VIII}=\frac{-2}{n^2} \sum_{i=1}^{[nu]}  \sum_{j=1}^{[nu]} \sum_{k=[nu]+1}^{n} h_{t}(Y_i,Y_j,Y_k).
\end{equation*}
Similar to the arguments for term $VII$, it is enough to show that $\sup_{u \in \idx_c} \sup_{v \in \idx_c} \sup_{t \in [0,\M]} \left\lvert \widetilde{VIII}'' \right\rvert = o_P(1)$ as $n \rightarrow \infty$ where
\begin{equation*}
    \widetilde{VIII}''=\frac{-2}{n^2} \sum_{i=1}^{[nu]}  \sum_{j=1}^{[nu]} \sum_{k=1}^{[nv]} h_{t}(Y_i,Y_j,Z'_k).
\end{equation*}
Define $f_{t,u}(x,y)=\frac{1}{n}\sum_{k=1}^{[nv]} h_t(x,y,Z'_k)$ such that $\widetilde{VIII}''=\frac{-2}{n}\sum_{i=1}^{[nu]}  \sum_{j=1}^{[nu]} f_{t,v}(Y_i,Y_j)$. Note that $\sup_{u \in \idx_c} \sup_{v \in \idx_c} \sup_{t \in [0,\M]} \left \lvert \frac{-2}{n} \sum_{j=1}^{[nu]} f_{t,v}(Y_j,Y_j) \right \rvert = o_P(1)$ as $n \rightarrow \infty$ by Lemma \ref{lma: 2sample_5}. Hence it is enough to show that $\sup_{u \in \idx_c} \sup_{v \in \idx_c} \sup_{t \in [0,\M]}  \left \lvert \frac{-2}{n} \sum_{i,j=1, j \neq i}^{[nu]} f_{t,v}(Y_i,Y_j) \right \rvert = o_{P}(1)$ as $n \rightarrow \infty$. Next see that for any $\epsilon > 0$,
\begin{align*}
    & \prob \left( \sup_{u \in \idx_c} \sup_{v \in \idx_c} \sup_{t \in [0,\M]}  \left \lvert \frac{-2}{n} \sum_{i,j=1, j \neq i}^{[nu]} f_{t,v}(Y_i,Y_j) \right \rvert > \epsilon \right) \\ = & \E_{Z'_1, \dots, Z'_n} \left \lbrace \prob_{|\mathcal{Z}'} \left( \sup_{u \in \idx_c} \sup_{v \in \idx_c} \sup_{t \in [0,\M]}  \left \lvert \frac{-2}{n} \sum_{i,j=1, j \neq i}^{[nu]} f_{t,v}(Y_i,Y_j) \right \rvert > \epsilon \right) \right\rbrace.
\end{align*}
Using Lemma 2.8 together with Corollary 2.15 in \citep{eich:01} conditionally on $\{Z_1', \dots, Z'_{n}\}$ one has for any $\epsilon > 0$ and constants $C_0,C_1,K_0,K_1 > 0$ that
\begin{align*}
    & \prob_{|\mathcal{Z}'} \left( \sup_{u \in \idx_c} \sup_{v \in \idx_c} \sup_{t \in [0,\M]}  \left \lvert \frac{-2}{n} \sum_{i,j=1, j \neq i}^{[nu]} f_{t,v}(Y_i,Y_j) \right \rvert > \epsilon \right) \\ \leq & C_0 \prob_{|\mathcal{Z}'}\left(  \sup_{t \in [0,\M]} \sup_{v \in \idx_c} \left\lvert \frac{2}{n} \sum_{i,j=1, j \neq i}^{n} f_{t,v}(Y_i,Y_j) \right\rvert > C_1 \epsilon  \right) \\ + & K_0 \prob_{|\mathcal{Z}'}\left(  \sup_{t \in [0,\M]} \sup_{v \in \idx_c} \left\lvert \frac{2}{n} \sum_{j=1}^{n} f_{t,v}(Y^1_j,Y^2_j) \right\rvert > K_1 \epsilon  \right)
\end{align*}
where $\prob_{|\mathcal{Z}'}(\cdot)=\prob(\cdot |Z'_1, \dots, Z'_n)$. Using the inequality 2.10 in Theorem 2.2 of \cite{gine:97} for some constant $D>0$ one has unconditionally
\begin{align*}
    & \prob \left( \sup_{u \in \idx_c} \sup_{v \in \idx_c} \sup_{t \in [0,\M]}  \left \lvert \frac{-2}{n} \sum_{i,j=1, j \neq i}^{[nu]} f_{t,v}(Y_i,Y_j) \right \rvert > \epsilon \right) \\ \leq & C_0 \prob\left(  \sup_{t \in [0,\M]} \sup_{v \in \idx_c} \left\lvert \frac{2}{n} \sum_{i,j=1, j \neq i}^{n} f_{t,v}(Y_i,Y_j) \right\rvert > C_1 \epsilon  \right) \\ + & K_0 D \  \prob \left(  \sup_{t \in [0,\M]} \sup_{v \in \idx_c} \left\lvert \frac{2}{n} \sum_{j=1}^{n} f_{t,v}(Y_j,Y_j) \right\rvert > K_1 \epsilon  \right)
\end{align*}
where $\prob \left(  \sup_{t \in [0,\M]} \sup_{v \in \idx_c} \left\lvert \frac{2}{n} \sum_{j=1}^{n} f_{t,v}(Y_j,Y_j) \right\rvert > K_1 \epsilon  \right) \rightarrow 0$ as $n \rightarrow \infty$ since by Lemma \ref{lma: 2sample_5} one has 
$$ \sup_{t \in [0,\M]} \sup_{\omega \in \Omega} \left \lvert \frac{1}{n}  \sum_{k=1}^{n}  h_t(\omega,\omega,Z'_k) \right \rvert =o_P(1),$$ as $n \rightarrow \infty$.  Hence it remains to show that $\prob\left(  \sup_{t \in [0,\M]} \sup_{v \in \idx_c} \left\lvert \frac{2}{n} \sum_{i,j=1, j \neq i}^{n} f_{t,v}(Y_i,Y_j) \right\rvert > C_1 \epsilon  \right) \rightarrow 0$ as $n \rightarrow \infty$. Using L\'evy's inequality on page 431 of \cite{well:96} conditional on $Y_1, \dots, Y_n$ one has
\begin{align*}
    & \prob_{Y_1, \dots, Y_n} \left(  \sup_{t \in [0,\M]} \sup_{v \in \idx_c} \left\lvert \frac{2}{n} \sum_{i,j=1, j \neq i}^{n} f_{t,v}(Y_i,Y_j) \right\rvert > C_1 \epsilon  \right) \\ \leq & 2 \prob_{Y_1, \dots, Y_n} \left(  \sup_{t \in [0,\M]} \left\lvert \frac{2}{n} \sum_{i,j=1, j \neq i}^{n} f_{t,1} (Y_i,Y_j) \right\rvert > C_1 \epsilon  \right)
\end{align*}
and therefore it is enough to show that $\prob\left(  \sup_{t \in [0,\M]}  \left\lvert \frac{2}{n} \sum_{i,j=1, j \neq i}^{n} f_{t,1}(Y_i,Y_j) \right\rvert > C_1 \epsilon  \right) \rightarrow 0$ as $n \rightarrow \infty$. Observe that $\E_{|\mathcal{Z}'}\left( f_{t,v}(Y_i,Y_j) |Y_i \right)=\E_{|\mathcal{Z}'}\left( f_{t,v}(Y_i,Y_j) |Y_j \right)=0$ almost surely and therefore conditionally on $Z'_1, \dots, Z'_n$, $\left \lvert \frac{1}{n(n-1)} \sum_{i,j=1, j \neq i}^{n} f_{t,1}(Y_i,Y_j) \right \rvert$ is a degenerate $U$-process of order two indexed by the function class $\mathcal{F}^{Z'}=\{(x,y) \mapsto f_{t,1}(x,y): t \in [0,\M]\}$. One has
\begin{align*}
  & \lVert f_{s,1}(Y',Y'')-f_{t,1}(Y',Y'') \rVert_{L_2(P_1 \times P_1)} \\ \leq\ & \sup_{\omega \in \Omega} \|h_s(Y,Y",\omega)-h_t(Y,Y",\omega)\|_{L_2(P_1 \times P_1)} \\ & + \sup_{\omega \in \Omega} \|\E_{Y}(h_s(Y,Y",\omega)-h_t(Y,Y",\omega))\|_{L_2(P_1 \times P_1)},
\end{align*}
where $Y$ $Y''$ are independently distributed as $P_1$ and are independent of the data $Y_1,\dots,Y_n$. Replicating the arguments in the proof of Lemma \ref{lma: 2sample_1} and using assumption A1 one has for some constant $C > 0$
\begin{equation*}
    \sup_{\omega \in \Omega} \|h_s(Y,Y",\omega)-h_t(Y,Y",\omega)\|_{L_2(P_1 \times P_1)} \leq  C\epsilon
\end{equation*}
whenever $0 < \epsilon \leq 1$ and $|s-t|^2 < \epsilon^2$ and therefore by Jensen's inequality,
\begin{equation*}
\lVert f_{s,1}(Y',Y'')-f_{t,1}(Y',Y'') \rVert_{L_2(P_1 \times P_1)} \leq C\epsilon. 
\end{equation*}
% \blu{Therefore the $L_2(P_1 \times P_1)$ covering number $N(\epsilon, \mathcal{H},L_2(P_1 \times P_1))$ of $\mathcal{H}$ with balls of radius $\epsilon$ is upper bounded by the covering number of $\mcM$ with balls of radius $\frac{\epsilon^2}{4(C'_U)^2\left \lbrace 3C_0 + 2 \sqrt{C_0} \right \rbrace^2}$, [original; to be removed]} \yc{
Therefore, conditioning on $Z'_1, \dots, Z'_n$, the packing number $D(\epsilon, \mathcal{F}^{Z'},L_2(P_1 \times P_1))$, i.e. the maximum number of $\epsilon$-separate elements in $\mathcal{F}^{Z'}$ is upper bounded by the packing number $D(\tfrac{\epsilon^2}{C^2},[0,\M],d_E)$, i.e. the maximum number of $\tfrac{\epsilon^2}{C^2}$-separate elements in $[0,\M]$, equal to a constant times $\epsilon^{-2}$, %} 
which implies that the class $\mathcal{F}^{Z'}$ is Euclidean conditioning on $Z'_1, \dots, Z'_m$. 
Note that an envelope function for $\mathcal{F}^{Z'}$ is given by the function $F = 2 \sup_{\omega \in \Omega} \sup_{t \in [0,\M]} |\tilde{F}_\omega(t)|$
with $\tilde{F}_\omega(t)$ defined as in the proof of Lemma \ref{lma: 2sample_5}.
Following the proof of Corollary 4 in \cite{sher:94} one has for some $0 < \alpha < 1$,
\begin{equation*}
  \E \left( {\sup_{t\in[0,\M]}} n \left \lvert \frac{1}{n(n-1)} \sum_{i,j=1, j \neq i}^{n} f_{t,1}(Y_i,Y_j) \right \rvert  \mid Z'_1, \dots, Z'_m \right) \leq {\mathrm{const.}} \left\lbrace 2 \sup_{\omega \in \Omega} \sup_{t \in [0,\M]} |\tilde{F}_\omega(t)| \right \rbrace^{\alpha}.
\end{equation*}
Utilizing this 
\begin{align*}
    & \prob\left(  \sup_{t \in [0,\M]}  \left\lvert \frac{2}{n} \sum_{i,j=1, j \neq i}^{n} f_{t,1}(Y_i,Y_j) \right\rvert > C_1 \epsilon  \right) \\ = & const. \ \frac{\E\left\lbrace  \sup_{\omega \in \Omega} \sup_{t \in [0,\M]} |\tilde{F}_\omega(t)| \right \rbrace^{\alpha}}{\epsilon}
\end{align*}
Theorem 5.2 in \cite{dube:24} and the continuous mapping theorem in conjunction with the fact that $\left\lbrace \sup_{\omega \in \Omega} \sup_{t \in [0,\M]}  |\tilde{F}_\omega(t)| \right\rbrace^{\alpha}\leq 1$, which implies the uniform integrability of  $\left\lbrace  \sup_{\omega \in \Omega} \sup_{t \in [0,\M]}  \right. \\ \left. |\tilde{F}_\omega(t)| \right \rbrace^{\alpha}$ $\leq 1$, one has $\E\left\lbrace  \sup_{\omega \in \Omega} \sup_{t \in [0,\M]} |\tilde{F}_\omega(t)| \right \rbrace^{\alpha}=o(1)$ as $n \rightarrow \infty$ which completes the proof of this segment, thereby completing the arguments to establish that $\sup_{t \in [0,\M]}\sup_{u \in \idx_c} \lvert U^3_n(t,u)\rvert=o_P(1)$ as $n \rightarrow \infty$.

\noindent \textbf{\underline{Part II:}} Here we derive the asymptotic weak limit of $(t,u) \mapsto \hat{T}^2_n(t,u)$ as $n \rightarrow \infty$ under $H_0$ and find the asymptotic distribution of $\hat{T}_n$ given by the asymptotic distribution of $\sup_{u \in \idx_c} \int_0^{\M} \hat{T}^2_n(t,u) dt = \max_{[nc] \leq k \leq (n-[nc])} \int_0^{\M} \hat{T}^2_n\left(t,\frac{k}{n}\right) dt$ under $H_0$. 
\\Recollect that $n b_n(u) = \frac{[nu](n-[nu])}{n}$ and
\begin{align*}
    & \hat{T}^2_n(t,u) \\ = &   \E_{Y} \left( \frac{\sqrt{n b_n(u)}}{[nu]} \sum_{j=1}^{[nu]}  \lbrace \ind[d(Y,Y_j) \leq t]-F_{Y}(t) \rbrace -\frac{\sqrt{n b_n(u)}}{(n-[nu])} \sum_{j=[nu]+1}^{n}  \lbrace \ind[d(Y,Y_j) \leq t]-F_{Y}(t) \rbrace \right)^2 \\ = & \E_{Y} \left( W^n_Y(t,u) \right)^2
\end{align*}
with $Y$ being a random object following $P_1$ that is independent of the data $Y_1, \dots, Y_n$ and $W_x(t,u)=\frac{\sqrt{n b_n(u)}}{[nu]} \sum_{j=1}^{[nu]}  \lbrace \ind[d(x,Y_j) \leq t]-F_{x}(t) \rbrace-\frac{\sqrt{n b_n(u)}}{(n-[nu])} \sum_{j=[nu]+1}^{n}  \lbrace \ind[d(x,Y_j) \leq t]-F_{x}(t) \rbrace$. Note that for each fixed $x \in \Omega$, the process $t \mapsto \frac{1}{\sqrt{n}} \sum_{j=1}^{n}  \lbrace \ind[d(x,Y_j) \leq t]-F_{x}(t) \rbrace $, $t \in [0,\M]$ is Donsker by Theorem 5.2 in \cite{dube:24} and converges weakly to a zero mean Gaussian process whose covariance is given by
\begin{align*}
    C_x \left( t_1, t_2 \right) = \mathrm{Cov}\left( \ind[d(x,Y')\leq t_1], \ind[d(x,Y')\leq t_2]\right). 
\end{align*}
By the functional Donsker's theorem for partial sum processes as per Theorem 2.12.1 in \cite{well:96}, for each fixed $x \in \Omega$, $(t,u) \mapsto \frac{\sqrt{n b_n(u)}}{[nu]} \sum_{j=1}^{[nu]}  \lbrace \ind[d(x,Y_j) \leq t]-F_{x}(t) \rbrace$, $t \in [0,\M], u \in \idx_c$ converges weakly to a zero mean Gaussian process whose covariance is given by
\begin{align*}
    C^{(1)}_x \left( (t_1,u_1), (t_2,u_2) \right) = \sqrt{\frac{(1-u_1)(1-u_2)}{u_1u_2}} \min(u_1,u_2) C_x \left( t_1, t_2 \right),
\end{align*}
and $(t,u) \mapsto \frac{\sqrt{n b_n(u)}}{(n-[nu])} \sum_{j=[nu]+1}^{n}  \lbrace \ind[d(x,Y_j) \leq t]-F_{x}(t) \rbrace$, $t \in [0,\M], u \in \idx_c$  converges weakly to a tight zero mean Gaussian process whose covariance is given by
\begin{align*}
    C^{(2)}_x \left( (t_1,u_1), (t_2,u_2) \right) = \sqrt{\frac{u_1u_2}{(1-u_1)(1-u_2)}} \min(1-u_1,1-u_2) C_x \left( t_1, t_2  \right).
\end{align*}
For each $x \in \Omega$ since the components $(t,u) \mapsto \frac{\sqrt{n b_n(u)}}{[nu]} \sum_{j=1}^{[nu]}  \lbrace \ind[d(x,Y_j) \leq t]-F_{x}(t) \rbrace$ and $(t,u) \mapsto \frac{\sqrt{n b_n(u)}}{(n-[nu])} \sum_{j=[nu]+1}^{n}  \lbrace \ind[d(x,Y_j) \leq t]-F_{x}(t) \rbrace$ are independent,  one has that for each $x \in \Omega$,
$(t,u) \mapsto W_x(t,u)$ converges weakly to a tight zero mean Gaussian process $(t,u) \mapsto G_x(t,u)$ in the space $\ell^\infty([0,\mathcal{M}] \times \idx_c)$ whose covariance is given by
\begin{align*}
    C^{\star}_x \left( (t_1,u_1), (t_2,u_2) \right) = c(u_1,u_2) C_x\left( t_1, t_2 \right),
\end{align*}
where $c(u_1,u_2)=\sqrt{\frac{(1-u_1)(1-u_2)}{u_1u_2}} \min(u_1,u_2)+\sqrt{\frac{u_1u_2}{(1-u_1)(1-u_2)}} \min(1-u_1,1-u_2)$ as $n \rightarrow \infty$. Since $C^{\star}_x \left( (t_1,u_1), (t_2,u_2) \right)$ is separable in $(u_1,u_2)$ and $(t_1,t_2)$, by Mercer's theorem one has
\begin{equation*}
     C^{\star}_x \left( (t_1,u_1), (t_2,u_2) \right) = \sum_{j=1}^\infty \sum_{l=1}^\infty \lambda_j^x \gamma_l \phi^x_j(t_1) \phi^x_j(t_2) \psi_l(u_1)\psi_l(u_2)
\end{equation*}
where $\lambda^x_1 \geq  \lambda^x_2 \geq \dots $ and $\phi^x_j(\cdot), j=1,2,\dots$ are the eigenvalues and the corresponding eigenfunctions of $C_x(t_1,t_2)$, where $\phi^x_j(\cdot), j=1,2,\dots$ forms an orthonormal basis for $\mathcal{L}^2([0,\M])$ and $\gamma_1 \geq  \gamma_2 \geq \dots $ and $\psi_l(\cdot), l=1,2,\dots$ are the eigenvalues and the corresponding eigenfunctions of $c(u_1,u_2)$ which form an orthonormal basis for $\mathcal{L}^2(\idx_c)$. By the Karhunen-Loève expansion one may write
\begin{equation}
\label{eq:gp}
    G_x(t,u)= \sum_{j=1}^\infty \sum_{l=1}^\infty Z_{jl} \sqrt{\lambda_j^x \gamma_l} \phi^x_j(t)\psi_l(u)
\end{equation}
where $Z_{jl}$ are uncorrelated $N(0,1)$ random variables. By repeating the previous arguments in conjunction with Theorem 5.2 in \cite{dube:24}, the process $(x,t,u) \mapsto W_x(t,u)$ converges weakly to $(x,t,u) \mapsto G_x(t,u)$ in the space $\ell^\infty(\Omega \times [0,\mathcal{M}] \times \idx_c)$. Together with the continuous mapping theorem this implies that the asymptotic weak limit of $\hat{T}^2_n(t,u) =\E_{Y} \left( W^n_Y(t,u) \right)^2$ is given by the law of $(t,u) \mapsto E_Y\left\lbrace \left(G_Y(t,u)\right)^2  \right\rbrace$. Therefore using the continuous mapping theorem again the asymptotic distribution of $\sup_{u \in \idx_c} \int_{0}^{\mathcal{M}} \hat{T}^2_n\left(t,u\right) dt$ turns out to be the same as the law of the random variable  $\sup_{u \in \idx_c} \int_{0}^{\mathcal{M}} E_Y\left\lbrace \left(G_Y\left(t,u\right)\right)^2  \right\rbrace dt $, with $G_x(t,u)$ as defined in \eqref{eq:gp}, which by the Fubini's theorem is the same as the law of $\sup_{u \in \idx_c} \mathbb{E}_Y\left\lbrace \int_{0}^{\mathcal{M}}  \left(G_Y\left(t,u\right)\right)^2  dt \right\rbrace$. Observe that
\begin{equation*}
    \left(G_Y(t,u)\right)^2= \sum_{l=1}^\infty \sum_{l'=1}^\infty\sum_{j=1}^\infty\sum_{j'=1}^\infty Z_{jl} Z_{j'l'}\sqrt{\lambda_j^Y \lambda_{j'}^Y \gamma_{l'}\gamma_l} \phi^Y_j(t)\phi^Y_{j'}(t)\psi_{l'}(u)\psi_l(u)
\end{equation*}
and 
\begin{align*}
    \int_0^{\M} \left(G_Y(t,u)\right)^2 dt = & \sum_{l=1}^\infty \sum_{l'=1}^\infty\sum_{j=1}^\infty Z_{jl} Z_{jl'} \lambda_j^Y \sqrt{\gamma_{l'}\gamma_l}\psi_{l'}(u)\psi_l(u) \\ = & \sum_{j=1}^\infty \lambda_j^Y \left\lbrace \sum_{l=1}^{\infty}Z_{jl} \sqrt{\gamma_l}\psi_l(u) \right\rbrace^2.
\end{align*}
By the Karhunen-Loeve expansion, the law of $u \mapsto \int_0^{\M} \left(G_Y(t,u)\right)^2 dt$ is the same as the law of $u \mapsto \sum_{j=1}^\infty \lambda_j^Y \mathcal{G}_j^2(u)$ where $\mathcal{G}_1, \mathcal{G}_2, \dots$ are independent replicates of the zero mean Gaussian process $u \mapsto \mathcal{G}(u)$ with covariance given by $c(u_1,u_2)$. By the dominated convergence theorem, the distribution of $u \mapsto \mathbb{E}_Y\left\lbrace \int_{0}^{\mathcal{M}}  \left(G_Y(t,u)\right)^2  dt \right\rbrace$ as $n \rightarrow \infty$ is given by the law of $u \mapsto \sum_{j=1}^\infty \mathbb{E}_Y\{\lambda_j^Y\}\mathcal{G}_j^2(u)$. Hence the asymptotic limit of the scan statistic $\hat{T}_n$ is given by the law of $\sup_{u \in \idx_c}\sum_{j=1}^\infty \E_{Y}\{\lambda_j^Y\} \mathcal{G}_j^2\left(u\right)$ which can be represented as the law of $\mathcal{T}$ where
\begin{equation*}
\label{eq:asy_law}
    \mathcal{T}=\sup_{u \in \idx_c}\sum_{j=1}^\infty \E_{Y}\{\lambda_j^Y\} \mathcal{G}_j^2\left(u\right).
\end{equation*}

\subsection*{Proof of Theorem 2}
Let $\alpha \in (0,1)$. One has that
\begin{align*}
    \beta_n = & \prob_{H_{1,n}} \left( p \leq \alpha \right) \\ = & \prob_{H_{1,n}} \left( \hat{T}_n \geq q_\alpha \right) \\ \geq & \prob_{H_{1,n}} \left( n \hat{T}_n(\tau) \geq q_\alpha \right)
\end{align*}
where the last inequality follows by observing that $\hat{T}_n=\sup_{u \in \idx_c} n \hat{T}_n(u) \geq n \hat{T}_n(\tau)$ under $H_1$ when $\tau \in \idx_c$. Observe that
\begin{equation*}
    n\hat{T}_n(\tau) \geq \min(\tau,(1-\tau)) T_{[n\tau],(n-[n\tau])} (X_1,X_2)+ d_n
\end{equation*}
where $X_1=\{Y_1, \dots, Y_{[n\tau]}\}$, $X_2=\{Y_{[n\tau]+1}, \dots, Y_{n}\}$ and $T_{[n\tau],(n-[n\tau])} (X_1,X_2)$ is the test statistic defined as per equation (24) in \cite{dube:24} between the samples $X_1$ and $X_2$ and $d_n=O\left( \frac{1}{n}\right)$ as $n \rightarrow \infty$. Since $\min(\tau,(1-\tau)) > 0$, using Theorem 6.2 in \cite{dube:24} one has $\prob_{H_{1,n}} \left( n\hat{T}_n(\tau) \geq q_\alpha \right) \rightarrow 1$ as $n \rightarrow \infty$ provided that $n a_n \rightarrow \infty$, which implies that $\beta^\alpha_n \rightarrow 1$ as $n \rightarrow \infty$ as long as $n a_n \rightarrow \infty$. 

\noindent For the next part, let $a(u)=u(1-u)$ and recall that as $n \rightarrow \infty$,
\begin{equation*}
    \hat{T}_n(u) = a(u) \left( \denom[n] \add[in] \intgrt  \left\lbrace \prehatF[it]-\posthatF[it]\right\rbrace^2 dt \right) + \rem. 
\end{equation*}
Define $b(u,\tau)=\frac{(1-\tau)^2}{(1-u)^2} \ind[u \leq \tau] + \frac{\tau^2}{u^2} \ind[u > \tau] $ and an oracle process $T_n(u)$ given by 
\begin{equation}
\label{eq: oracle}
    T_n(u) =  a(u) b(u,\tau) \left( \denom[n] \add[in] \intgrt  \left\lbrace \Fone[Y_i](t)-\Ftwo[Y_i](t)\right\rbrace^2 dt \right).
\end{equation}
By Lemma \ref{lma: aux_lemma_1}, under $H_{1,n}$ and assumptions A1 and A2, as $n \rightarrow \infty$, 
\begin{equation}
\label{eq: bound_oh_That}
      \sup_{u \in \idx_c} \lvert \hat{T}_n(u) -T_n(u) \rvert = O_P\left(\frac{1}{n}+\sqrt{\frac{a_n}{n}}\right) 
\end{equation}
almost surely. %Let $\tilde{T}_n = \sup_{u \in \idx_c} \hat{T}_n(u)$ and $\sup_{u \in \idx_c} T_n(u) = T_n(\tau)$. Therefore one has $\lvert \tilde{T}_n - T_n(\tau) \rvert = O_P\left(\frac{1}{n}+\sqrt{\frac{a_n}{n}}\right)$. 
Now observe that
\begin{align}
\label{eq: perm_power}
    \tilde{\beta}^\alpha_n = & \prob_{H_{1,n}} \left( \hat{p}_K \leq \alpha \right) \nonumber \\ \geq & \prob_{H_{1,n}} \left(  \sup_{u \in \idx_c}  \hat{T}_n(u) >  T_n(\tau)/2, \ \sup_{u \in \idx_c}  \hat{T}^{\pi_k}_n(u) \leq T_n(\tau)/2 \ \text{for all} \ k=1, \dots, K \right).
\end{align}
We prove the rest in two steps. First, see that
\begin{align*}
    \prob_{H_{1,n}} \left( \sup_{u \in \idx_c}  \hat{T}_n(u) >  T_n(\tau)/2\right) & \geq  \prob_{H_{1,n}} \left(   \hat{T}_n(\tau) >  T_n(\tau)/2\right)  \\ & = \prob_{H_{1,n}} \left( \hat{T}_n(\tau) - T_n(\tau) >  -T_n(\tau)/2 \right) \\ & \geq  \prob_{H_{1,n}} \left( \hat{T}_n(\tau) - T_n(\tau) >  -\frac{a(\tau)\min(\tau,1-\tau) \Delta}{4} \right) 
\end{align*}
where the last step follows from \eqref{eq: T_n_bound}. Therefore as $n \rightarrow \infty$
\begin{align}
\label{eq: power_one}
     & \prob_{H_{1,n}} \left( \sup_{u \in \idx_c}  \hat{T}_n(u) >  T_n(\tau)/2\right)  \nonumber \\  \geq &  \prob_{H_{1,n}} \left( \left(n + \sqrt{\frac{n}{a_n}}\right)\{\hat{T}_n(\tau) - T_n(\tau)\} >  -\frac{a(\tau)\min(\tau,1-\tau)}{4} \left(n + \sqrt{\frac{n}{a_n}}\right)\Delta \right) \rightarrow 1
\end{align}
since $\left(n + \sqrt{\frac{n}{a_n}}\right)\{\hat{T}_n(\tau) - T_n(\tau)\} = O_P(1)$ under $H_{1,n}$ by \eqref{eq: bound_oh_That} and $\left(n + \sqrt{\frac{n}{a_n}}\right)\Delta \rightarrow \infty$ provided $n a_n \rightarrow \infty$ with $\Delta=a_n$ under $H_{1,n}$. Finally using exchangeability of the permutations $\pi_1, \dots, \pi_K$ given the data one has
\begin{align*}
    & \prob_{H_{1,n}} \left( \sup_{u \in \idx_c}  \hat{T}^{\pi_k}_n(u) \leq T_n(\tau)/2 \ \text{for all} \ k=1, \dots, K \right) \\ = & 1- \prob_{H_{1,n}} \left(\sup_{u \in \idx_c}  \hat{T}^{\pi_k}_n(u) > T_n(\tau)/2 \ \text{for some} \ k=1, \dots, K \right)  \\ \geq & 1 - K \times \prob_{H_{1,n}} \left( \sup_{u \in \idx_c}  \hat{T}^{\pi_1}_n(u) > T_n(\tau)/2 \right)  \\ \geq  & 1 - K \times \prob_{H_{1,n}} \left( \sup_{u \in \idx_c}  \hat{T}^{\pi_1}_n(u) >  \frac{a(\tau) \min(\tau,1-\tau) \ \Delta}{4}  \right)
\end{align*}
where the last step follows from \eqref{eq: T_n_bound}. If we can show that under $H_{1,n}$, $\frac{ \sup_{u \in \idx_c}  \hat{T}^{\pi_1}_n(u) }{\Delta} = o_P\left( 1\right)$ provided that $n a_n \rightarrow \infty$   as $n \rightarrow \infty$, then one has 
\begin{equation}
\label{eq: power_two}
\prob_{H_{1,n}} \left( \sup_{u \in \idx_c}  \hat{T}^{\pi_k}_n(u) \leq T_n(\tau)/2 \ \text{for all} \ k=1, \dots, K \right)
 \rightarrow 1
\end{equation}
 as $n \rightarrow \infty$. Hence \eqref{eq: power_two} in conjunction with \eqref{eq: power_one} and \eqref{eq: perm_power} implies that $\tilde{\beta}^\alpha_n \rightarrow 1$ as $n \rightarrow \infty$. Therefore to wrap up, we establish that under $H_{1,n}$, $\frac{ \sup_{u \in \idx_c}  \hat{T}^{\pi_1}_n(u) }{\Delta} = o_P\left( 1\right)$ provided that $n a_n \rightarrow \infty$   as $n \rightarrow \infty$. We will show that under $H_{1,n}$, $  \sup_{u \in \idx_c}  \ n\hat{T}^{\pi_1}_n(u) = O_P\left( 1\right)$ which implies the above and completes the proof.
 
\noindent For the sake of simplicity of notation, we denote the random permutation using $\pi$ instead of $\pi_1$. For any $x \in \Omega$,
define $\hat{F}^{(u,\pi)}_x(t) = \frac{1}{[nu]} \sum_{j=1}^{[nu]} \ind[d(x,Y_{\pi(j)}) \leq t] $ and  $\hat{F}^{(1-u,\pi)}_x(t) = \frac{1}{(n-[nu])} \sum_{j=[nu]+1}^{n} \ind[d(x,Y_{\pi(j)}) \leq t] $. Observe that
\begin{align*}
    & \hat{F}^{(u,\pi)}_x(t) - \hat{F}^{(1-u,\pi)}_x(t) = \add[jn] s_j(u) \ind[d(x,Y_{\pi(j)}) \leq t]
\end{align*}
where 
\begin{equation*}
    s_j(u) = \begin{cases} \frac{1}{[nu]} \quad \text{for} \quad 1 \leq j \leq [nu] \\ \frac{-1}{(n-[nu])} \quad \text{for} \quad [nu]+1 \leq j \leq n
    \end{cases}. 
\end{equation*}
Using the coupling construction in section 5.3 of \cite{chun:13} we obtain $\bar{Y}_{\pi_0}=( \bar{Y}_{\pi_0(1)}, \dots, \bar{Y}_{\pi_0(n)})$ such that $Y_i$ and $\bar{Y}_{\pi_0(i)}$ agree for all $i$ except for a random number $D$ such that $\E\left( \frac{D}{n} \right) = O\left( \frac{1}{\sqrt{n}}\right)$ (see 5.8 in \cite{chun:13}) and $\bar{Y}_{\pi_0(i)}, i=1, \dots, n$ are i.i.d. observations from $\bar{P}$ where $\bar{P}=\tau P_1 +(1-\tau) P_2$ is the mixture distribution of $P_1$ and $P_2$ with weights $\tau$ and $1-\tau$.  With this coupling construction observe that
\begin{align*}
& \hat{F}^{(u,\pi)}_x(t) - \hat{F}^{(1-u,\pi)}_x(t) \\ &  = \add[jn] s_j(u) \ind[d(x,Y_{\pi(j)}) \leq t] \\ & = \add[jn] s_j(u) [\ind[d(x,Y_{\pi(j)}) \leq t] - \ind[d(x,\bar{Y}_{\pi\pi_0(j)}) \leq t]]+ \add[jn] s_j(u)\ind[d(x,\bar{Y}_{\pi\pi_0(j)}) \leq t].
\end{align*}
Note that $$\sup_{x \in \Omega, u \in \idx_c, t \in [0,\mathcal{M}]} \left\lvert  \add[jn] s_j(u) [\ind[d(x,Y_{\pi(j)}) \leq t] - \ind[d(x,\bar{Y}_{\pi\pi_0(j)}) \leq t]] \right\rvert = O_P\left( \frac{1}{\sqrt{n}} \right)$$ independently of $\pi$ by the construction of $\bar{Y}_{\pi_0}$.  It also turns out 
 by Theorem 5.2 in \cite{dube:24} that $\sup_{x \in \Omega, u \in \idx_c, t \in [0,\mathcal{M}]} \left\lvert  \add[jn] s_j(u)\ind[d(x,\bar{Y}_{\pi\pi_0(j)}) \leq t] \right\rvert = O_P\left( \frac{1}{\sqrt{n}} \right)$ independently of $\pi$. Together they imply that $\sup_{x \in \Omega, u \in \idx_c, t \in [0,\mathcal{M}]} \left\lvert  \hat{F}^{(u,\pi)}_x(t) - \hat{F}^{(1-u,\pi)}_x(t) \right\rvert = O_P\left( \frac{1}{\sqrt{n}} \right)$ as $n \rightarrow \infty$ independently of $\pi$, which in turn implies that $\sup_{u \in \idx_c} \hat{T}^\pi_n(u) = O_P\left( \frac{1}{{n}} \right) $ since $\sup_{u \in \idx_c} \hat{T}^\pi_n(u) \leq \mathcal{M} \left( \sup_{x \in \Omega, u \in \idx_c, t \in [0,\mathcal{M}]} \left\lvert  \hat{F}^{(u,\pi)}_x(t) - \hat{F}^{(1-u,\pi)}_x(t) \right\rvert \right)^2$ thereby completing the proof.

\subsection*{Proof of Theorem 3} First observe that since $\hat{\tau}$ is a maximizer of $\hat{T}_n(\cdot)$, $\hat{T}_n(\hat{\tau}) -  \hat{T}_n(\tau) \geq 0$. Next see that $\tau$ is the unique maximizer of $T_n(\cdot)$ \eqref{eq: oracle} almost surely. Moreover by the definition in \eqref{eq: oracle} and the bounds in Lemma \ref{lma: aux_lemma_1} in \eqref{eq: T_n_bound}, $T_n(\tau) -T_n(\hat{\tau}) \geq C \lvert \hat{\tau}-\tau \rvert \Delta$ for some constant $C > 0$. Hence one has almost surely that 
\begin{equation}
\label{eq: local_curvature}
    \{\hat{T}_n(\hat{\tau}) -  \hat{T}_n(\tau)\} - \{T_n(\hat{\tau})-T_n(\tau) \} \geq  C \lvert \hat{\tau}-\tau \rvert \Delta. 
\end{equation}

\noindent Let $M > 0$ and $\beta_n > 0$ be a sequence such that $\frac{n}{\beta_n} \rightarrow \infty$ as $n \rightarrow 0$. Observe that
\begin{align*}
   &  \prob \left( \beta_n \lvert \hat{\tau} - \tau  \rvert  \geq 2^M \right) \leq    \sum_{j=M}^\infty \prob  \left( \frac{2^j}{\beta_n} \leq \lvert \hat{\tau} - \tau  \rvert  < \frac{2^{j+1}}{\beta_n} \right).
\end{align*}
Now whenever $\frac{2^j}{\beta_n} \leq \lvert \hat{\tau} - \tau  \rvert  < \frac{2^{j+1}}{\beta_n}$,
\begin{align*}
    & \sup_{\frac{2^j}{\beta_n} \leq \lvert u-\tau \rvert \leq \frac{2^{j+1}}{\beta_n}} \lvert \{\hat{T}_n(u) -  \hat{T}_n(\tau)\} - \{T_n(u)-T_n(\tau) \}  \rvert \\  & \geq\{\hat{T}_n(\hat{\tau}) -  \hat{T}_n(\tau)\} - \{T_n(\hat{\tau})-T_n(\tau) \} \\ & \geq  C \Delta \frac{2^j}{\beta_n}
\end{align*}
and therefore 
\begin{align*}
       \prob \left( \beta_n \lvert \hat{\tau} - \tau  \rvert  \geq 2^M \right) \leq   &   \sum_{j=M}^\infty \prob  \left( \sup_{\frac{2^j}{\beta_n} \leq\lvert u-\tau \rvert \leq \frac{2^{j+1}}{\beta_n}} \lvert \{\hat{T}_n(u) -  \hat{T}_n(\tau)\} - \{T_n(u)-T_n(\tau) \}  \rvert \geq  C \Delta \frac{2^j}{\beta_n} \right) \\ \leq & \sum_{j=M}^\infty \frac{ \beta_n \E \left( \sup_{\frac{2^j}{\beta_n} \leq \lvert u-\tau \rvert \leq \frac{2^{j+1}}{\beta_n}}  \lvert \hat{T}_n(u) -\hat{T}_n(\tau)-T_n(u)+T_n(\tau) \rvert \right) }{C \Delta 2^j}.
\end{align*}
where the last step follows using Markov's inequality. Here on we consider 2 cases, fixed $H_1$ and local alternatives $H_{1,n}$ where $\Delta = a_n$. Under fixed $H_1$ by Lemma \ref{lma: aux_lemma_2},
\begin{align*}
    \E \left( \sup_{\frac{2^j}{\beta_n} \leq \lvert u-\tau \rvert \leq \frac{2^{j+1}}{\beta_n}}  \lvert \hat{T}_n(u) -\hat{T}_n(\tau)-T_n(u)+T_n(\tau) \rvert \right) = O \left( \frac{2^{j/2}}{\sqrt{n\beta_n}}\right)
\end{align*}
which implies that for some constant $K > 0$,
\begin{align*}
       \prob \left( \beta_n \lvert \hat{\tau} - \tau  \rvert  \geq 2^M \right) \leq    & K \sum_{j=M}^\infty \frac{ \sqrt{\beta_n} }{\sqrt{n} 2^{j/2}}.
\end{align*}
Setting $\beta_n=\frac{n}{\log(n)}$ establishes the result under fixed $H_1$. 

\noindent Under $H_{1,n}$ by Lemma \ref{lma: aux_lemma_2},
\begin{align*}
    \E \left( \sup_{\frac{2^j}{\beta_n} \leq \lvert u-\tau \rvert \leq \frac{2^{j+1}}{\beta_n}}  \lvert \hat{T}_n(u) -\hat{T}_n(\tau)-T_n(u)+T_n(\tau) \rvert \right) = O \left( \frac{2^{j/2}\sqrt{a_n}}{\sqrt{n\beta_n}}\right)
\end{align*}
and therefore for some constant $K > 0$,
\begin{align*}
       \prob \left( \beta_n \lvert \hat{\tau} - \tau  \rvert  \geq 2^M \right) \leq    & K \sum_{j=M}^\infty \frac{ \sqrt{\beta_n} }{\sqrt{n a_n} 2^{j/2}}
\end{align*}
which implies that setting $\beta_n = n a_n$ works under the local alternatives $H_{1,n}$.

\section{Auxiliary results and their proofs}
\label{sec: aux}

\begin{Lemma}
\label{lma: 2sample_1}
Under $H_0$ and assumption A1 one has as $n \rightarrow \infty$,
\begin{equation*}
    \sup_{t \in [0,\M]} \left \lvert \frac{1}{n(n-1)} \sum_{i,j=1, j \neq i}^{n}   h_{t}(Y_i,Y_j,Y_j) \right \rvert = o_P(1)
\end{equation*}
and 
\begin{equation*}
    \sup_{t \in [0,\M]} \left \lvert \frac{1}{n(n-1)} \sum_{i,j=1, j \neq i}^{n}   h_{t}(Y_i,Y_i,Y_j) \right \rvert = o_P(1)
\end{equation*}
where  $h_{t}(x,y,z)= \lbrace \ind[d(x,y) \leq t]-F_x(t) \rbrace\lbrace \ind[d(x,z) \leq t]-F_x(t) \rbrace - \E_{Y}( \lbrace \ind[d(Y,y) \leq t]-F_Y(t) \rbrace\lbrace \ind[d(Y,z) \leq t]-F_Y(t) \rbrace )$ for $t \in [0,\M]$ and $Y \sim P_1$ is independent of the data $Y_1, \dots, Y_n$.
\end{Lemma}
\noindent \textbf{Proof.} Observe that  $\left\lbrace \frac{1}{n(n-1)} \sum_{i,j=1, j \neq i}^{n}   h_{t}(Y_i,Y_j,Y_j) \right\rbrace_{t \in [0,\M]}$  is a non-degenerate $U$-process of order two indexed by the function class $\mathcal{G}=\{g_t\colon \Omega\times\Omega\rightarrow\mathbb{R}\mid t \in [0,\M]\}$ with $g_t(x,y)=h_t(x,y,y)$ for $x,y\in\Omega$. One can show that
\begin{align*}
   |g_s(x,y)-g_t(x,y)| \leq G_{st}(x,y)
\end{align*}
where 
\begin{align*}
\label{eq: G_st}
  & G_{st}(x,y) \\
  = & C_G \left \lbrace |\ind[d(x,y)\leq s]-\ind[d(x,y)\leq t]| + \sup_{x \in \Omega} |F_x(s)-F_x(t)|+\E_{Y} \left( |F_Y(s)-F_Y(t)|\right) \right \rbrace
\end{align*}
for some constant $C_G>0$. Moreover
given any $\epsilon > 0$ as long as $|s-t|<\epsilon$ using assumption A1 one has $\|F_{Y}(s)- F_{Y}(t)\|_{L_1(P_1)} \leq C_0 \epsilon$ for some constant $C_0>0$ which leads to $G_{st}(x,y) \leq G_s(x,y)$ where
\begin{align}
  G_{s}(x,y) = & C \left \lbrace |\ind[d(x,y)\leq s+\epsilon]-\ind[d(x,y)\leq s-\epsilon]| + 2\epsilon \right \rbrace
 \end{align}
for some constant $C>0$. By the property of indicator functions and using assumption A1 again one has
%\begin{align*}
  %& \lVert \ind[d(X,X')\leq u]-\ind[d(X,X')\leq v] \rVert_{L_1(P_1 \times P_1)} \\ = & \left|\prob(d(X,X')\leq u)-\prob(d(X,X')\leq v) \right| \\ \leq & C_0 \epsilon.
%\end{align*}
%This leads to
\begin{align*}
   \|G_{s}(Y,Y') \|_{L_1(P_1 \times P_1)} < 4C \epsilon.
\end{align*}
Let $s_1, \dots, s_{N_\epsilon}$ denote an $\frac{\epsilon}{4C}$-net of $[0,\M]$ where $N_\epsilon$ is the covering number of $[0,\M]$ with $\frac{\epsilon}{4C}$-radius balls. %If $|t_{i-1},t_i|>2\epsilon$, it is guaranteed to find $v=v_j$ such that $|u-v_j|<\epsilon$. If $|t_{i-1},t_i|< 2\epsilon$, choose $v=\frac{t_i+t_{i-1}}{2}$. 
Then $\mathcal{I} = \{s_1,\dots,s_{N_\epsilon}\}$ %\yc{(should be $\{v_1,\dots,v_{N_\epsilon}\}$)} %and $(t_{i-1}+t_i)/2$ for intervals $[t_{i-1},t_i] \in \mcM$ 
forms a collection of indices such that for any $t \in [0,\M]$, there exists $s_k \in \mathcal{I}$ such that $|g_t(x,y)-g_{s_k}(x,y)| \leq G_{s_k}(x,y)$ and $ \|G_{s_k}(Y,Y') \|_{L_1(P_1 \times P_1)} < \epsilon$. As per the definition of brackets in page 1512 of \cite{arco:93}, the bracketing number $N^{(1)}_{[]}(\epsilon, \mathcal{G}, P_1 \times P_1) $ is upper bounded by the cardinality of $\mathcal{I}$, i.e. $N_\epsilon$, and therefore finite. By Corollary 3.5 in \cite{arco:93}, $ \sup_{t \in [0,\M]} \left \lvert \frac{1}{n(n-1)} \sum_{i,j=1, j \neq i}^{n}   h_{t}(Y_i,Y_j,Y_j) \right \rvert = o_P(1)$. 

Similar to the first part, $\left\lbrace \frac{1}{n(n-1)} \sum_{i,j=1, j \neq i}^{n}   h_{t}(Y_i,Y_j,Y_j) \right\rbrace_{t \in [0,\M]}$ is a non-degenerate U-process of order 2 indexed by the function class $\mathcal{G}'=\{g'_t\colon \Omega\times\Omega\rightarrow\mathbb{R}\mid t \in [0,\M]\}$ where $g'_t(x,y)=h_t(x,x,y)$ for $x,y \in \Omega$. Observe that for some constant $C'>0$
\begin{equation*}
    \lvert g'_{s}(x,y)-g'_{t}(x,y) \rvert \leq C' G_{st}(x,y)
\end{equation*}
where $G_{s,t}(x,y)$ is as defined in \eqref{eq: G_st}. This leads to $ \sup_{t \in [0,\M]} \left \lvert \frac{1}{n(n-1)} \sum_{i,j=1, j \neq i}^{n}   h_{t}(Y_i,Y_j,Y_j) \right \rvert = o_P(1)$ by repeating the same arguments as the first part which completes our proof.

\begin{Lemma}
\label{lma: 2sample_2}
Under $H_0$ and assumption A1 one has as $n \rightarrow \infty$,
\begin{equation*}
    \sup_{t \in [0,\M]} \left \lvert \frac{1}{n(n-1)(n-2)} \sum_{i,j,k=1, i\neq j \neq k}^{n}   h_{t}(Y_i,Y_j,Y_k)  \right \rvert = o_P\left( \frac{1}{n} \right)
\end{equation*}
where  $h_{t}(x,y,z)= \lbrace \ind[d(x,y) \leq t]-F_x(t) \rbrace\lbrace \ind[d(x,z) \leq t]-F_x(t) \rbrace - \E_{Y}( \lbrace \ind[d(Y,y) \leq t]-F_Y(t) \rbrace\lbrace \ind[d(Y,z) \leq t]-F_Y(t) \rbrace )$ for $t \in [0,\M]$ and $Y \sim P_1$ is independent of the data $Y_1, \dots, Y_n$.
\end{Lemma}
\noindent \textbf{Proof.} Note that $\left \lbrace \frac{1}{n(n-1)(n-2)} \sum_{i,j,k=1, i\neq j \neq k}^{n}   h_{t}(Y_i,Y_j,Y_k)  \right \rbrace_{t \in [0,\M]}$ is a degenerate $U$-process of order three \citep{sher:94} indexed by the function class %\yc 
$\mathcal{H}=\{(x,y,z) \mapsto h_t(x,y,z)\colon \Omega\times\Omega\times\Omega\rightarrow \mathbb{R}\mid t \in [0,\M]\}$. Note that $\sup_{t \in [0,\M]} \sup_{x,y,z \in \Omega} |h_t(x,y,z)| \leq 4$ the constant function $4$ is an envelope function for the function class $\mathcal{H}$. For $s,t \in [0,\M]$ one has
\begin{align*}
  & |h_t(x,y,z)-h_s(x,y,z)| \leq  \lvert H_{st}(x,y,z)\rvert + \lvert \E_{Y}(H_{st}(Y,y,z)) \rvert
\end{align*}
where 
\begin{equation}
\begin{aligned}
    &H_{st}(x,y,z) \\ 
    = & C_H \Big\{ |\ind[d(x,y)\leq s]-\ind[d(x,y)\leq t]| \\
+&  |\ind[d(x,z)\leq s]-\ind[d(x,z)\leq t]|\\
+&   \sup_{x \in \Omega} |F_x(s)-F_x(t)|\Big\} 
\end{aligned}
\label{eq: hst}
\end{equation}
% \begin{equation}
% \begin{split}
% &H_{st}(x,y,z) \nonumber \\ = & C_H \Big\{ |\ind[d(x,y)\leq s]-\ind[d(x,y)\leq t]|  \\
% +&  |\ind[d(x,z)\leq s]-\ind[d(x,z)\leq t]|
%  \\
% +&   \sup_{x \in \Omega} |F_x(s)-F_x(t)|\Big\} 
% \end{split}
% \end{equation}
for some constant $C_H > 0$. In what follows let $Y$,$Y'$ and $Y''$ be independent copies of $P_1$ also independent of the data. Notice that by the property of indicator functions,
\begin{align*}
  & \left \lVert \ind[d(Y,Y')\leq s]-\ind[d(Y,Y')\leq t] \right \rVert_{L_2(P_1 \times P_1)} \\ = & \sqrt{\left|\prob(d(Y,Y')\leq s)-\prob(d(Y,Y')\leq t) \right|} \\ = & \sqrt{\left|\E_{Y'}\left\lbrace \prob(d(Y',Y)\leq s\mid Y') \right\rbrace-\E_{Y'}\left\lbrace \prob(d(Y',Y)\leq t\mid Y')\right\rbrace \right|} \\ \leq & \sqrt{\E_{Y'} \left\lvert \prob(d(Y',Y)\leq s\mid X')-\prob(d(Y',Y)\leq t\mid Y') \right \rvert }
\end{align*}
where the last step follows using Jensen's inequality. By the independence of $Y$ and $Y'$ and assumption A1, one has for some constant $C_0 > 0$
\begin{equation}
\label{eq: indicators}
  \left \lVert \ind[d(Y,Y')\leq s]-\ind[d(Y,Y')\leq t] \right \rVert_{L_2(P_1 \times P_1)} \leq \sqrt{C_0 \epsilon}
\end{equation}
whenever $|s-t|<\epsilon$. Therefore for any given $t \in [0,\M]$ by picking $s \in [0,\M]$ such that $|s-t|<\epsilon$ %and $u,v \in [t_{i-1},t_i]$ for some $i$, 
it turns out that %\pd
{$\lVert H_{st}(Y,Y',Y'')\rVert_{L_2(P_1 \times P_1 \times P_1)} < C_1 \sqrt{\epsilon}$ for some constant $C_1 > 0$.} %\pd
{By Jensen's inequality, 
\begin{align*}
  & \lVert \E_{Y}(H_{st}(Y,Y',Y'') \rVert_{L_2(P_1 \times P_1 )} \leq \lVert H_{st}(Y,Y',Y'') \rVert_{L_2(P_1 \times P_1 \times P_1)} %\yc{\ [\E_{X'}\text{ should be removed}]} 
   <  C_1 \sqrt{\epsilon}.
\end{align*}
}This implies that %\pd
{$\lVert h_t(Y,Y',Y'') -h_s(Y,Y',Y'') \rVert_{L_2(P_1 \times P_1 \times P_1)} < 2 C_1 \sqrt{\epsilon}$ whenever $|s-t|<\epsilon$.} %\yc{[$\lVert f_u -f_v \rVert_{L_2(P_1 \times P_1 \times P_1)} < 2 C \sqrt{\epsilon}$ for any $u,v\in\mcM$ such that $|u-v|<\epsilon$.]} 
% \blu{Therefore, the covering number of the function class $\mathcal{F}$ with balls of radius $\epsilon$, $N(\epsilon,\mathcal{F},L_2(P_1 \times P_1 \times P_1))$, is upper bounded by the %sum of the 
% covering number of $\mcM$ with balls of radius $\frac{\epsilon^2}{4C^2}$. % and the number of intervals $[t_{i-1},t_i]$. 
% Hence $N(\epsilon,\mathcal{F},L_2(P_1 \times P_1 \times P_1)) = O \left(\epsilon^{-2}\right)$. By the equivalence of covering numbers and packing numbers, [original; to be commented out. I think the radius should be $\tfrac{\epsilon^2}{16C^2}$; specifically, 
% $N(\epsilon,\mathcal{F},L_2(P_1 \times P_1 \times P_1)) \le D(\tfrac{\epsilon}{2},\mathcal{F},L_2(P_1 \times P_1 \times P_1)) \le D(\tfrac{\epsilon^2}{16C^2}, \mcM, d_E) \le N(\tfrac{\epsilon^2}{16C^2}, \mcM, d_E)$. Alternatively, I suggest we rewrite the entire sentence as in the following in red.] (Once checked, please do not delete this but leave this as a comment in latex.)} \yc{
Therefore, the packing number $D(\epsilon,\mathcal{H},L_2(P_1 \times P_1 \times P_1))$, i.e. the maximum number of $\epsilon$-separate elements in $\mathcal{H}$ endowed with $L_2(P_1 \times P_1 \times P_1)$ metric, is upper bounded by the packing number $D(\tfrac{\epsilon^2}{4C_1^2}, [0,\M], d_E)$, i.e. the maximum number of $\tfrac{\epsilon^2}{4C_1^2}$-separate elements in $[0,\M]$ endowed with the Euclidean metric $d_E$. Note that $D(\tfrac{\epsilon^2}{4C_1^2}, [0,\M], d_E) = O\left(\epsilon^{-2}\right)$, whence %} 
$\mathcal{H}$ is a Euclidean class as per Definition 3 of \cite{sher:94}. By Corollary 4 in \cite{sher:94}, $n^{3/2}  \sup_{t \in [0,\M]} \left \lvert \frac{1}{n(n-1)(n-2)} \sum_{i,j,k=1, i\neq j \neq k}^{n}   h_{t}(Y_i,Y_j,Y_k)  \right \rvert = O_P(1) $ which completes the proof. 

\begin{Lemma}
\label{lma: 2sample_3}
Under $H_0$ and assumption A1 one has as $n \rightarrow \infty$,
\begin{equation*}
   \sup_{t \in [0,\M]} \sup_{x,y\in \Omega} \left\lvert \frac{1}{n}\sum_{i=1}^{n} h_t(Y_i,x,y) \right\rvert = o_P(1)
\end{equation*}
where  $h_{t}(x,y,z)= \lbrace \ind[d(x,y) \leq t]-F_x(t) \rbrace\lbrace \ind[d(x,z) \leq t]-F_x(t) \rbrace - \E_{Y}( \lbrace \ind[d(Y,y) \leq t]-F_Y(t) \rbrace\lbrace \ind[d(Y,z) \leq t]-F_Y(t) \rbrace )$ for $t \in [0,\M]$ and $Y \sim P_1$ is independent of the data $Y_1, \dots, Y_n$.
\end{Lemma}
\noindent \textbf{Proof.} Consider the process $\left\lbrace \frac{1}{n}\sum_{i=1}^{n} h_t(Y_i,\omega_1,\omega_2)\right\rbrace_{(\omega_1,\omega_2)\in\Omega\times\Omega,t \in [0,\M]}$ and the function class $\mathcal{L}=\{l_{\omega_1,\omega_2,t}(x): \omega_1,\omega_2 \in \Omega, t \in [0,\M] \}$ where $l_{\omega_1,\omega_2,t}(x)=h_t(x,\omega_1,\omega_2)$. Observe that for some constant $C_L > 0$
\begin{align*}
  & |l_{\omega_1,\omega_2,t}(x)-l_{\omega'_1,\omega'_2,s}(x)|\\ 
  \leq & C_L \left \lbrace H_{st}(x,\omega_1,\omega_2) + |\ind[d(x,\omega_1)\leq s]-\ind[d(x,\omega'_1)\leq s]| \right. \\
  +& \left.|\ind[d(x,\omega_2)\leq s]-\ind[d(x,\omega'_2)\leq s]|\right \rbrace
\end{align*}
where $H_{st}(\cdot,\cdot,\cdot)$ is defined in \eqref{eq: hst}.
\noindent %\pd{
Let $|s-t|<\epsilon$, $d(\omega_1,\omega_1')<\epsilon$ and $d(\omega_2,\omega_2')<\epsilon$. Observe that if $d(x,\omega_1) \leq s$, then by triangle inequality $d(x,\omega_1') \leq s+\epsilon$. Similarly if $d(x,\omega_1') \leq s-\epsilon$, then by triangle inequality $d(x,\omega_1) \leq s$. Hence if $d(\omega_1,\omega_1')<\epsilon$ and $d(\omega_2,\omega_2')<\epsilon$ then one has
\begin{align*}
  & |\ind[d(x,\omega_1)\leq s]-\ind[d(x,\omega'_1)\leq s]| \leq |\ind[d(x,\omega_1')\leq s-\epsilon]-\ind[d(x,\omega'_1)\leq s+2\epsilon]|
\end{align*}
and
\begin{align*}
  & |\ind[d(x,\omega_2)\leq s]-\ind[d(x,\omega'_2)\leq s]| \leq |\ind[d(x,\omega_2')\leq s-\epsilon]-\ind[d(x,\omega'_2)\leq s+\epsilon]|.
\end{align*}
Moreover when $|s-t|<\epsilon$, $d(\omega_1,\omega_1')<\epsilon$ and $d(\omega_2,\omega_2')<\epsilon$, $H_{st}(x,\omega_1,\omega_2) \leq H_{s}(x,\omega'_1,\omega'_2)$ where
\begin{align*}
&H_{s}(x,\omega'_1,\omega'_2)  \\ = & C_3 \left \lbrace |\ind[d(x,\omega'_1)\leq s+2\epsilon]-\ind[d(x,\omega'_1)\leq s-2\epsilon]| \right. \\ + & \left. |\ind[d(x,\omega'_2)\leq s+2\epsilon]-\ind[d(x,\omega'_2)\leq s-2\epsilon]|+ \epsilon \right \rbrace 
\end{align*}
for some constant $C_3 > 0$. This implies that 
\begin{align*}
   & |l_{\omega_1,\omega_2,t}(x)-l_{\omega'_1,\omega'_2,s}(x)| \leq b_{\omega_1',\omega_2',s}(x) 
\end{align*}
with
\begin{align*}
  b_{\omega_1',\omega_2',s}(x) =\ & C_4 \left \lbrace  H_{s}(x,\omega'_1,\omega'_2) + |\ind[d(x,\omega_1')\leq s-\epsilon]-\ind[d(x,\omega'_1)\leq s+\epsilon]| \right. \\ & + \left. |\ind[d(x,\omega_2')\leq s-\epsilon]-\ind[d(x,\omega'_2)\leq s+\epsilon]| \right \rbrace.
\end{align*}
Furthermore assumption A1 implies that
\begin{align*}
  & \|b_{\omega_1',\omega_2',u'}(X)\|_{L_1(P_1)} \leq C_5 \epsilon
\end{align*}
for some $C_5 > 0$. Let $s_1, s_2, \dots, s_{N'_\epsilon}$ be an $\frac{\epsilon}{2C_5}$-net of $[0,\M]$ and $\omega_1, \dots, \omega_{M'_\epsilon}$ be an $\frac{\epsilon}{2C_5}$-net of $\Omega$, where $N'_{\epsilon}$ and $M'_{\epsilon}$ are the covering numbers of $[0,\M]$ and $\Omega$ with balls of radius $\frac{\epsilon}{C_5}$ respectively. Then the brackets $\left\lbrace l_{\omega_j,\omega_k,s_l} \pm b_{\omega_j,\omega_k,s_l} \right\rbrace_{j,k\in[M'_{\epsilon}],\,l\in[N'_{\epsilon}]}$ cover $\mathcal{L}$ and the $L_1(P_1)$ width of each bracket, that is $2 \|b_{\omega_1',\omega_2',u'}(X)\|_{L_1(P_1)}$, is upper bounded by $\epsilon$. %} 
% \noindent By Assumption \ref{ass:assumption_weights}, $\E_{X}|w_{X}(u)- w_X(u')| \leq L_w |u-u'|$ and by Assumption \ref{ass:assumption_lipschitz} and the mean value theorem, $\E_{X'}|F^X_{X'}(u)-F^X_{X'}(u')| \leq L_X |u-u'|$. Observe that 
% \begin{align*}
%   & |F^X_{\omega_1}(u)-F^X_{\omega'_1}(u')| \\ \leq & |F^X_{\omega_1}(u)-F^X_{\omega_1}(u')| + |F^X_{\omega_1}(u')-F^X_{\omega'_1}(u')|,
% \end{align*}
% and a similar inequality holds for $|F^X_{\omega_2}(u)-F^X_{\omega'_2}(u')|$. By Assumption \ref{ass:assumption_lipschitz} and the mean value theorem $|F^X_{\omega_1}(u)-F^X_{\omega_1}(u')| \leq L_X |u-u'|$. Furthermore
% \begin{equation*}
%   \prob(d(X,\omega_1') \leq u'-d(\omega_1,\omega'_1)) \leq \prob(d(X,\omega_1) \leq u') \leq \prob(d(X,\omega_1') \leq u'+d(\omega_1,\omega'_1)).
% \end{equation*}
% This implies that $|F^X_{\omega_1}(u')-F^X_{\omega'_1}(u')| \leq |F^X_{\omega'_1}(u'+d(\omega_1,\omega'_1))-F^X_{\omega'_1}(u'-d(\omega_1,\omega'_1))|\leq 2 L_X d(\omega_1,\omega'_1)$ also by Assumption \ref{ass:assumption_lipschitz} and the mean value theorem. Combining these arguments and reasoning out similarly for $|F^X_{\omega_2}(u)-F^X_{\omega'_2}(u')|$ one has
% \begin{align*}
%   & \|l_{\omega_1,\omega_2,u}(X)-l_{\omega'_1,\omega'_2,u'}(X)\|_{L_1(P_1)} \\ \leq & C_U \left \lbrace (L_w+3L_X) |u-u'| + 4L_X d(\omega_1,\omega'_1) \right \rbrace \\ \leq & C_L \left \lbrace |u-u'| + d(\omega_1,\omega'_1) \right\rbrace
% \end{align*}
% for some constant $C_L>0$. 
Hence for any $\epsilon>0$ the $L_1(P_1)$-bracketing entropy of $\mathcal{L}$, $N_{[]}(\epsilon,\mathcal{L},L_1(P_1))$, is upper bounded by $N'_\epsilon {M'_\epsilon}^2$ and therefore $N_{[]}(\epsilon,\mathcal{L},L_1(P_1)) < \infty$ for any $\epsilon > 0$. %\yc{[I replaced $N_\epsilon$ and $M_\epsilon$ with $N'_\epsilon$ and $M'_\epsilon$, respectively as $N_\epsilon$ has been used to denote another cover numbering.]} 
%\yc{[The previous inequality is not sufficient to imply the blue part. Rather, we need there exists a function $L\colon\Omega\rightarrow\mathbb{R}$ such that $|l_{\omega_1,\omega_2,u}(x)-l_{\omega'_1,\omega'_2,u'}(x)| \le L(x)\{|u-u'| + d(\omega_1,\omega'_1)\}$, for all $x,\omega_1,\omega'_1\in\Omega$, $u,u'\in\mcM$ (see Theorem~2.7.11 of \cite{well:96}). This does not hold under current assumptions.]} 
By Theorem 2.4.1 in \cite{well:96}, $\mathcal{L}$ is a Glivenko--Cantelli class of functions which implies that 
\begin{equation*}
\label{eq: glivenko_cantelli}
  \sup_{t \in [0,\M]} \sup_{x,y\in \Omega} \left\lvert \frac{1}{n}\sum_{i=1}^{n} h_t(Y_i,x,y) \right\rvert = o_P(1).
\end{equation*}

\begin{Lemma}
\label{lma: 2sample_4}
Under $H_0$ and assumption A1 conditionally on $Z'_1, \dots, Z'_n$ one has for any $\epsilon >0$ that as $n \rightarrow \infty$,
\begin{equation*}
  \prob_{|\mathcal{Z}'}\left(  \sup_{t \in [0,\M]}  \left\lvert \frac{1}{n} \sum_{j,k=1, j \neq k}^{n} g^{Z'}_{t,0}(Y_j,Y_k)  \right\rvert > \epsilon , \sup_{\omega_1,\omega_2 \in \Omega} \sup_{v \in \idx_c}\sup_{t \in [0,\M]} g^{Z'}_{t,v}(\omega_1,\omega_2) \leq t \right) \rightarrow 0
\end{equation*}
where $g^{Z'}_{t,0}(Y_j,Y_k) = \frac{1}{n} \sum_{i=1}^{n} h_t(Z'_i,Y_j,Y_k)$, $h_{t}(x,y,z)= \lbrace \ind[d(x,y) \leq t]-F_x(t) \rbrace\lbrace \ind[d(x,z) \leq t]-F_x(t) \rbrace - \E_{Y}( \lbrace \ind[d(Y,y) \leq t]-F_Y(t) \rbrace\lbrace \ind[d(Y,z) \leq t]-F_Y(t) \rbrace )$ for $t \in [0,\M]$ with $Y \sim P_1$ independent of the data $Y_1, \dots, Y_n$ and $Z'_1, \dots, Z'_n \sim P_1$ and are independent of $Y_1, \dots, Y_n$.
\end{Lemma}
\noindent \textbf{Proof.} Observe that one has almost surely $ \left\lbrace \frac{1}{n} \sum_{j,k=1, j \neq k}^{n} g^{Z'}_{t,0}(Y_j,Y_k) \right\rbrace_{t \in [0,\M]}$ is a degenerate $U$-process of order two indexed by the function class $\mathcal{V}^{Z'}=\{(y,z) \mapsto g^{Z'}_{t,0}(y,z): t \in [0,\M]\}$ conditionally on $Z'_1, \dots, Z'_n$. In addition, note that
\begin{align*}
  & \left|g^{Z'}_{t,0}(y,z)-g^{Z'}_{s,0}(y,z)\right| \\ = & \left| \frac{1}{n} \sum_{i=1}^n h_s(Z'_i,y,z)- \frac{1}{n} \sum_{i=1}^n h_t(Z'_i,y,z) \right| \\ \leq & \denom[n]\add[in]  \left\lvert h_s(Z'_i,y,z)-h_t(Z'_i,y,z) \right\rvert.
\end{align*}
Using the bound on the functions $\lvert h_s(Z'_i,y,z)-h_t(Z'_i,y,z)\rvert$ similarly to what was derived in the proof of Lemma \ref{lma: 2sample_2} one has 
\begin{align*}
  & \denom[n]\add[in] \left|h_s(Z'_i,y,z)-h_t(Z'_i,y,z) \right| \\ \leq & \denom[n]\add[in] \lvert H_{st}(Z'_i,y,z)\rvert + \lvert \E_{Y}(H_{st}(Y,y,z)) \rvert
\end{align*}
whereby tracing the steps in the proof of Lemma \ref{lma: 2sample_2} conditionally on $Z'_1,\dots Z'_n$ and utilizing the independence of $Y_1, \dots, Y_n$ and $Z'_1,\dots Z'_n$ we obtain 
\begin{align*}
  \left\| h_s(Z'_i,Y',Y'')-h_t(Z'_i,Y',Y'') \right\|_{L_2(P_2 \times P_2)} \leq C_0 \sqrt{\epsilon}
\end{align*}
whenever $|s-t|<\epsilon$ for any $0<\epsilon<1$ and some constant $C_0>0$. This in turn implies that 
\begin{equation*}
  \lVert g^{Z'}_{t,0}(Y,Y'')-g^{Z'}_{s,0}(Y,Y'') |\rVert_{L_2(P_2 \times P_2)} \leq 2 C_0 \sqrt{\epsilon}. %\yc{\ [\text{should be }4 C_U (C_0+\sqrt{C_0}) \sqrt{\epsilon}]}
\end{equation*}
$|s-t|<\epsilon$. Conditional on $Z'_1, \dots, Z'_n$, %\yc{
the packing number $D(\epsilon,\mathcal{V}^{Z'},L_2(P_2 \times P_2))$, i.e. the maximum number of $\epsilon$-separate elements in $\mathcal{V}^{Z'}$ is upper bounded by the maximum number of $\tfrac{\epsilon^2}{4C^2_0}$-separate elements in $[0,\M]$, which equals %} 
a constant times $\epsilon^{-2}$. This implies that $\mathcal{V}^{Z'}$ is a Euclidean class as per Definition 3 of \cite{sher:94}. %by the equivalence of covering numbers and packing numbers. 
On the event $\left\lbrace \sup_{\omega_1,\omega_2 \in \Omega} \sup_{v \in \idx_c}\sup_{t \in [0,\M]} g^{Z'}_{t,v}(\omega_1,\omega_2) \leq t \right\rbrace$ an envelope function for $\mathcal{V}^{Z'}$ is the constant function $c$. By following the proof of Corollary 4 in \cite{sher:94} and using Markov's inequality, it turns out that for some $\alpha \in(0,1)$,
\begin{align*}
  &    \prob_{|\mathcal{Z}'}\left(  \sup_{t \in [0,\M]}  \left\lvert \frac{1}{n} \sum_{j,k=1, j \neq k}^{n} g^{Z'}_{t,0}(Y_j,Y_k)  \right\rvert > \epsilon , \sup_{\omega_1,\omega_2 \in \Omega} \sup_{t \in [0,\M]} g^{Z'}_{t,0}(\omega_1,\omega_2) \leq c \right)\\  
  \leq\ & \frac{\E_{|\mathcal{Z}'}\left(\sup_{t \in [0,\M]} \left\lvert \frac{1}{n} \sum_{j,k=1, j \neq k}^{n} g^{Z'}_{t,0}(Y_j,Y_k)  \right\rvert  \ind[\sup_{\omega_1,\omega_2 \in \Omega} \sup_{t \in [0,\M]} g^{Z'}_{t,0}(\omega_1,\omega_2) \leq c]\right)}{\epsilon} 
  \\ \leq\ & \mathrm{const.} \frac{c^\alpha}{\epsilon}, 
\end{align*}
almost surely. 

\begin{Lemma}\label{lma: 2sample_5}
Under $H_0$ and assumption A1 one has for any $\epsilon >0$ that as $n \rightarrow \infty$,
\begin{equation*}
  \prob\left(  \sup_{u \in \idx_c} \sup_{v \in \idx_c} \sup_{t \in [0,\M]}  \left \lvert \frac{1}{n^2}  \sum_{j=1}^{[nu]} \sum_{k=1}^{[nv]}  h_t(Y_j,Y_j,Z'_k) \right \rvert > \epsilon \right) \rightarrow 0
\end{equation*}
where $h_{t}(x,y,z)= \lbrace \ind[d(x,y) \leq t]-F_x(t) \rbrace\lbrace \ind[d(x,z) \leq t]-F_x(t) \rbrace - \E_{Y}( \lbrace \ind[d(Y,y) \leq t]-F_Y(t) \rbrace\lbrace \ind[d(Y,z) \leq t]-F_Y(t) \rbrace )$ for $t \in [0,\M]$.
\end{Lemma}
\noindent \textbf{Proof.} Note that 
\begin{align*}
    &  \prob\left(  \sup_{u \in \idx_c} \sup_{v \in \idx_c} \sup_{t \in [0,\M]}  \left \lvert \frac{1}{n^2}  \sum_{j=1}^{[nu]} \sum_{k=1}^{[nv]}  h_t(Y_j,Y_j,Z'_k) \right \rvert > \epsilon \right) \\ = &  \E_{Z'_1, \dots, Z'_n} \left\lbrace \prob_{Z'_1, \dots, Z'_n} \left(\sup_{u \in \idx_c} \sup_{v \in \idx_c} \sup_{t \in [0,\M]}  \left \lvert \frac{1}{n^2}  \sum_{j=1}^{[nu]} \sum_{k=1}^{[nv]}  h_t(Y_j,Y_j,Z'_k) \right \rvert > \epsilon \right) \right\rbrace \\ \leq & 2 \E_{Z'_1, \dots, Z'_n} \left\lbrace \prob_{Z'_1, \dots, Z'_n} \left( \sup_{v \in \idx_c} \sup_{t \in [0,\M]}  \left \lvert \frac{1}{n^2}  \sum_{j=1}^{n} \sum_{k=1}^{[nv]}  h_t(Y_j,Y_j,Z'_k) \right \rvert > \epsilon \right) \right\rbrace \\ \leq & 2 \E_{Z'_1, \dots, Z'_n} \left\lbrace \prob_{Z'_1, \dots, Z'_n} \left( \sup_{v \in \idx_c} \sup_{t \in [0,\M]} \sup_{\omega \in \Omega}  \left \lvert \frac{1}{n}  \sum_{k=1}^{[nv]}  h_t(\omega,\omega,Z'_k) \right \rvert > \epsilon \right) \right\rbrace \\ = & 2 \prob \left( \sup_{v \in \idx_c} \sup_{t \in [0,\M]}  \sup_{\omega \in \Omega} \left \lvert \frac{1}{n}  \sum_{k=1}^{[nv]}  h_t(\omega,\omega,Z'_k) \right \rvert > \epsilon \right)  \\ \leq & 4 \prob \left( \sup_{t \in [0,\M]} \sup_{\omega \in \Omega} \left \lvert \frac{1}{n}  \sum_{k=1}^{n}  h_t(\omega,\omega,Z'_k) \right \rvert > \epsilon \right)
\end{align*}
where we use L\'evy's inequality on page 431 of \cite{well:96} first on $u \in \idx_c$ and then on $v \in \idx_c$. Hence it is enough to show that $\prob \left( \sup_{t \in [0,\M]} \sup_{\omega \in \Omega} \left \lvert \frac{1}{n}  \sum_{k=1}^{n}  h_t(\omega,\omega,Z'_k) \right \rvert > \epsilon \right) \rightarrow 0$ as $n \rightarrow \infty$. Define $\tilde{F}_x(t)= \frac{1}{n} \sum_{k=1}^n \lbrace \ind[d(x,Z'_k) \leq t]-F_x(t) \rbrace$. With this observe that 
\begin{align*}
    &  \sup_{t \in [0,\M]} \sup_{\omega \in \Omega} \left \lvert \frac{1}{n}  \sum_{k=1}^{n}  h_t(\omega,\omega,Z'_k) \right \rvert \\ \leq & \sup_{t \in [0,\M]} \sup_{\omega \in \Omega} \left \lvert(1-F_\omega(t)) \tilde{F}_\omega(t) \right \rvert + \sup_{t \in [0,\M]} \sup_{\omega \in \Omega} \left \lvert\E_{Y} \lbrace \ind[d(Y,\omega) \leq t]-F_\omega(t) \rbrace\tilde{F}_Y(t)\rbrace  \right \rvert \\ \leq & 2 \sup_{t \in [0,\M]} \sup_{\omega \in \Omega} \left \lvert \tilde{F}_\omega(t) \right \rvert.
\end{align*}
By Theorem 5.2 in \cite{dube:24} one has $\sup_{t \in [0,\M]} \sup_{\omega \in \Omega} \left \lvert \tilde{F}_\omega(t) \right \rvert=o_P(1)$ as $n \rightarrow \infty$ which completes our proof.

\begin{Lemma}
\label{lma: aux_lemma_1}
    Suppose assumptions A1 and A2 hold. Then as $n \rightarrow \infty$, $\sup_{u \in \idx_c} \lvert \hat{T}_n(u) -T_n(u) \rvert = o_P(1)$. Under $H_{1,n}$, $\sup_{u \in \idx_c} \lvert \hat{T}_n(u) -T_n(u) \rvert = O_P\left(\frac{1}{n}+\sqrt{\frac{a_n}{n}}\right)$ as $n \rightarrow \infty$. 
\end{Lemma}
\noindent \textbf{Proof.} There are two cases, one when $u \leq \tau$ and two, when $u > \tau$.  We establish the steps for $u \leq \tau$, and by symmetry the arguments carry over for $u > \tau$. When $u \leq \tau$, observe that almost surely
\begin{align}
\label{eq: part1}
     (\prehatF[it]-\posthatF[it])-\frac{1-\tau}{1-u} (\Fone[Y_i](t)-\Ftwo[Y_i](t))
     = \Delta^1_{Y_i}(u,t) + \rem
\end{align}
where for $u < \tau$,
\begin{align}
    \label{eq: delta1}
    \Delta^1_{Y_i}(u,t) & = \frac{1}{[nu]}\sum_{j=1}^{[nu]} \{\ind[d(Y_i,Y_j) \leq t]-\Fone[Y_i](t)\} \nonumber \\ & - \frac{1}{1-u} \left\lbrace \frac{1}{n}\sum_{j=[nu]+1}^{[n\tau]} \{\ind[d(Y_i,Y_j) \leq t]-\Fone[Y_i](t)\} \right\rbrace \nonumber 
    \\ & - \frac{1-\tau}{1-u} \left\lbrace \frac{1}{n-[n\tau]}\sum_{j=[n\tau]+1}^{n} \{\ind[d(Y_i,Y_j) \leq t]-\Ftwo[Y_i](t)\}\right\rbrace,
\end{align}
and 
\begin{align*}
    \Delta^1_{Y_i}(\tau,t) & = \frac{1}{[n\tau]}\sum_{j=1}^{[n\tau]} \{\ind[d(Y_i,Y_j) \leq t]-\Fone[Y_i](t)\} \nonumber \\ & -  \left\lbrace \frac{1}{n-[n\tau]}\sum_{j=[n\tau]+1}^{n} \{\ind[d(Y_i,Y_j) \leq t]-\Ftwo[Y_i](t)\}\right\rbrace.
\end{align*}
Hence almost surely for $u \leq \tau$
\begin{align}
\label{eq: part2}
     & (\prehatF[it]-\posthatF[it])+\frac{1-\tau}{1-u} (\Fone[Y_i](t)-\Ftwo[Y_i](t)) \nonumber 
     \\ & = \Delta^1_{Y_i}(u,t) + \Delta^2_{Y_i}(u,t) +\rem
\end{align}
where 
\begin{equation}
 \label{eq: delta1}
    \Delta^2_{Y_i}(u,t) = 2\left( \frac{1-\tau}{1-u} \right) (\Fone[Y_i](t)-\Ftwo[Y_i](t)).
\end{equation}
Observe that $\sup_{u \in [c,\tau], t \in [0,\mathcal{M}]} \lvert \Delta_{Y_i}^1(u,t) \rvert$ and $\sup_{u \in [c,\tau], t \in [0,\mathcal{M}]} \lvert \Delta_{Y_i}^2(u,t) \rvert$ are uniformly upper bounded almost surely. Therefore combining the parts in equations \eqref{eq: part1} and \eqref{eq: part2} for $u \leq \tau$ one has almost surely that 
\begin{align*}
    & \hat{T}_n(u)-T_n(u) = a(u) \intgrt \denom[n] \add[in] \left\lbrace (\Delta^1_{Y_i}(u,t))^2 + \Delta^1_{Y_i}(u,t) \Delta^2_{Y_i}(u,t) \right\rbrace \ dt + \rem. 
\end{align*}
First see that 
\begin{align}
\label{eq: first term}
     \intgrt \left\lbrace \denom[n] \add[in]  (\Delta^1_{Y_i}(u,t))^2 \right\rbrace dt \leq \mathcal{M} \left( \sup_{\omega \in \Omega, u \in [c,\tau], t \in [0,\mathcal{M}]} \lvert \Delta_\omega^1(u,t) \rvert \right)^2.
\end{align}
Next see that by applying the Cauchy-Schwarz inequality one has
\begin{align*}
    & \denom[n]\add[in] \intgrt \lvert \Delta^1_{Y_i}(u,t) \Delta^2_{Y_i}(u,t) \rvert dt \\  \leq & \denom[n]\add[in] \sqrt{\intgrt (\Delta^1_{Y_i}(u,t))^2 dt} \sqrt{\intgrt (\Delta^2_{Y_i}(u,t))^2  dt} \\ \leq & 2 \left( \frac{1-\tau}{1-u} \right)  \sqrt{\mathcal{M}} \sup_{\omega \in \Omega, u \in [c,\tau], t \in [0,\mathcal{M}]} \lvert \Delta_\omega^1(u,t) \rvert  \left\lbrace \denom[n]\add[in] \sqrt{\intgrt (\Fone[Y_i](t)-\Ftwo[Y_i](t))^2  \ dt } \right \rbrace
    \\ \leq & 2 \left( \frac{1-\tau}{1-u} \right)  \sqrt{\mathcal{M}} \sup_{\omega \in \Omega, u \in [c,\tau], t \in [0,\mathcal{M}]} \lvert \Delta_\omega^1(u,t) \rvert   \sqrt{ \intgrt  \left\lbrace \denom[n]\add[in](\Fone[Y_i](t)-\Ftwo[Y_i](t))^2   \right \rbrace \ dt }.
\end{align*} 
For controlling the term $\intgrt  \left\lbrace \denom[n]\add[in](\Fone[Y_i](t)-\Ftwo[Y_i](t))^2   \right \rbrace \ dt$ observe that almost surely
\begin{align*}
  & \intgrt  \left\lbrace \denom[n]\add[in](\Fone[Y_i](t)-\Ftwo[Y_i](t))^2   \right \rbrace \ dt \\ = & \denom[n]\sum_{i=1}^{[n\tau]} \left[ \intgrt (\Fone[Y_i](t)-\Ftwo[Y_i](t))^2 dt- \tau \E_{Y_i \sim P_1}  \left\{\intgrt (\Fone[Y_i](t)-\Ftwo[Y_i](t))^2 dt\right\}\right] \\ & + \denom[n]\sum_{i=[n\tau]+1}^{n} \left[ \intgrt (\Fone[Y_i](t)-\Ftwo[Y_i](t))^2 dt- (1-\tau) \E_{Y_i \sim P_2}  \left\{\intgrt (\Fone[Y_i](t)-\Ftwo[Y_i](t))^2 dt\right\}\right] \\ & + \tau \E_{Y_i \sim P_1}  \left\{\intgrt (\Fone[Y_i](t)-\Ftwo[Y_i](t))^2 dt\right\} \\ & + (1-\tau) \E_{Y_i \sim P_2}  \left\{\intgrt (\Fone[Y_i](t)-\Ftwo[Y_i](t))^2 dt\right\}   + \rem.
\end{align*}
By the weak law of large numbers $$\left \lvert \denom[{n}]\sum_{i=1}^{[n\tau]} \left[ \intgrt (\Fone[Y_i](t)-\Ftwo[Y_i](t))^2 dt- \E_{Y_i \sim P_1}  \left\{\intgrt (\Fone[Y_i](t)-\Ftwo[Y_i](t))^2 dt\right\}\right] \right \rvert = O_P \left( \frac{1}{\sqrt{n}}\right)$$ and $$\left \lvert \denom[n]\sum_{i=[n\tau]+1}^{n} \left[ \intgrt (\Fone[Y_i](t)-\Ftwo[Y_i](t))^2 dt- \E_{Y \sim P_2}  \left\{\intgrt (\Fone[Y_i](t)-\Ftwo[Y_i](t))^2 dt\right\}\right] \right \rvert = O_P \left( \frac{1}{\sqrt{n}}\right).$$Hence for sufficiently large $n$ one has almost surely that
\begin{equation}
\label{eq: T_n_bound}
    \frac{\min(\tau,1-\tau) \Delta}{2} \leq \intgrt  \left\lbrace \denom[n]\add[in](\Fone[Y_i](t)-\Ftwo[Y_i](t))^2   \right \rbrace \ dt \leq 2 \max(\tau,1-\tau) \Delta 
\end{equation}
where $\Delta$ is as defined in equation 6 in the paper. Combining these arguments one has almost surely for some constant $C_1>0$ and sufficiently large $n$,
\begin{equation}
\label{eq: second term}
    \sup_{u \in [c,\tau]}\denom[n]\add[in] \intgrt \lvert \Delta^1_{Y_i}(u,t) \Delta^2_{Y_i}(u,t) \rvert dt  \ \leq \ C_1 \sup_{\omega \in \Omega, u \in [c,\tau], t \in [0,\mathcal{M}]} \lvert \Delta_\omega^1(u,t) \rvert \sqrt{\Delta} . 
\end{equation}
Finally see that combining \eqref{eq: first term} and \eqref{eq: second term} one has almost surely that for some constant $C_2 > 0$ and sufficiently large $n$,
\begin{align}
\label{eq: final term}
   &  \sup_{u \in [c,\tau]} \lvert \hat{T}_n(u)-T_n(u) \rvert \nonumber \\ \leq  & C_2  \left[ \left( \sup_{\omega \in \Omega, u \in [c,\tau], t \in [0,\mathcal{M}]} \lvert \Delta_\omega^1(u,t) \rvert \right)^2 + \sup_{\omega \in \Omega, u \in [c,\tau], t \in [0,\mathcal{M}]} \lvert \Delta_\omega^1(u,t) \rvert \sqrt{\Delta}  \right] + \rem.
\end{align}
Under assumptions A1 and A2 using Theorem 5.2 in \citep{dube:24} one has $$\sup_{\omega \in \Omega, t \in [0,\mathcal{M}]} \lvert \Delta_\omega^1(\tau,t) \rvert = O_P\left(\frac{1}{\sqrt{n}}\right).$$ This in conjunction with the L\'evy's inequality on page 431 of \cite{well:96} implies that $\sup_{\omega \in \Omega, u \in [c,\tau), t \in [0,\mathcal{M}]} \lvert \Delta_\omega^1(u,t) \rvert = O_P\left(\frac{1}{\sqrt{n}}\right)$. Plugging this into \eqref{eq: final term} immediately completes the proof as under $H_{1,n}$, $\Delta = a_n$.

\begin{Lemma}
\label{lma: aux_lemma_2}
    Suppose that assumptions A1 and A2 hold and $\delta_n \rightarrow 0$ be a sequence such that  $n\delta_n \rightarrow \infty$ as $n \rightarrow \infty$. Then one has $\E \left( \sup_{|u-\tau| \leq \delta_n} \lvert \hat{T}_n(u) -\hat{T}_n(\tau)-T_n(u)+T_n(\tau) \rvert \right) = O\left( \sqrt{\frac{\delta_n}{n}}\right)$ under $H_1$, and  \ $\E \left( \sup_{|u-\tau| \leq \delta_n} \lvert \hat{T}_n(u) -\hat{T}_n(\tau)-T_n(u)+T_n(\tau) \rvert \right) = O\left( \sqrt{\frac{\delta_n a_n}{n}} \right)+\rem$ under $H_{1,n}$ as $n \rightarrow \infty$. 
\end{Lemma}
\noindent \textbf{Proof.} To simplify notations we will use $C$ as a generic symbol for constants used in this proof. Like in the proof of Lemma \ref{lma: aux_lemma_1}, there are two cases, one when $u \leq \tau$ and two, when $u > \tau$.  We establish the steps for $u \leq \tau$, and by symmetry the arguments carry over for $u > \tau$. Borrowing the notations from the proof of Lemma \ref{lma: aux_lemma_1} one has almost surely that
\begin{align*}
    & \left \lvert \hat{T}_n(u) -\hat{T}_n(\tau)-T_n(u)+T_n(\tau) \right\rvert \\ = & \left \lvert a(u) \intgrt \denom[n] \add[in] \left\lbrace (\Delta^1_{Y_i}(u,t))^2 + \Delta^1_{Y_i}(u,t) \Delta^2_{Y_i}(u,t) \right\rbrace \ dt  \right. \\ & \left. - a(\tau) \intgrt \denom[n] \add[in] \left\lbrace (\Delta^1_{Y_i}(\tau,t))^2 + \Delta^1_{Y_i}(\tau,t) \Delta^2_{Y_i}(\tau,t) \right\rbrace \ dt   \right \rvert + \rem \\ \leq & I + II  + \rem
\end{align*}
using the triangle inequality where 
\begin{equation*}
    I = \left \lvert a(u) \intgrt \denom[n] \add[in]  (\Delta^1_{Y_i}(u,t))^2 \ dt  - a(\tau) \intgrt \denom[n] \add[in]  (\Delta^1_{Y_i}(\tau,t))^2 \ dt  \right \rvert 
\end{equation*}
and 
\begin{equation*}
    II = \left \lvert \intgrt \denom[n] \add[in] \left \lbrace a(u) \Delta^1_{Y_i}(u,t) \Delta^2_{Y_i}(u,t) - a(\tau) \Delta^1_{Y_i}(\tau,t) \Delta^2_{Y_i}(\tau,t) \right\rbrace \ dt   \right\rvert.
\end{equation*}
For controlling term I, one has that $\sup_{u \in [\tau-\delta_n,\tau]} I \leq C \left( \sup_{\omega \in \Omega, u \in [c,\tau], t \in [0,\mathcal{M}]} \lvert \Delta_\omega^1(u,t) \rvert \right)^2$ by using \eqref{eq: first term}. Using the arguments in the last paragraph of Lemma \ref{lma: aux_lemma_1}, $\sup_{u \in [\tau-\delta_n,\tau]} I = O_P \left( \frac{1}{n} \right)$. Next we will study the behavior of the term $\sup_{u \in [\tau-\delta_n,\tau]} II$ under fixed $H_1$ and under $H_{1,n}$ as $n \rightarrow \infty$. 

\noindent Observe that using the triangle inequality term $II$ can be upper bounded as
\begin{equation*}
    II \leq III + IV + V
\end{equation*}
where 
\begin{equation*}
    III = \left \lvert \intgrt \denom[n] \add[in] \left \lbrace a(u) - a(\tau) \right \rbrace  \Delta^1_{Y_i}(u,t) \Delta^2_{Y_i}(u,t)  \ dt   \right\rvert,
\end{equation*}
\begin{equation*}
    IV = \left \lvert \intgrt \denom[n] \add[in]  a(\tau) \Delta^1_{Y_i}(u,t) \left \lbrace   \Delta^2_{Y_i}(u,t) -  \Delta^2_{Y_i}(\tau,t) \right\rbrace \ dt   \right\rvert
\end{equation*}
and 
\begin{equation*}
    V = \left \lvert \intgrt \denom[n] \add[in]  a(\tau) \Delta^2_{Y_i}(\tau,t)\left \lbrace  \Delta^1_{Y_i}(u,t)  - \Delta^1_{Y_i}(\tau,t)  \right\rbrace \ dt   \right\rvert. 
\end{equation*}
We will study the terms $III, IV$ and $V$ under fixed $H_1$ and under $H_{1,n}$ as $n \rightarrow \infty$ to complete the proof.

\noindent \underline{Term III:} Observe that
\begin{align*}
     \sup_{u \in [\tau-\delta_n,\tau]} III & \leq  \sup_{u \in [\tau-\delta_n,\tau]} \lvert a(u)-a(\tau) \rvert \sup_{u \in [\tau-\delta_n,\tau]} \left \lvert \intgrt \denom[n] \add[in]  \Delta^1_{Y_i}(u,t) \Delta^2_{Y_i}(u,t)  \ dt   \right\rvert \\ & \leq  C \delta_n \sup_{\omega \in \Omega, u \in [c,\tau], t \in [0,\mathcal{M}]} \lvert \Delta_\omega^1(u,t) \rvert \sqrt{\Delta}
\end{align*}
using \eqref{eq: second term} and the fact that $\sup_{u \in [\tau-\delta_n,\tau]} \lvert a(u)-a(\tau) \rvert = O(\delta_n)$. Using arguments in the last paragraph of Lemma \ref{lma: aux_lemma_1}, $\sup_{\omega \in \Omega, u \in [c,\tau], t \in [0,\mathcal{M}]} \lvert \Delta_\omega^1(u,t) \rvert = O_P\left( \frac{1}{\sqrt{n}}\right)$. Combining all the steps under fixed $H_1$, $\sqrt{\Delta}$ is a constant, and therefore $ \sup_{u \in [\tau-\delta_n,\tau]} III = O_P\left( \frac{\delta_n}{\sqrt{n}}\right)$. Under $H_{1,n}$, $\sqrt{\Delta}=\sqrt{a_n}$, and therefore $ \sup_{u \in [\tau-\delta_n,\tau]} III = O_P\left( \frac{\delta_n\sqrt{a_n}}{\sqrt{n}}\right)$.

\noindent \underline{Term IV:} Observe that
\begin{align*}
     \Delta^2_{Y_i}(u,t) -  \Delta^2_{Y_i}(\tau,t)  & = 2 \left \lbrace \frac{1-\tau}{1-u} - \frac{1-\tau}{1-\tau}\right \rbrace \left( \Fone[Y_i](t)-\Ftwo[Y_i](t)\right) \\ & = \frac{1}{1-\tau} \left( u-\tau \right) \Delta^2_{Y_i}(u,t)
\end{align*}
and therefore
\begin{align*}
     \sup_{u \in [\tau-\delta_n,\tau]}  IV & = \sup_{u \in [\tau-\delta_n,\tau]}  \left \lvert a(\tau) \frac{1}{1-\tau} \left( u-\tau \right) \intgrt \denom[n] \add[in]   \Delta^1_{Y_i}(u,t) \Delta^2_{Y_i}(u,t) \ dt   \right\rvert \\ & \leq C \delta_n \sup_{u \in [\tau-\delta_n,\tau]} \left \lvert \intgrt \denom[n] \add[in]  \Delta^1_{Y_i}(u,t) \Delta^2_{Y_i}(u,t)  \ dt   \right\rvert \\ & \leq  C \delta_n \sup_{\omega \in \Omega, u \in [c,\tau], t \in [0,\mathcal{M}]} \lvert \Delta_\omega^1(u,t) \rvert \sqrt{\Delta}
\end{align*}
using \eqref{eq: second term}. Hence the asymptotic behavior of term $IV$ is same as that of term $III$.

\noindent \underline{Term V:} 
Observe that
\begin{align*}
    V \leq a(\tau) \sup_{\omega \in \Omega, t \in [0,\M]} \left \lvert  \Delta^1_{\omega}(u,t)  - \Delta^1_{\omega}(\tau,t)  \right\rvert \intgrt \denom[n] \add[in]    \left \lvert  \Delta^2_{Y_i}(\tau,t)\right\rvert dt.
\end{align*}
Next see that using the Cauchy-Schwarz inequality
\begin{align*}
     \intgrt \denom[n] \add[in]    \left \lvert  \Delta^2_{Y_i}(\tau,t)\right\rvert dt & = 2   \denom[n] \add[in]  \intgrt   \left \lvert  \Fone[Y_i](t)-\Ftwo[Y_i](t) \right\rvert dt \\ & \leq 2 \sqrt{\M} \denom[n] \add[in]  \sqrt{\intgrt   \left \lbrace  \Fone[Y_i](t)-\Ftwo[Y_i](t) \right\rbrace^2 dt} \\ & \leq 2 \sqrt{\M} \sqrt{\denom[n] \add[in]  \intgrt   \left \lbrace  \Fone[Y_i](t)-\Ftwo[Y_i](t) \right\rbrace^2 dt} \\ & \leq 2 \sqrt{\M} \sqrt{2\max(\tau,1-\tau) \Delta} 
\end{align*}
by \eqref{eq: T_n_bound}. Hence one has
\begin{equation*}
    \sup_{u \in [\tau-\delta_n,\tau]} V \leq C \sqrt{\Delta} \sup_{u \in [\tau-\delta_n,\tau],\omega \in \Omega, t \in [0,\M]} \left \lvert  \Delta^1_{\omega}(u,t)  - \Delta^1_{\omega}(\tau,t)  \right\rvert.
\end{equation*}
Let $h_{\omega,t}(x)= \ind[d(\omega,x)\leq t]-F_\omega(t)$. With this for any $u \leq \tau$,
\begin{equation*}
    \Delta^1_{\omega}(u,t) = \frac{1}{[nu]} \sum_{j=1}^{[nu]} h_{\omega,t}(Y_j)-\frac{1}{(n-[nu])} \sum_{j=[nu]+1}^{n} h_{\omega,t}(Y_j).
\end{equation*}
Therefore
\begin{align*}
    & \left \lvert  \Delta^1_{\omega}(u,t)  - \Delta^1_{\omega}(\tau,t)  \right\rvert \\  =  & \left \lvert \left(\frac{1}{[nu]} - \frac{1}{[n\tau]}\right) \sum_{j=1}^{[nu]} h_{\omega,t}(Y_j) -\frac{1}{[n\tau]} \sum_{j=[nu]+1}^{[n\tau]} h_{\omega,t}(Y_j) \right. \\ & - \left. \frac{1}{(n-[nu])} \sum_{j=[nu]+1}^{[n\tau]} h_{\omega,t}(Y_j) -  \left(\frac{1}{(n-[nu])} - \frac{1}{(n-[n\tau])}\right) \sum_{j=[n\tau]+1}^{n} h_{\omega,t}(Y_j)\right \rvert \\ & \leq \rem + \left \lvert \tau-u\right \rvert \left \lvert  \frac{1}{n} \sum_{j=1}^{[nu]} h_{\omega,t}(Y_j) + \frac{1}{n} \sum_{j=[n\tau]+1}^{n} h_{\omega,t}(Y_j)\right\rvert + C \left \lvert  \frac{1}{n} \sum_{j=[nu]+1}^{[n\tau]} h_{\omega,t}(Y_j) \right\rvert.
\end{align*}
Using L\'evy's inequality on page 431 of \cite{well:96} together with Theorem 5.2 in \citep{dube:24} one has that 
\begin{equation*}
    \sup_{u \in [\tau-\delta_n,\tau]}\sup_{\omega \in \Omega, t \in [0,\M]} \left \lvert  \frac{1}{n} \sum_{j=1}^{[nu]} h_{\omega,t}(Y_j) + \frac{1}{n} \sum_{j=[n\tau]+1}^{n} h_{\omega,t}(Y_j)\right\rvert = O_P\left( \frac{1}{\sqrt{n}}\right)
\end{equation*}
and therefore as $n \rightarrow \infty$
\begin{equation*}
     \sup_{u \in [\tau-\delta_n,\tau]}\sup_{\omega \in \Omega, t \in [0,\M]}  \left \lvert \tau-u\right \rvert \left \lvert  \frac{1}{n} \sum_{j=1}^{[nu]} h_{\omega,t}(Y_j) + \frac{1}{n} \sum_{j=[n\tau]+1}^{n} h_{\omega,t}(Y_j)\right\rvert  = O_P\left( \frac{\delta_n}{\sqrt{n}}\right).
\end{equation*}
Next we claim that $ \sup_{u \in [\tau-\delta_n,\tau]}\sup_{\omega \in \Omega, t \in [0,\M]}  \left \lvert  \frac{1}{n} \sum_{j=[nu]+1}^{[n\tau]} h_{\omega,t}(Y_j) \right\rvert= O_P\left( \frac{\sqrt{\delta_n}}{\sqrt{n}}\right)$. We will show that 
\begin{align*}
    \begin{split}
    \prob \left( \sup_{u \in [\tau-\delta_n,\tau]}\sup_{\omega \in \Omega, t \in [0,\M]}\left \lvert \frac{1}{\sqrt{n\delta_n}} \sum_{j=[nu]+1}^{[n\tau]} h_{\omega,t}(Y_j) \right\rvert > L \right) \rightarrow 0
    \end{split}
\end{align*}
as $n \rightarrow \infty$ and $L \rightarrow \infty$. Note that since $Y_1, Y_2, \dots, Y_{[n\tau]}$ are i.i.d,
%$\prob \left( \sup_{u \in [\tau-\delta_n,\tau]}\sup_{\omega \in \Omega, t \in [0,\M]}\\  \left \lvert \frac{1}{\sqrt{n\delta_n}} \sum_{j=[nu]+1}^{[n\tau]} h_{\omega,t}(Y_j) \right\rvert > L \right) \rightarrow 0$ as $n \rightarrow \infty$ and $L \rightarrow \infty$. Note that since $Y_1, Y_2, \dots, Y_{[n\tau]}$ are i.i.d, 
\begin{align*}
    & \prob \left( \sup_{u \in [\tau-\delta_n,\tau]}\sup_{\omega \in \Omega, t \in [0,\M]}  \left \lvert  \frac{1}{\sqrt{n\delta_n}} \sum_{j=[nu]+1}^{[n\tau]} h_{\omega,t}(Y_j) \right\rvert > L \right)   \\ = & \prob \left( \sup_{u \in [0,\delta_n]}\sup_{\omega \in \Omega, t \in [0,\M]}  \left \lvert  \frac{1}{\sqrt{n\delta_n}} \sum_{j=1}^{[nu]} h_{\omega,t}(Y_j) \right\rvert > L \right) \\ = & \prob \left( \max_{1 \leq k \leq [n\delta_n]}\sup_{\omega \in \Omega, t \in [0,\M]}  \left \lvert  \frac{1}{\sqrt{n\delta_n}}\sum_{j=1}^{k} h_{\omega,t}(Y_j) \right\rvert > L \right).
\end{align*}
By L\'evy's inequality on page 431 of \cite{well:96} one has
\begin{align*}
    & \prob \left( \sup_{u \in [\tau-\delta_n,\tau]}\sup_{\omega \in \Omega, t \in [0,\M]}  \left \lvert  \frac{1}{\sqrt{n\delta_n}} \sum_{j=[nu]+1}^{[n\tau]} h_{\omega,t}(Y_j) \right\rvert > L \right)    \\ \leq & 2 \prob \left( \sup_{\omega \in \Omega, t \in [0,\M]}  \left \lvert  \frac{1}{\sqrt{n\delta_n}} \sum_{j=1}^{[n\delta_n]} h_{\omega,t}(Y_j) \right\rvert > L \right) \rightarrow 0
\end{align*}
as $n \rightarrow \infty$ and $L \rightarrow \infty$ by Theorem 5.2 in \citep{dube:24}. Hence one has
\begin{equation*}
   \sup_{u \in [\tau-\delta_n,\tau]}\sup_{\omega \in \Omega, t \in [0,\M]}  \left \lvert  \Delta^1_{\omega}(u,t)  - \Delta^1_{\omega}(\tau,t)  \right\rvert = O_P\left( \frac{\sqrt{\delta_n}}{\sqrt{n}}\right)
\end{equation*}
as when $n \rightarrow \infty$, $\delta_n \rightarrow 0$ and $n \delta_n \rightarrow \infty$, $\frac{\sqrt{\delta_n}}{\sqrt{n}}$ dominates both $\frac{\delta_n}{\sqrt{n}}$ and $\frac{1}{n}$. Combining all the steps under fixed $H_1$, $\sqrt{\Delta}$ is a constant, and therefore $ \sup_{u \in [\tau-\delta_n,\tau]} V = O_P\left( \frac{\sqrt{\delta_n}}{\sqrt{n}}\right)$. Under $H_{1,n}$, $\sqrt{\Delta}=\sqrt{a_n}$, and therefore $ \sup_{u \in [\tau-\delta_n,\tau]} V = O_P\left( \frac{\sqrt{\delta_n}\sqrt{a_n}}{\sqrt{n}}\right)$.

\noindent Combining the conclusions for terms $III, IV$ and $V$ together with term $I$ completes the proof. 

\bibliographystyle{abbrvnat}
\bibliography{JASA_template/reference}